\documentclass[preprint,12pt]{aastex}

\usepackage{booktabs}

\makeatletter
\newsavebox{\@tabnotebox}
\providecommand\tnote{}
\newenvironment{tabularwithnotes}[3][c]
  {\long\def\@tabnotes{#3}%
   \renewcommand\tnote[2][a]{\textsuperscript{\itshape##1}\,##2\par}
   \begin{lrbox}{\@tabnotebox}
   \begin{tabular}{#2}}
  {\end{tabular}\end{lrbox}%
   \parbox{\wd\@tabnotebox}{
     \usebox{\@tabnotebox}\par
     \smallskip\@tabnotes
   }%
  }
\makeatother

\begin{document}

\title{Rings and Radial Waves in the Disk of the Milky Way}

\author{
Yan Xu\altaffilmark{\ref{NAOC},\ref{RPI}},
Heidi Jo Newberg\altaffilmark{\ref{RPI}},
Jeffrey L. Carlin\altaffilmark{\ref{RPI}}, 
Chao Liu\altaffilmark{\ref{NAOC}},
Licai Deng\altaffilmark{\ref{NAOC}},
Jing Li\altaffilmark{\ref{SHAO}},
Ralph Sch{\"o}nrich \altaffilmark{\ref{Oxford}}, \&
Brian Yanny\altaffilmark{\ref{FNAL}}
}

\altaffiltext{1}{National Astronomical Observatories, Chinese Academy of Sciences, Beijing 100012, PR China;xuyan@bao.ac.cn\label{NAOC}}
\altaffiltext{2}{Dept. of Physics, Applied Physics and Astronomy, Rensselaer Polytechnic Institute Troy, NY 12180\label{RPI}}
\altaffiltext{3}{Key Laboratory for Research in Galaxies and Cosmology, Shanghai Astronomical Observatory, Chinese Academy of Sciences,80 Nandan Road, Shanghai 200030, China\label{SHAO}}
\altaffiltext{4}{Rudolf-Peierls Centre for Theoretical Physics, University of Oxford, 1 Keble Road, OX1 3NP, UK\label{Oxford}}
\altaffiltext{5}{Experimental Astrophysics Group, Fermi National Accelerator Laboratory,PO Box 500, Batavia, IL 60510\label{FNAL}}


\shortauthors{Xu et al.}

\begin{abstract}

We show that in the anticenter region, between Galactic longitudes of $110^\circ<l<229^\circ$, there is an  oscillating asymmetry in the main sequence star counts on either side of the Galactic plane using data from the Sloan Digital Sky Survey.  This asymmetry oscillates from more stars in the north at distances of about 2 kpc from the Sun to more stars in the south at 4-6 kpc from the Sun to more stars in the north at distances of 8-10 kpc from the Sun.  We also see evidence that there are more stars in the south at distances of 12-16 kpc from the Sun.  The three more distant asymmetries form roughly concentric rings around the Galactic center, opening in the direction of the Milky Way's spiral arms.  The northern ring, 9 kpc from the Sun, is easily identified with the previously discovered Monoceros Ring.  Parts of the southern ring at 14 kpc from the Sun (which we call the TriAnd Ring) have previously been identified as related to the Monoceros Ring and others have been called the Triangulum Andromeda Overdensity.  The two nearer oscillations are approximated by a toy model in which the disk plane is offset by of the order 100 pc up and then down at different radii.  We also show that the disk is not azimuthally symmetric around the Galactic anticenter and that there could be a correspondence between our observed oscillations and the spiral structure of the Galaxy.   Our observations suggest that the TriAnd and Monoceros Rings (which extend to at least 25 kpc from the Galactic center) are primarily the result of disk oscillations. 

\end{abstract}

\section{Introduction}

The disk of the Milky Way galaxy is typically modeled with two double exponential density functions - one for the thin disk and one for the thick disk.  This density function is a good fit to the star counts regardless of whether there are two distinct components, or a continuum from lower scale height and higher metallicity to higher scale height and lower metallicity.  However, recent studies of the disk component of the Milky have uncovered disk warps, asymmetries in the star counts above and below the disk, and bulk motion of the disk that is inconsistent with our exponential disk profile.  

More than a decade ago, an apparent ring of stars was discovered at low Galactic latitude near the Galactic anti-center, at an estimated distance of 18 kpc from the Galactic center \citep{2002ApJ...569..245N, 2003ApJ...588..824Y, 2012ApJ...757..151L}; this structure has been called the Monoceros Ring, but some authors have called it the Monoceros Overdensity or the Galactic Anticenter Stellar Stream \citep{2003ApJ...594L.119C, 2003ApJ...594L.115R}.  After more than a decade, it is still uncertain whether this structure is associated with the disk (a warp, flare, wave perturbation or spiral arm) or the halo component of the Milky Way (dwarf galaxy accretion).  Initially, the community was divided over the origin of the Monoceros Ring, which was thought to be associated with an accreting dwarf galaxy \citep{2004MNRAS.348...12M, 2005ApJ...626..128P}, or the normal Galactic warp/flare profiles \citep{2004A&A...421L..29M, 2006A&A...451..515M}.  

More recently the possibility that the structures are a result of an encounter with a massive satellite \citep{2008ApJ...688..254K, 2008ApJ...676L..21Y} has been put forward; it is suggested that the massive satellite could be the Sagitarius dwarf galaxy \citep{2011Natur.477..301P}.  Evidence for oscillations of the disk in Sloan Digital Sky Survey Data (SDSS; York et al. 1999) was put forward by \citet{2012MNRAS.423.3727G, 2012ApJ...750L..41W, 2013ApJ...777...91Y}.  Velocity substructure in the disk that could be caused by a wavelike perturbation of the disk was also seen by \citet{2013MNRAS.436..101W} in data from the RAdial Velocity Experiment \citep{2006AJ....132.1645S}, and by \citet{2013ApJ...777L...5C} in data from the Large Area Multi-Object Spectroscopic Telescope (LAMOST; Cui et al. 2012, Deng et al. 2012, Zhao et al. 2012) survey.  The debate over the origin of the velocity structure has just begun; \citet{2014MNRAS.440.2564F} have recently suggested that both radial and vertical flows, as seen by these surveys, can be induced by Galactic spiral arms.   However, their models produce vertical flows from heating rather than vertical motion of the disk midplane.  This contrasts with the affects of satellites; satellites passing through the disk with lower infall velocity produce bending modes while satellites with larger infall velocities produce breathing modes \citep{2014MNRAS.440.1971W}.  Bending modes would cause bulk vertical motion of the disk midplane.

At the same time, the debate over the nature of the Monoceros Ring still goes on.  \citet{2011ApJ...730L...6S} look at deep imaging with the CFHT MegaCam, and conclude that the density is not consistent with extragalactic models; thus they favor an accretion origin for the structure.  \citet{2012ApJ...754..101C} look at the star counts with deep imaging from Subaru, and claim to see a wall of stars at 10 kpc from the Sun, with a distance that does not change with longitude.  They claim that the Monoceros ring, at ([Fe/H]$\sim -1.0$), is more metal-rich than nearby stars, and that the perturbed disk model does not reproduce the observed density profile.  Since they expect a flare would produce too high a metallicity, they therefore also favor a satellite origin for the structure.  One difficulty with a satellite origin is that it is difficult to get an accreting satellite into a circular orbit.  However, \citet{2011MNRAS.414L...1M} found a way to put the progenitor of the Monoceros ring into a circular orbit by having it collide with the Sagittarius dwarf (or similar) galaxy.  On the other hand, \citet{2014A&A...567A.106L} argue that the disk flares can explain the Monoceros Ring, and also explain why the disk appears to end 15 kpc from the Galactic center.  In their words, ``the hypothesis to interpret the Monoceros ring in terms of a tidal stream of a putative accreted dwarf galaxy is not only unnecessary because the observed flare explains the overdensity in the Monoceros ring observed in SDSS fields, but it appears to be inappropriate."

The Monoceros Ring is not the only unexplained density structure near the Galactic anticenter.  \citet{2004ApJ...615..732R, 2004ApJ...615..738M} identified a density structure (in M giants and main sequence stars, respectively) at low (absolute) latitude, south of the Galactic plane, with Galactic longitude of $100^\circ<l<150^\circ$.  Dubbed the Triangulum Andromeda Stream (TriAnd), this structure is 20-30 kpc from the Sun and further from the Galactic center than the Monoceros Ring.  Martin et al. (2007) identify two faint main sequences, in the foreground of the Andromeda galaxy, and dub the more distant, lower contrast one ``TriAnd2."  Although it has been suggested that the Monoceros Ring and TriAnd might be different parts of the same tidal stream \citep{2005ApJ...626..128P}, \citet{2011ApJ...731L..30C} find different metallicities for the two substructures.  

Recently, several papers have presented evidence for substructure within TriAnd.  \citet{2014ApJ...787...19M} map five structures in the direction of the Andromeda galaxy using the Pan-Andromeda Archaeological Survey (PAndAS; McConnachie et al. 2009).  In addition to structures that they identify as the Monoceros Ring (Ibata et al. 2003; though we argue below that this is {\it not} the Monoceros Ring), the TriAnd Stream (Rocha-Pinto et al. 2003), TriAnd2 (Martin et al. 2007), the Pisces globular cluster stream (Bonaca et al. 2012; Martin et al. 2013); they find two odd substructures that look like a narrow stellar stream (the PAndAS MW stream) and a ``blob" (the PAndAS NE blob) at the same distance as the TriAnd Stream.  \citet{2014ApJ...793...62S} find considerably different ages (and possibly metallicity) from main sequence fitting of TriAnd and TriAnd2 (using 2MASS M giants and optical main sequence /turnoff features), and suggest that they could be debris from the same dwarf galaxy that was torn off on different pericentric passages near the Galactic center.  \citet{2014MNRAS.444.3975D} use data from the Spectroscopic and Photometric Landscape of Andromeda's Stellar Halo (SPLASH;Guhathakurta et al. 2006;Gilbert et al. 2012) and Sloan Extension for Galactic Understanding and Exploration (SEGUE; Yanny et al. 2009) spectroscopic surveys to suggest that TriAnd may be the result of group infall to the Milky Way.  They note that the PAndAS stream and the SEGUE 2 dwarf galaxy \citep{2009MNRAS.397.1748B} have the same position and line-of-sight velocity as TriAnd.

Substructure is also apparent in the Monoceros Ring \citep{2012ApJ...757..151L}.  \citet{2006ApJ...651L..29G} finds at least three substructures in the Monoceros Ring, including two or more narrow substreams as well as a broader, diffuse component.  He suggests ``the progenitor of the stream was likely a dwarf galaxy of significant size and mass."
\citet{2011ApJ...738...98G} find an eccentric orbit and possible progenitor for the Eastern Banded Structure (EBS), that is also at low Galactic latitude and at the same distance as the Monoceros Ring.  More recently, data from Pan-STARRS1 (K. C. Chambers et al., in preparation) was used to create maps of the stellar density near the Galactic anticenter, showing that the Monoceros Ring is observed over $130^\circ$ of Galactic longitude, in the range $-25^\circ<b<30^\circ$ \citep{2014ApJ...791....9S}.  They show that this structure exhibits complex structure; the southern part is $\sim 5.5$ kpc from the Sun, while the northern part is $\sim 9.5$ kpc from the Sun.  They show that current models of tidal disruption and disk distortion do not reproduce the observed substructure. However, it is possible that fine tuning of the parameters in these models to match the new data could yield better agreement.

The evidence for tidal debris in the Monoceros and TriAnd structures, especially with the suggestion that tidal debris may have come into the Milky Way in groups, leads some authors to conclude these larger structures are formed through accretion.  Clearly, other authors are adamant that at least the Monoceros Ring is due to a flaring of the disk.  And more recently evidence is put forward that disk wiggles (which could themselves be influenced by dwarf galaxy accretion or spiral arms) could be associated with the Monoceros Ring.  

It is with this history that we re-examine a mystery that existed in the original paper that showed for the first time that the halo of the Milky Way is lumpy, and that there is an apparent ring of stars 20 kpc from the center of the Milky Way (the Monoceros Ring).
The wedge plot in Figure 1 of Newberg et al. (2002) also shows other structure near the Galactic plane.  This figure shows the density of turnoff stars with $0.1<(g-r)_0<0.3$ on the Celestial Equator.  In particular, if one looks in the direction $(l,b)=(223^\circ, 20^\circ)$ in that figure, there are two apparent ``lumps" standing out from the smooth background: the one at $g_0 \sim 19.5$ is associated with the Monoceros Ring, but there is another similar-looking overdensity at $g_0 \sim 16.5$.  In the direction $(l,b)=(200^\circ,-24^\circ)$, there is an apparent overdensity at $g_0 \sim 17.5$, in addition to the structure at $g_0 \sim 19.8$ that is identified in this paper as possibly part of what is now called the Monoceros Ring. 


In the Newberg et al. (2002) paper, the nearer overdensities were assumed to be somehow associated with the disk of the Milky Way, while the more distant Monoceros Ring was not.  This is because a reasonable thick disk model could produce a peak in the star counts at bright magnitudes and low Galactic latitudes; it could not reproduce the Monoceros Ring in part because the disk was not thought to extend past 15 kpc from the Galactic center.  Additionally, a specially tuned flaring of the disk at 20 kpc from the Galactic center was necessary to match the high scale height and narrow apparent magnitude distribution of the stars in the ring.

In this paper, we set out to determine whether the nearer low Galactic latitude structures, both above and below the Galactic plane, could be explained by a thin or thick disk.  We found that, like the Monoceros structure, they did not fit in with a standard disk model.  In particular, there is an asymmetry in the star counts above and below the Galactic plane, and that asymmetry shifts with distance from the Galactic center.

This result is introduced in Section 2, where we show that if one subtracts a Hess diagram of the stellar population south of the Galactic plane from a similar Hess diagram of the stellar population at the same angle above the Galactic plane, a pattern of alternating black and white stripes is produced.  Each stripe is aligned in the direction expected for main sequence stars.  Where the stripes are white, there is an overdensity of main sequence stars in the north, and where they are black there is an overdensity of main sequence stars in the south.  The alternating overdensities with apparent magnitude suggest that the asymmetry alternates with distance from us.  We name the new overdensities the ``north near structure," and the ``south middle structure."  Two other overdensities are identified with previously named structures: ``Monoceros" in the north and ``TriAnd" in the south.

Sections 3, 4, and 5 describe the metallicity, spatial extent, and velocity distribution of the stars in the north near and south middle structures.  We conclude that the stars in these structures are consistent with our expectations for thin and thick disk stars, even though their spatial distribution of the stars is asymmetric.  

In Section 6, we experiment with an oscillating vertical perturbation of a double exponential model to see if we can match the observed density distribution of stars.  The best axisymmetric solution we found perturbs the Galactic plane up by 70 pc at 10.5 kpc from the Galactic center and 170 pc down at 14 kpc from the Galactic center in a somewhat sinusoidal pattern.  To fit the star counts, though, we would need to use a model that is not axisymmetric.

In section 7, we compare the shape and positions of the overdensities with the shapes and positions of the Milky Way spiral arms, and find a rough agreement.  One surprise is that the spiral arms, which trace Galactic structure much closer to the Galactic plane than our stellar tracers, seem to show vertical asymmetry in the opposite sense to the asymmetry we observe; at Galactocentric distances where we see more stars in the north, tracers of the spiral structure are seen with greater frequency in the south, and vice versa.  We suggest this could be consistent if there are simultaneously radial waves (as we observe in this paper) and vertical waves (as found by previous authors).  Finally, in section 8 we attempt to fit our observations that suggest radial waves in the disk to previous observational evidence from the Monoceros and TriAnd Rings.

\section{The North-South Asymmetry of the Disk}

\begin{figure}
\includegraphics[scale=1.0]{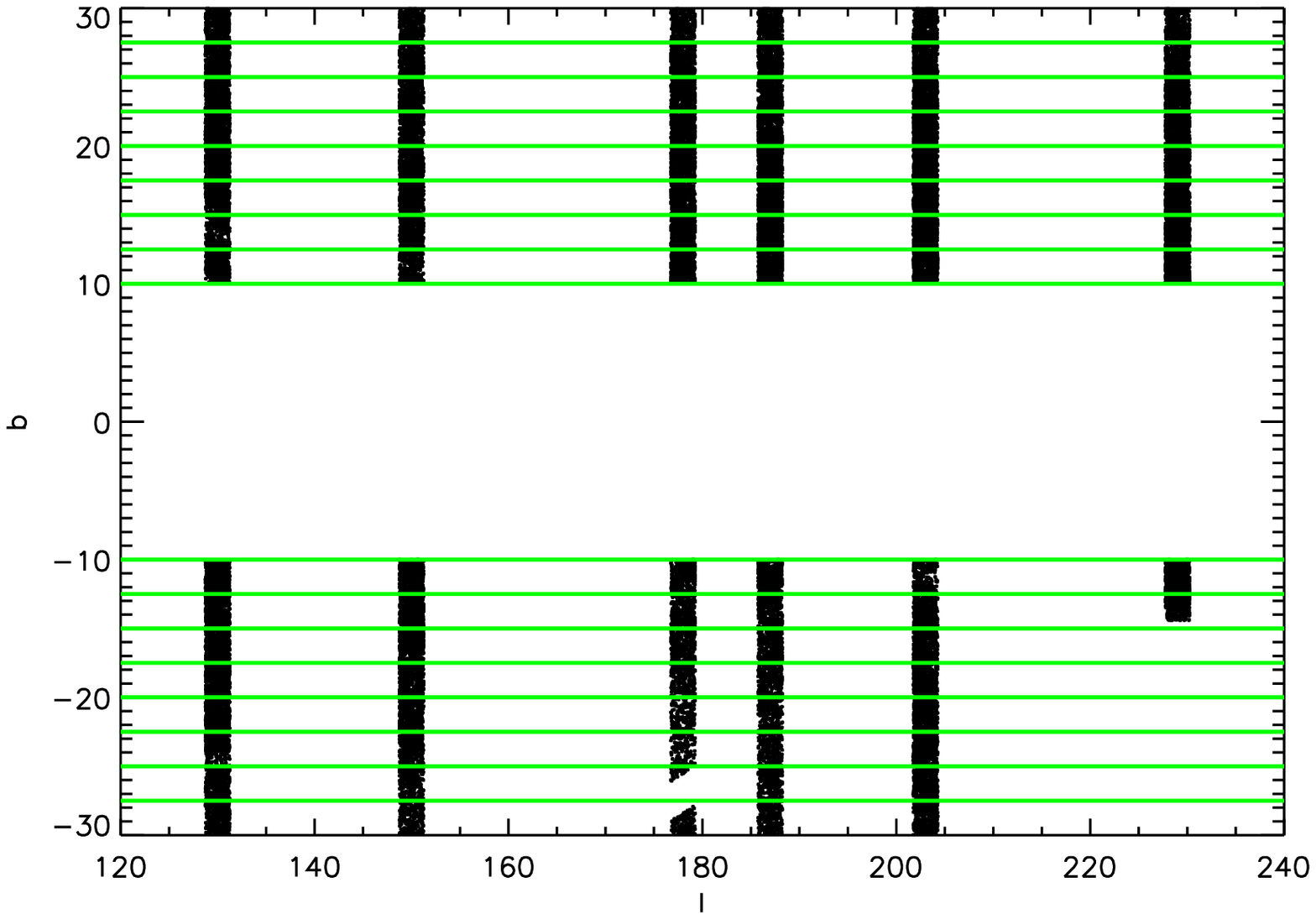}
\caption[lbmap] {
\footnotesize
 Sky coverage of photometric data of SDSS DR8 which is used in this work.  The green lines show the patches into which the data was divided.
}\label{skycoverage}
\end{figure}

We select photometric data from SDSS DR8 (Aihara et al. 2011) in the region $110^\circ<l<230^\circ$ and $10^\circ<|b|<30^\circ$.   Within this region there are seven $2.5^\circ$-wide, constant longitude stripes that extend through the low Galactic latitude region ($l=110^\circ,130^\circ, 150^\circ, 178^\circ, 187^\circ,$ $203^\circ$, and $229^\circ$).  The locations of the SDSS stripes are shown in Figure \ref{skycoverage}. Latitudes lower than $|b|<10^\circ$ are excluded due to poor photometry and high dust extinction in these crowded fields.  We divide the data along these stripes into bins of $2.5^\circ$ in latitude.  Because the low latitude SDSS photometry follows constant longitude, we can select sky areas that are exactly symmetric around the Galactic plane.

The SDSS magnitudes were corrected for extinction according to Schlafly and Finkbeiner (2011).  This was done by first extracting the extinction from the SDSS database; the SDSS extinction is calculated from: $A_u=5.155 E(B-V), A_g=3.793 E(B-V), A_r=2.751 E(B-V), A_i=2.086 E(B-V)$ and $A_z=1.479 E(B-V)$ (Stoughton et al. 2002), where the $E(B-V)$ reddening values are from \citet{1998ApJ...500..525S}.  We then adjusted the values so that they were instead $A'_u=4.239 E(B-V), A'_g=3.303 E(B-V), A'_r=2.285 E(B-V), A'_i=1.698 E(B-V)$ and $A'_z=1.263 E(B-V)$ (Schlafly and Finkbeiner 2011, using $R_V=3.1$).  These modified extinction values were subtracted from the measured apparent magnitudes to determine the extinction-corrected magnitudes, denoted with subscript $_0$.  

Table 1 gives the average E(B-V) values for each patch. From Table 1, we can see that extinction is very uneven.  High extinction ($E(B-V)>0.25$) is found at $l=110^\circ$,$10^\circ<b<20^\circ$; $l=130^\circ$,$10^\circ<b<17.5^\circ$; $l=150^\circ,10^\circ<b<15^\circ$; and in the southern parts of stripes $l=178^\circ,187^\circ$. There are two potential problems with the extinction direction of the low galactic latitude area.  The first one is that the extinction could be overestimated in very high reddening areas \citep{2012ApJ...757..166B, 2014MNRAS.443.1192C}.  The second one is that bright(apparent magnitude) and low Galactic latitude stars could be embedded in the dust plane, so the 3D extinction is smaller than the maximum cumulative value.

Neither of these effects should influence the result of comparison of the main sequence star counts of north and south sky.  Although the extinction could be overestimated or underestimated in some lines of sight, the direction of the extinction vector in the CMD is almost aligned with the direction of the main sequence.  Incorrect reddening values will move stars up and down on the main sequence, but the assumed distance (as measured from the apparent magnitude at a particular color) to the stars will not change.  We will discuss this point in more detail later.  As we will see, the stellar density patterns at low Galactic latitude change little with Galactic longitude while the extinction is a strong function of Galactic longitude.  There is higher extinction in the north when $l<180^\circ$ and higher extinction in south when $l>180^\circ$.  Regarding the effects of 3D extinction, we tried to plot the relation of E(B-V) vs. distance to the Sun in a region of the sky, (l,b)=$(178^\circ,-15^\circ)$, with very high reddening.  The majority of the low latitude extinction is within 1 kpc of the Sun along this line of sight.  We will show later that the nearest identified overdensity is 2 kpc from the Sun at this latitude.  In addition, the 3D extinction would have to be pathologically consistent as a function of Galactic latitude and longitude to produce the same apparent density substructure (as a function of distance) in every direction.  In summary, the results in this section are fairly robust to inaccuracies in the reddening correction.

\begin{deluxetable}{crrrrrrr}
\tabletypesize{\scriptsize}
\tablecolumns{8}
\footnotesize
\tablecaption{Mean E(B-V) values for each patch}
\tablewidth{0pt}
\tablehead{}
\startdata
$b \setminus l$     &  $110^\circ$  &  $130^\circ$ &  $150^\circ$ & $178^\circ$ & $187^\circ$ & $203^\circ$ &$229^\circ$ \\
$27.5^\circ\sim30^\circ$ & 0.061 & 0.046 & 0.049 & 0.045 & 0.051 & 0.043&0.037 \\
$25^\circ\sim27.5^\circ$ & 0.073 & 0.061 & 0.047& 0.050 & 0.056 & 0.039 & 0.025 \\
$22.5^\circ\sim25^\circ$ & 0.110 & 0.082& 0.072& 0.061&0.046 &0.035 & 0.026\\
$20^\circ\sim22.5^\circ$ & 0.265&0.133 &0.119 &0.062 &0.059 & 0.042&0.032\\
$17.5^\circ\sim20^\circ$ &0.575 & 0.200& 0.157& 0.088& 0.067&0.035 &0.039 \\
$15^\circ\sim17.5^\circ$ &0.676 & 0.375&0.178 &0.114 &0.083 &0.059 &0.078 \\
$12.5^\circ\sim15^\circ$ &0.917 &0.652 &0.380 &0.153 &0.079 &0.099 & 0.129 \\
$10^\circ\sim12.5^\circ$ &0.559 &0.810 &0.694 &0.233 &0.104 &0.093 & 0.112 \\
$-10^\circ\sim-12.5^\circ$ &0.195 &0.274&0.231&0.564&0.360&0.945&0.282\\
$-12.5^\circ\sim-15^\circ$ &0.162 &0.160&0.504&0.423&0.500&0.208&\\
$-15^\circ\sim-17.5^\circ$ &0.120 &0.094&0.171&0.456&0.419&0.275&\\
$-17.5^\circ\sim-20^\circ$ &0.130 &0.073&0.154&0.507&0.397&0.168&\\
$-20^\circ\sim-22.5^\circ$ &0.122 &0.060&0.130&0.558&0.462&0.188&\\
$-22.5^\circ\sim-25^\circ$ &0.140 &0.056&0.153&0.567&0.297&0.111&\\
$-25^\circ\sim-27.5^\circ$ &0.078 &0.056&0.194&0.454&0.247&0.064&\\
$-27.5^\circ\sim-30^\circ$ &0.050 &0.056&0.143&0.321&0.223&0.043&\\
\enddata
\end{deluxetable}

For each $2.5^\circ$ patch of each of the five stripes, we present a $g_0$ vs. $(g-r)_0$ Hess diagram.  Figure 2 shows the Hess diagrams for the north Galactic cap, and Figure 3 shows the Hess diagrams for the south Galactic cap.  We expect to see few stars bluer than $(g-r)_0=0.4$, since this is approximately the turnoff of the thick disk.  Redder than that, we expect to see main sequence stars smoothly distributed in distance, and therefore smoothly distributed in apparent magnitude.  At faint magnitudes we expect to see a smaller number of halo stars with a somewhat bluer turnoff than the disk.  On the very red side of the diagram, we see primarily disk M dwarf stars.  Because M dwarf stars with a variety of masses pile up at the same $(g-r)_0$ color, the density of stars is much higher on the red side of the main sequence.

We see in Figure 2 that near the Galactic plane the high reddening tends to smear out the main sequence stars, and completely obscure the fainter blue stars.  As we move away from the Galactic plane we see that the disk stars are more concentrated in distance (and therefore apparent magnitude) than we expect.  All but a few very low latitude panels and other directions of high extinction show a narrow main sequence with a blue turnoff near $g_0 \sim 19.5$.  This is consistent with the properties of the Monoceros Ring.  There is also a second, possibly broader main sequence with a turnoff near $g_0 \sim 16.5$.  We will refer to this brighter apparent overdensity as the ``north near structure."  The turnoff color of the north near structure is significantly redder than that of Monoceros for $b>20^\circ$.  Above $b=20^\circ$, this pattern can be seen at all longitudes.  

In the south (Figure 3), one sees a very broad distribution of stars, with a turnoff in the vicinity of $g_0 =18$ (half way between the two sequences in the north) and with a turnoff color similar to the nearer northern stars.  In some locations, the Hess diagrams give the impression of more than one main sequence feature extending from the turnoff, e.g. in plate $(178^\circ,-27.5^\circ)$. 
In many of the panels (particularly those further from the Galactic plane), there is a narrow main sequence with a bluer turnoff that is a little fainter than Monoceros in the north - maybe $g_0 \sim 20$.  In particular, this fainter main sequence is apparent in all of the panels with $b>22.5^\circ$.  We will later associate this fainter main sequence with the TriAnd Ring.  It is likely that the extinction in the low latitude fields makes it impossible to see the fainter structure in many of the southern fields, even though it might extend to lower latitudes.

The difference between the star counts in the north and the south can be seen in dramatic fashion by subtracting the panels in Figure 3 from the corresponding panel in Figure 2.  In addition to the six panels with $l=229^\circ$ for which there is no data, there are three southern panels in Figure 3 for which some data is missing.  SDSS covers only $83.85\%$ of the patch at $(l,b)=(130^\circ, -22.5^\circ)$, only $22.03\%$ of the patch at $(178^\circ, -25^\circ)$, and only $62.86\%$ of the patch at $(178^\circ, -27.5^\circ)$.  The counts in these three panels was divided by the fraction observed before subtracting to create Figure 4; the noise in the star counts in these panels is correspondingly larger.  

If the Galactic disk is symmetric around the Galactic plane, the difference panels should be nearly zero. Because the Sun is about 27 pc (about a tenth the scale height of the thin disk) above the midplane, we would expect a very small asymmetry, in the sense that there should be slightly fewer stars in the north.  The closest stars in our sample are about 1.5 kpc from the Sun, and the lowest latitude of these are about one thin disk scale height from the plane.  Since the Sun's height above the plane is about 10\% of the disk scale height, the expected north-south asymmetry in thin disk star counts is at most 20\% (since $e^{-0.9}/e^{-1.1}=1.2$). The asymmetry in the thick disk star counts should be much smaller, since the Sun's height above the plane is less than 4\% of the thick disk scale height.  The asymmetry should decrease monotonically with distance, as the observed stars are farther and farther from the Galactic plane.

In contrast, the difference plots in Figure 4 show that there is a large north-south asymmetry in the star counts in the Milky Way, and the sign of the asymmetry changes as a function of distance.  There are white-black-white main sequence patterns in each panel, with the exception of a few panels at $(110^\circ,10^\circ),(130^\circ,10^\circ),(150^\circ,10^\circ),$ and $(229^\circ,10^\circ)$, where the Hess diagram is smeared out by serious extinction.  White means that there are more stars on north side; black means that there are more stars on the south side.  By comparison with Figure 2 and Figure 3, the brightest white feature corresponds to north near structure. The middle black feature corresponds to the south middle structure. The faintest white feature corresponds to Monoceros Ring. 

Figure 5 shows an expanded view of the difference Hess diagram for $(l,b)=(178^\circ,\pm 15^\circ)$, which represents the Galactic longitude closest to the anticenter, at the lowest latitude that does not suffer enormous extinction. For this example difference diagram, we also show a numerical representation of the $0.4<(g-r)_0<0.5$ star counts in the north, south, and subtracted diagrams for this portion of the sky.  The direction of the reddening vector, as calculated from Schlafly and Finkbeiner (2011) is indicated by the arrow.  The data is undoubtedly affected by inaccuracies in the correction for dust as a function of position in the sky, in part due to the 3D nature of the dust.  An estimate for the color and magnitude errors associated with reddening can be obtained from the equations for the extinction due to dust:
$A_g=1.068*R_v*E(B-V)$, $A_r=0.737*R_v*E(B-V)$ and $A_{(g-r)} = 0.331*R_v*E(B-V)$.  The precision of $R_v$ (estimated to be about 3.1) is 0.1-0.2 \citep{2012ApJ...757..166B}, and the error in determining the extinction is about 20\% in high extinction areas.  The error in E(B-V) dominates the errors in extinction correction, resulting in extinction corrections that are uncertain by about 20\%.  Since the E(B-V) at $(l,b)=(178^\circ,15^\circ)$ is 0.114, we estimate $\delta A_g = 0.08$, $\delta A_r = 0.05$, and $\delta A_{(g-r)} = 0.02$.  At $(l,b)=(178^\circ,-15^\circ)$, $E(B-V)$ is 0.456.  In this region, $\delta A_g = 0.3$, $\delta A_r = 0.2,$ and $\delta A_{(g-r)} = 0.09$.
  However, the observed asymmetries cannot be erased by adjusting the reddening values because moving stars along the reddening direction in each panel of Figure 4 will not make the apparent banding go away.  Inaccuracies in the reddening correction cannot cause the apparent asymmetry.

Notice in the lower panel of Figure 5 that the subtracted counts are not a small fraction of the star counts in this panel.  The apparent magnitudes of the peaks differ by a full magnitude. Alternatively, one could describe the difference in star counts; at some apparent magnitudes the number of stars differs by a factor of two.  This large difference in the star counts for sky regions located at the same Galactic longitude and the same distance from the Galactic plane shows quite clearly that the disk is not symmetric about the Galactic plane.  Although there could be radial or vertical age or metallicity gradients in the disk, these would have to be quite contrived (and asymmetric) to account for the strong asymmetry that we observe. 

We will briefly consider our observed density structure in the context of the known Galactic warp or flare.  The maximum warp of M giants \citep{2006A&A...451..515M} appears at $l=120^\circ$ in northern sky and $l=240^\circ$ in the southern sky at both distances 3 kpc and 7 kpc. But in this paper, both the north near structure and south middle
structure extend from $l=130^\circ$ to $229^\circ$ without switching sides of the Galactic plane. The warp might explain an azimuthal dependence to the observed stellar density, but does not explain an asymmetry that changes with distance and is evident at all Galactic longitudes.  Regarding the possibility of a disk flaring, \citet{2014ApJ...794...90K} review the flaring of stars in different sky surveys. The scale height of the disk appears to increase with the distance from the Galactic center.  The flare of the thin disk and thick disk show different characteristics, but there is no evidence given for whether the flare is asymmetric above and below the midplane.  We discuss possible connections with spiral arms in section 7.

\section{Metallicities of the structures}

Figure 6 shows the locations of SDSS stellar spectra in the region around the Galactic anticenter.  The red crosses show the centers of the plates observed in the stripes with $l=110^\circ, 130^\circ, 150^\circ, 178^\circ, 187^\circ, 203^\circ,$ and $229^\circ$.  There are 26 plates selected in the north and 20 plates selected in the south.  This selection of plates will be used in this section to explore the stellar populations of the overdensities, and will also be used later to explore the velocity substructure.

Figure 7 shows how spectra were selected to coincide with the north near and south middle structures.  The top panel shows a sample Hess diagram of stars in the north Galactic cap, and the lower panel shows the corresponding panel in the south Galactic cap.  Polynomials were fit to the ridgelines of the north near structure and the south middle structure. The red line in the upper panel of Figure 7 represents the ridgeline of the north near structure, and the green line of the lower panel represents ridgeline of the south middle structure. Spectra around the ridgelines (within the dashed lines) were analyzed to determine the properties of each structure.  Control samples are selected at the same position in the Hess diagram from the sky area mirror reflected across the Galactic plane. For instance, the spectra around the red line in the lower panel are selected as the control sample for the north near structure. The spectra around the green line in the upper panel are selected as the control sample for the south middle structure. 

In the top panel of this figure, notice that the majority of the points on the north near ridgeline are green, indicating metallicities close to [Fe/H]$=-0.4$, which is representative of the upper part of the thin disk.  The brighter (closer) stars are typically more metal-rich and fainter (further) stars are more metal-poor.  This is the expected trend; more distant stars in this sample are not only farther from the Galactic center but also further above the plane of the Galaxy, and are expected to have lower metallicity.  The same trend is observed in the southern field.  However, there is a different distribution of stars with distance; in the south there are many more stars just below the red curves, with metallicities closer to -0.7.  The SDSS selection criteria determining for which stars spectra were obtained is also very different between these northern and southern fields.

The distribution of metallicities in the north near structure is shown quantitatively in Figure 8. 
 The panels show histograms of metallicity for six different Galactic longitudes; all of the panels are near $b=15^\circ$.  We expect metallicities of [Fe/H]$>-0.6$ for thin disk stars, $-1.2<$[Fe/H]$<-0.6$ for thick disk stars, and [Fe/H]$<-1.2$ for halo stars \citep{2012ApJ...757..151L}. These panels show that the majority of the spectra at each longitude have the expected metallicities for thin disk stars. Figure 9 shows histogram of stars at similar locations below the Galactic plane.  The metallicities above and below the plane are very similar, despite the fact that the stellar densities are very different.

Figure 10 shows that for stars in the south middle structure, the metallicity varies as a function of Galactic latitude.  Unfortunately, there are many fewer SDSS stellar spectra in the southern Galactic hemisphere, and we show here all of the directions for which we have a significant number of spectra in this fainter region of the Hess diagram in the south.  Stars in the south middle structure are about 6 kpc from the Sun, so at $|b|=30^\circ$, they are 3.0 kpc from the Galactic plane, where as at $|b|=15^\circ$, they are 1.6 kpc above the plane.  At Galactic latitudes more than $15^\circ$ from the plane, the stars look like thick disk stars with [Fe/H]$\sim -0.8$ dex, while closer to the plane the metallicity is about $-0.6$ dex.  This metallicity difference as a function of Galactic latitude is not apparent for the north near structure because, as we shall see in Section 4, the distance to the north near structure as calculated from the ridgeline changes with latitude so that this structure is identified at approximately the same height above the disk at all Galactic latitudes.

There are very few SDSS spectra of stars in the farther (Monoceros and TriAnd) structures, but the metallicities of stars in these structures are measured from previous papers describing the Monoceros Ring and the Triangulum-Andromeda Stream.  The Monoceros ring has a metallicity of [Fe/H]$\sim -1$ \citep{2012ApJ...757..151L}.  M giants in the Triangulum-Andromeda overdensity have a metallicity of [Fe/H]$\sim -0.6$ (Sheffield et al. 2014).  However, \citet{2014MNRAS.444.3975D} measure metallicities of giant and A-type stars at more like [Fe/H]$\sim -0.8$ (see Figure 8 of that paper), and fit an isochrone with [Fe/H]$\sim -1.5$ to the TriAnd photometry.  Possibly, the M giant stars in TriAnd are preferentially more metal rich than other stellar populations.

\section{Distances to low latitude stellar rings}

In Figures 2, 3, and 4 we established that at least in the Galactic longitude range $110^\circ<l<230^\circ$, there is an asymmetry in the number of stars, that oscillates as a function of Galactocentric radius from more stars to the north of the plane, then more stars to the south of the plane, then more stars to the north of the plane (this last corresponding to the location of the Monoceros Ring).  The two main sequences in the north appear to be narrower than the one in the south.  There appears to be another narrow main sequence in the south that is slightly farther away than the Monoceros Ring, that is seen at Galactic latitudes further from the plane where there is less extinction (see faint, blue portions of lower panels in Figure 3).

We note that the apparent magnitude difference between the two northern main sequences in Figure 4 changes as a function of Galactic latitude.  The more distant Monoceros Ring remains at about the same apparent magnitude regardless of the Galactic latitude, while the nearer apparent structure is fainter (implying more distant) at lower Galactic latitudes.  In fact, the implied distance above the plane for the north near structure remains constant as a function of Galactic latitude.  The apparent variation of the distance to the ``north near" structure makes it more difficult to interpret this structure as a physical ring of stars at a particular Galactocentric distance.


When observing the star counts from within a disk with an exponential density profile, the number of stars of a given type in a given solid angle of sky first increases with distance as the volume increases, and then decreases since the density of the disk decreases exponentially with distance from the Galactic plane.  If the apparent ``north near" main sequence is due to the tradeoff between volume surveyed and density decline, then it is reasonable that the apparent distance of the main sequence would increase at lower latitudes since the volume sampled at low latitude remains closer to the Galactic plane.   In fact, for a plane-parallel disk with a density that falls off exponentially with distance from the plane, as viewed from a point on the plane, the distance to the density maximum in a particular solid angle is exactly inversely proportional to the sine of the Galactic latitude.  Therefore, the north near structure gives the appearance of a stellar population that stretches over many kiloparsecs along our line of sight, and falls off exponentially as a function of distance from the Galactic plane.
 The important thing to remember about the north near structure, though, is that there are more stars in the north than in the south a couple of kiloparsecs further from the Galactic center than the Sun.

The apparent oscillatory pattern in Figure 4 suggests that the position of the midplane of the disk might change as a function of distance.  To test whether the oscillatory pattern is centered on the Galactic center, we use isochrone fitting to determine the distance to each of the main sequences found in Figures 2 and 3, at the five latitude ranges $12.5^\circ<|b|<15^\circ$, $15^\circ<|b|<17.5^\circ$, $17.5^\circ<|b|<20^\circ$, $25^\circ<|b|<27.5^\circ$, and $27.5^\circ<|b|<30^\circ$.

The fact that the metallicity is different for each of the structures means that the determination of the distance to each structure requires a different isochrone (and for the south middle structure, the stellar population additionally varies with angle below the plane).  Additionally, the nature of the observed structures that we are attempting to fit is different enough that we elected to use a different procedure to fit each one.

All of the isochrones that were fit to the substructures were taken from An et al. (2009).  In this paper, $ugriz$ isochrones were generated from the Yale Rotating Evolutionary Code (Sills et al. 2000, Delahaye \& Pinsonneault 2006), using MARCS (Gustafsson et al.
2008) model atmospheres.  These model atmospheres are compared with fiducial sequences fit to open clusters and globular clusters observed with the SDSS and corrected for reddening using values from \citet{2003PASP..115..143K}.  They then make empirical corrections to the $(g-r)_0$ vs. $T_{\rm eff}$ relationship so that the theoretical isochones matched the fiducial main sequences observed in star clusters.  We checked these isochrones against those of Girardi et al. (2004) and find that they are in good agreement for main sequence stars wit $(g-r)<0.8$, so our distances are apparently not very model dependent. 

To fit an isochrone to the near north structure, stars in each $2.5^\circ$ by $2.5^\circ$ patch of sky were divided into samples with a 0.1 magnitude wide range of $(g-r)_0$ color.  A histogram of $g_0$ magnitude was generated for each of these samples, and the maximum of the histogram was selected by eye.  Then the central color and magnitude of each histogram peak was plotted on a color-magnitude diagram; for example, see the upper black plus signs in the top panel of Figure 11.  The distance to an 8 Gyr isochrone with [Fe/H]=-0.5 was then varied so that the closest fit to the plus signs was achieved.  Because  the internal color and absolute magnitude of the bluest point of the turnoff of the 8 Gyr isochrone is $((g-r)_0,g_0)=(0.316,4.11)$,  the absolute magnitude ($M_{g_0}$) of the stars with $(g-r)_0<0.316$  was assumed to be 4.11.  With this assumption, we could fit stars even if they were bluer than the turnoff.  At a fixed metallicity, the main sequence of isochrones will be the same between $0.3 < (g-r) < 1.2$ for different ages, so the age of the isochrone is not important for fitting the main sequence.  The apparent positions of the near north structure, projected onto the Galactic plane, are shown in the top panel Figure 12.

Without an actual substructure with a significantly different density from the surroundings, the peak distance of the star counts will depend on the rate at which the density declines along a particular line of sight and the increase in the volume of a sphere with distance.  If the region of the Galaxy probed is small compared to the scale length of the disk and in the absence of significant reddening variations, the peak of the star counts distribution will be the same distance from the Sun at all Galactic longitudes and depend strongly on Galactic latitude.  The north near structure shows peaks in the distribution that depend on Galactic latitude and are roughly concentric around the position of the Sun.  In particular, they are further than 10 kpc from the Galactic center at $l=180^\circ$, and closer than 10 kpc at both higher and lower Galactic longitudes; the higher latitude distances are particularly concentric around the Sun and do not follow circles centered on the Galactic center (see Figure 12).  It is not clear that the north near structure represents a ring; however, it is still true that near the Sun there appear to be more stars above the plane than below the plane, so there must be some local structure near this position. 

We had particular difficulty fitting an isochrone to the south middle structure, since the stars appeared to be very spread out in distance.  At the faint end, the turnoff is often slightly bluer than the  bright end, and sometimes there is the appearance of more than one main sequence.  In the end, we decided to fit the center of the larger substructure, and disregard the individual smaller features.  We selected all of the stars bluer than $(g-r)_0<0.4$ in each patch of sky.  We then created histograms in $g_0$.  We then selected the peak of the distribution, and calculated the distance to the south middle structure by comparing that distribution peak with the turnoff of an interpolated isochrone from An et al. (2009).  We used the set of isochrones with ages of 12.5 Gyr.  Although this age seems old for the disk, younger isochrones were not a reasonable fit to the turnoff color of the south middle structure.  We generated an interpolated isochrone with the measured metallicity at different Galactic latitudes.  For stars with $|b|<15^\circ,$ we created an isochrone with [Fe/H]$=-0.44$ and turnoff absolute magnitude $M_{g_0}(TO)=5.06$.  The color of the interpolated isochrone turnoff is $(g-r)_0(TO)=0.403.$  For stars with $15^\circ < |b| < 20^\circ$, we created an isochrone for [Fe/H]$=-0.61$, $M_{g_0}(TO)=4.87$, and $(g-r)_0(TO)=0.384$.  For stars with $20^\circ < |b| < 25^\circ$, we created an isochrone for [Fe/H]$=-0.73$, $M_{g_0}(TO)=4.75$, and $(g-r)_0(TO)=0.360$.  For stars with $25^\circ < |b| < 30^\circ$, we created an isochrone for [Fe/H]$=-0.88$, $M_{g_0}(TO)=4.58$, and $(g-r)_0(TO)=0.328$.  The apparent positions of the south middle structure, calculated as described above and projected onto the Galactic plane, are shown by plus signs in the lower panel of Figure 11.

As is apparent from Figure 12, the brown and black plus signs in the south middle structure, which are at the larger angles from the Galactic plane, form concentric circles around the Sun's position and not around the Galactic center.  The latitudes closer to the Galactic plane are ambiguous as to whether they are concentric around the Sun or the Galactic center; they are somewhat in between.  

We also tried a second method for determining the distance to the south middle structure.  For a range of 0.1 magnitude wide $(g-r)_0$ color bins, we plotted a histogram of the number counts as a function of $g_0$ magnitude in a low Galactic latitude ($|b|$) field, and then subtracted a histogram of the number counts as a function of $g_0$ magnitude for the $-30^\circ< b < -27.5^\circ$ field at the same Galactic latitude. This was a logical thing to try, since the high ($|b|$) latitude field did not seem to contain the excess population, so the difference might be more indicative of the excess.  The distance to the peak of the difference of the histograms is plotted as diamonds in Figure 12.  These distances are about 300 to 500 pc further than the distances derived by the previous method, when similar sky positions are compared.  They also appear to be roughly Galactocentric, but with slightly larger Galactocentric distances at larger Galactic longitudes.

The more distant structures in both the north and the south were fit in a similar fashion to each other.  We used the An et al. fiducial sequence for the globular cluster M5 ([Fe/H]$\sim -1.2$ dex).  We made histograms of $(g-r)_0$ at constant $g_0$, which cuts through the H-R diagram in the opposite direction from what was used in measuring the near north structure.  We preferred to cut through the diagram horizontally so that we did not use the fainter, noisier data to determine the center of the main sequences.  Figure 11 shows the positions of the centers of the main sequence measured for each 0.1 magnitude wide range of $g_0$ for a sample region of sky.  The apparent positions of the distant northern structure, projected onto the Galactic plane, are shown in the upper panel of Figure 12; and the positions of the distant southern structure, projected onto the Galactic plane, are shown in the lower panel of Figure 12.

Figure 12 shows that the further structures (associated with the Monoceros Ring and the Triangulum Andromeda Stream) form approximately concentric rings around the Galactic center (taken to be 8 kpc from the Sun).  The south middle, far north, and far south structures all seem to be a little closer to the Galactic center in the second quadrant than they are in the third quadrant.  

\section{Velocity Substructure of North Near and South Middle Structure}

In the previous section, we identified three or four ring-like features near the plane of the Milky Way.  The Monoceros Ring and Triangulum-Andromeda Stream have been studied for more than a decade without agreement on the nature of these objects, which could be tidal debris or a warp/flare/oscillation of the outer disk (see the introduction for discussion and references).  Here, we inquire as to the nature of the nearer substructures.  We will further discuss the nature of all four structures in sections 7 \& 8.   We first compare the velocity distribution to a simple disk model in which the thick disk lags the thin disk by 50 km s${^-1}$, and then compare only the stars with disk metallicities with a better model that includes asymmetric drift but does not apply to a portion of our data that is too far from the Galactic plane.

 In Figure 13 we present velocity histograms of the stars in the north near structure at six different Galactic longitudes; all of the panels are near $b=15^\circ$ and are the same stars for which we presented metallicity histograms in Figure 8.  The line-of-sight, Galactic standard of rest velocity, $V_{gsr}$, for a particular star is calculated with:  $V_{gsr} = V_{helio} + 13.84 \cos b \cos l + 250 \cos b \sin l + 6 \sin b$ (Sch{\"o}nrich 2012). To compute the expected velocity distribution of stars in each panel, we use the average distance to the stars in the field along with the $(l,b)$ coordinates of the center of the field to determine the position of these stars in the Milky Way.  The distance to each individual star was determined from the apparent magnitude of the star and the absolute magnitude of stars with the same color in the isochrone that was fit to the near north structure.  Then, we compute the expected $V_{gsr}$ for the thin disk and thick disk at that location.  The thin disk stars are assumed to have a Gaussian velocity distribution with dispersion $\sigma=25$ km s$^{-1}$ centered around the circular speed 238 km s$^{-1}$ \citep{2012MNRAS.427.274S,2011MNRAS.414.2446M}. 
 We adopt a thick disk velocity that lags the thin disk by 50 km s$^{-1}$ \citep{2003A&A...398..141S}, and a thick disk velocity dispersion of $\sigma=50$ km s$^{-1}$. The velocity dispersions of the disk components are also taken from \citet{2003A&A...398..141S}, and although they are measured at the solar position, they are assumed to be approximately correct even 1 kpc above the Galactic plane and 2 kpc from the Sun.  The halo stars are assumed to have zero mean velocity, with a sigma of 100 km s$^{-1}$ around the mean \citep{2012ApJ...757..151L}.

Given the position of the stars and the mean velocity of each Galactic component at that position, we then calculate the component of the velocity along our line of sight for the thin and thick disks (the mean velocity of the halo is always zero).  The number of stars assigned to each component was determined from the number of stars in each of the three metallicity ranges in Figure 8.  In Figure 13, the red curve is the theoretical $V_{gsr}$ distribution of the thin disk, the blue curve is the theoretical $V_{gsr}$ distribution of the thick disk, the yellow line is the theoretical $V_{gsr}$ distribution of the halo stars, and the black curve is the sum of the three.  The dotted line shows the position of a 2.5 sigma excess in each bin, calculated from Poisson statistics. 
 The median line-of-sight velocity error is 4.84 km s$^{-1}$, which is much smaller than the binsize in our histograms.

Figure 14 shows the similarly calculated velocity distribution of south middle structure. Because it is located further than the north near structure, there are not sufficient spectra to trace this structure in most of the regions for which we have photometry.  For the six directions with a significant number of spectra, we fit the observational $V_{gsr}$ with our standard kinematic model; the fraction of each component was derived from the fractions of stars in each of the metallicity bins in Figure 10.  The stars we observe in the south middle structure are about 1.5 kpc above the Galactic plane, so most stars in the south middle structure belong to the thick disk.  The position of the peak velocity is consistent with expectations for thick disk stars.  However, most of the observational histograms have a smaller velocity dispersion than the model predicts.

 What we learn from these velocity histograms compared to the simple disk model is that the north near structure is expected to be primarily thin disk stars, and the data is roughly consistent with that; the south middle structure is expected to be a mix of thin and thick disk stars, and the data is also roughly consistent with the model.  However, there is a systematic shift in the model and observed velocity distribution, in the sense that there are too many low velocities observed at lower Galactic longitude in the second quadrant and too many high velocities observed at higher Galactic longitude in the third quadrant.  We next show that this systematic shift can be explained by properly including the asymmetric drift.  We found no evidence for a group of stars with velocities and/or metallicities that were unlike those of the Galactic disk.  If we had found such a population, that would be evidence that the star counts asymmetry might be due to satellite accretion.


\citet{2012MNRAS.419.1546S} derive an analytical formula that describes the fact that the azimuthal components of the velocities of stars in a galactic disk are skew, with many more stars lagging the disk than leading it (asymmetric drift).  Equation 29 from that paper gives the expected distribution of velocities as a function of position in the Milky Way.  We used the exact model from that paper; their model includes a thin disk with scale height 300 pc, a thick disk with scale height 1000 pc, and a scale length of 2.5 kpc for both components.  They also use 8 kpc from the Sun to the Galactic center.  We used the values in Table 3 of \citet{2012MNRAS.419.1546S} to get the disk velocity dispersion ($\sigma_0$), the scale length of the velocity dispersion ($R_\sigma$), and the local scale height ($h_0$), as a function of height above the disk.  Because this table only reaches to 2 kpc above the disk, we cannot apply it to the higher latitude data in the south middle structure.

The output of the \citet{2012MNRAS.419.1546S} model is a histogram of $V_\phi$ with a bin size of 1 km s$^{-1}$, for the particular location being modeled.  We then used a Monte Carlo simulation to select $V_\phi$ from the resulting histogram and $V_r$ from a Gaussian with a mean of 0 km s$^{-1}$ and a sigma of 35 km s$^{-1}$.  These values were then combined and converted to the line-of-sight $V_{gsr}$.  Multiple pulls from these two distributions were built up to calculate the expected $V_{gsr}$ histogram.  Our value for the width of the $V_r$ Gaussian was derived from the width of our velocity distribution at $l=178^\circ$, where there should be no contribution from the azimuthal velocity.

In Figure 15, we show the velocity distribution of the disk stars only, as determined from the metallicity cut at [Fe/H]$>-1.2$ (see Figure 8).  The solid red curve gives the predicted velocity distribution from the \citet{2012MNRAS.419.1546S} model, and the dotted blue curve shows the 2.5 sigma error limit for each bin.
The velocity distribution including asymmetric drift produces a much improved fit to the observed velocity distribution of disk stars.  

 There are a couple of single bins that are unusually high in the velocity histograms.  In particular, the $V_{gsr}=170$ km s$^{-1}$ peak at $(l,b)=(110^\circ, 16^\circ)$ appears quite significant.  Possibly there is another peak and at $V_{gsr}=80$ km s$^{-1}$  at $(l,b)=(150^\circ, 15^\circ)$.  We suspect that these peaks are due to the presence of disrupted star clusters, which will be further explored in a separate publication.

The improved disk model, plus a few dissolved open clusters, is a good match to the velocity distribution.  We do not see a substantial portion of the population at a different metallicity and/or velocity than the disk in either the near north or the south middle structures, as we would have expected if the disk asymmetry was due to satellite accretion.  We therefore conclude that the near north structure is due to a distortion of the stars in the disk component(s) of the Milky Way.

Figure 16 shows the velocity distribution of disk stars in the south middle structure, selected by metallicity, compared to the Sch{\"o}nrich \& Binney model.  Note that only three of the panels in Figure 10 are within $|z|<2$ and therefore accessible to the Sch{\"o}nrich \& Binney model.  Although there are fewer stars, and therefore fewer velocities, available for the south middle structure, we have the sense that the situation is similar; the metallicities and velocities do not show a deviation from expectations for the thin and thick disk, and the velocity distribution is as expected if one includes the effects of asymmetric drift.  We therefore conclude that the south middle structure is also likely due to a distortion of the stars in the disk component(s) of the Milky Way.

The two more distant structures, which we will call the Monoceros and TriAnd Rings, could also be disk structures.  The disk oscillation might explain why the disk appeared to suddenly end at 15 kpc from the Galactic center in previous studies.  Alternatively, the Monoceros and TriAnd Rings could be due to one or more accretion events (or more likely the accretion of a group of substructures), which would explain why there are so many stream-like substructures observed in SDSS and Pan-STARRS1 data.

\section{Fitting a Model Disk to the Star Counts}

We now attempt to fit a simple model to the star counts in the near north and south middle structures.  We select early K-type stars by color ($0.6<(g-r)_0<0.7$) from SDSS DR8 and fit a disk density profile to the observed star counts.  The vast majority of these stars are main sequence stars, that are assumed to have an absolute magnitude of $M_{g_0}=6.776$.  This absolute magnitude was determined from interpolation of the isochrone that was used to fit the north near structure for the color $(g-r)_0=0.65$.  For comparison, the isochrones that were used to fit the south middle structure have absolute magnitude $M_{g_0}=(6.723, 6.829, 6.911, 7.051)$ for metallicites [Fe/H]=$(-0.44, -0.61, -0.73, -0.88)$, respectively.  The absolute magnitude of the isochrone fit to the Monoceros and TriAnd Rings has an absolute magnitude of $M_{g_0}=7.272$ at $(g-r)_0=0.65$.  The use of the wrong absolute magnitude  (for example using $M_{g_0}=6.776$ instead of $M_{g_0}=7.272$, the difference between absolute magnitudes of the K stars from the outermost and innermost rings, and nearly the largest difference of the isochrones listed) for these outer rings could have introduced a 26\% distance error; since we are not fitting the structures out that far this was not a problem.  The distances derived from this model for the south middle structure are likely too large by about 4\% because we did not vary the absolute magnitude of our tracers with stellar population.

 We expect that lower Galactic latitudes, where both the star counts and the gradient in the star counts are higher, are more sensitive to the disk structure; therefore we use only directions with $10^\circ<|b|<20^\circ$.  At low latitudes, though, the larger extinction makes it impossible to detect the fainter stars in TriAnd and Monoceros Rings.  Even when these rings are visible, the completeness of the star counts is low.  Therefore, we only attempt to fit the disk star counts model to the near north and south middle structures.  We excluded some regions of the sky in fitting the model due to large reddening.

In Figure 17, we present histograms of the color-selected K-type star counts for each of our $2.5^\circ$ by $2.5^\circ$ sky patches.  We then compare the observed star counts with expectations from a very simple Galactic structure model with two exponential disks: a thin disk and a thick disk, and a cored, power law spheroid.  Note that the initial mass function of the stellar populations is not important here, since we are selecting only a small color range of primarily main sequence stars, which can be thought of as a single tracer population.

The number of stars expected in a given distance range, $R_1< R < R_2$, is given by:
\[A=\int \int_{R_1}^{R_2} \rho(R(r,z)) R^2 \mathrm{d}R \mathrm{d}\Omega,\]
where R is the distance from the Sun, and (r,z) are Galactocentric cylindrical coordinates, and $\rho$ is the stellar density.  The integral was performed numerically in the software package IDL, for each sky position and distance range.  The sky area $\mathrm{d}\Omega$ is the size of the patch of sky observed; the integral is only over the distance range of the stars in the bin.  The expected number of stars was calculated for eight bins in the range $16.5<g_0<20.5$, that are each half a magnitude wide. Given $M_{g_0}=6.776$, the distance range covered within the magnitude range is about 0.8 kpc to 5.6 kpc. In this magnitude range, we are fitting only the north near structure and the south middle structure.  Since these stars are selected in a narrow range of color, the errors in color near the limits of the survey will result in sampling a different population \citep{2011ApJ...743..187N}, so it is not advisable to fit the more distant Monoceros and TriAnd Rings.

The parameters for the thin and thick disk were chosen from the ranges presented by \citet{2001APJ...553L..184C}.  Because the star counts are sensitive to the scale height of the disk and the fraction of the stars in the thick disk, the numbers we use were chosen from the range given in \citet{2001APJ...553L..184C}, but were fit with a process that will be explained in detail later in this section.  The Sun's height above the Galactic plane, $z_0$, is 27 pc.  The scale height, $h_d$, and scale length, $l_d$ of the thin disk are 250 pc and 2250 pc, respectively.  The scale height, $h_{td}$, and scale length, $l_{td}$ of the thick disk are 700 pc and 3500 pc, respectively.  The local fraction of thick disk stars in the local volume is 8\% and the local fraction of halo stars is 0.125\%.
The thin disk and thick disk densities are given by:
\[\rho_d(r,z) \propto e^{(-r/h_d)} e^{(-|z|/h_d)},\]
\[\rho_{td}(r,z) \propto e^{(-r/h_{td})} e^{(-|z|/h_{td})}.\]
Here, $(r,z)$ are Galactocentric cylindrical coordinates, with z in the direction of the north Galactic pole.

The spheroid density function was adapted from Reid (1993), but we use a flattening of $q=0.8$ from, for example, Robin (1986).  The spheroid function is:
\[\rho_s(d(r,z)) \propto \frac{1}{a_0^n + d^n},\]
where $d=\sqrt{r^2+(z/q)^2}$, $n=3.5$, and $a_0=1000$ pc.

The total density is therefore:
\[\rho = \rho_0 (0.91875 \rho_d + 0.08 \rho_{td} + 0.00125 \rho_s), \]
where $\rho_0$ is the density of G-type stars at the solar position.  This number was adjusted so that the total number of stars in all of the panels in Figure 17 with model lines shown with solid lines, equaled the total area under the model curves in the same figure. For this normalization, both north and south panels were included in the calculation of a single overall normalization.   A summary of the model parameters is given in the top section of Table 2.

A comparison of the star counts to our simple model is shown in Figure 17.  The panels in which the model is shown as a dotted line are the panels that were not fit in the optimization of the model due to concern over high extinction.  Note that not a single panel portrays a good fit between the model and the observed shape of the apparent magnitude distribution.  Looking particularly at $l=203^\circ$, one sees that the peak in magnitude counts is closer to us than in the model for the northern fields, and farther than our model for the southern fields.  

Because we have already shown that the star counts are not symmetric above and below the Galactic plane, it is not surprising that the simple, symmetric model is a poor match to the data. Thus we attempt a simple modification to the density model as a demonstration of the kind of structure that would be required to fit the star counts we observe.  We modified the thin and thick disks so that the highest density (which was originally in the Galactic plane at $z=0$) oscillates up and down.  It was not sufficient to put in a simple sinusoidal oscillation.  To mimic the data we put in the positive half of a sine wave with a particular wavelength, amplitude and offset to model the north near structure.  We then put in the negative half of a sine wave, with an offset so that it started at zero where the sine wave for the north near structure went to zero, with a larger wavelength and amplitude to model the south middle structure.  This model is completely ad hoc, but does illustrate the type of modification that is necessary to fit the data. 

 Figure 18 shows the geometry of the wave we have added to the disk to attempt to match the star counts.  The density functions of the disks are modified to be:
\[\rho_d(r,z) \propto e^{(-r/h_d)} e^{(-|z-z_w|/h_d)},\]
\[\rho_{td}(r,z) \propto e^{(-r/h_{td})} e^{(-|z-z_w|/h_{td})},\]
where
\[z_w = A_w \sin(2 \pi (r - \phi_w)/\lambda_w),\]
$A_w$ is the amplitude of the oscillation.  The subscript ``$w$" indicates that the parameter is related to the wave that we are introducing. $r$ is the Galactocentric cylindrical radius, $\phi_w$ is the offset of the oscillation from the Galactic center, and $\lambda_w$ is the wavelength of the oscillation.  We use a different wavelength and offset for the part of the wave near the north near structure and for the part of the wave near the south middle structure.   The wavelength and offset of the north near structure have subscript ``$n$" and the wavelength and offset of the south middle structure have subscript ``$s$".  These four quantities can be derived from the three parameters $r_s, \lambda_s,$ and $\lambda_n$ from the equations:
\[r_s = r_s,\]
\[\phi_s = r_s - {\rm int} \left \{ \frac{r_s - \frac{3}{4} \lambda_s}{\lambda_s} \right \} \lambda_s - \frac{3}{4} \lambda_s, \]
\[r_n = r_s - \frac{1}{4} \lambda_s - \frac{1}{4} \lambda_n,\]
\[\phi_n = r_n - {\rm int} \left\{ \frac{r_n - \frac{1}{4} \lambda_n}{\lambda_n} \right \} \lambda_n - \frac{1}{4} \lambda_s.\]
Here, ``int" refers to the integer part of the quantity in brackets; this term assures that the offset will be less than one wavelength.
In the region where $r<r_n + \lambda_n/4$, we use $A_w=A_n$, $\phi_w=\phi_n$, $\lambda_w=\lambda_n$.  In the region where $r<r_n + \lambda_n/4$, we use $A_w=A_s$, $\phi_w=\phi_s$, $\lambda_w=\lambda_s$.

We now optimize the parameters in the model to best match the data.  The parameters that we fixed are five parameters for the oscillation: $A_n$, $A_s$, $\lambda_n$, $\lambda_s$, and $r_s$; and two parameters for the disk: $h_d$, and the local fraction of thick disk stars.  This optimization was accomplished using a grid search.  It turns out the fit is quite sensitive to the amplitudes $A_n$ and $A_s$, so these were fit first to be 70 pc and 170 pc, respectively.  The two wavelengths were fit between 3.0 and 7.8 in steps of 0.3 kpc.  The distance to the peak of the south middle structure was fit between 11.4 and 14 kpc, in steps of 0.2 kpc.  The thin disk scale height was fit between 90 and 350 pc in steps of ~50 pc, and the thick disk fraction was fit between 6.5\% and 13\% in steps of 1.5\%.  

The best fit values were $\lambda_n=6.3$ kpc, $\lambda_s=7.8$ kpc, $r_s=14$ kpc, $h_d=250$ pc, and the thick disk fraction was 8\%.  One can calculate from these numbers that $r_n=10.5$ kpc.  The distance to the near north structure is similar to the near north position in Figure 12, but the distance to the south middle structure is about 2 kpc farther away than in Figure 12.  The correction in distance from using an absolute magnitude that is brighter than we should have for this population only makes a difference of order 0.2 kpc.  The distances in Figure 12 were derived from the peak of the apparent magnitude histogram, which would make sense if we were measuring the distance to a narrow overdensity.  Instead this peak results from a combination of changing volume with distance, exponential decline in the disk density, and vertical oscillation of the disk midplane.


\begin{table}
\centering
\caption{ Summary of parameters in star counts models}
\medskip
\begin{tabularwithnotes}{crrr}
 {
  \tnote[]{ This table summarizes the parameters in our star counts models. In the top section, we present the parameters for the simple star counts model with exponential disks and a cored, power law stellar spheroid.  The model that includes wavelike oscillations uses all of the parameters above the line, plus the wave parameters in the lower section.}
 }
\toprule
parameter & fixed or not & range,step & value\\
\midrule
$Z_0$ & fixed & --- & 27(pc) \\
$h_d$ & not fixed & 90$\sim$350(pc),50(pc) & 250(pc)\\
$l_d$ & fixed & --- & 2250(pc)\\
$h_{td}$ & fixed & --- & 700(pc) \\
$\rho_{0td}$ & not fixed & $6.5\% \sim 13\%$, $1.5\%$ & 8\%\\
$\rho_{0halo}$ & fixed & --- & 0.125\%\\
q & fixed & --- & 0.8 \\
n & fixed & --- & 3.5 \\
\hline
$A_n$ & not fixed & $50\sim200(pc)$,10(pc) & 70(pc) \\
$A_s$ & not fixed & $50\sim200(pc)$,10(pc) & 170(pc) \\
$\lambda_n$ & not fixed & $3.0\sim7.8$,0.3(kpc) & 6.3(kpc)\\
$\lambda_s$ & not fixed & $3.0\sim7.8$,0.3(kpc) & 7.8(kpc)\\
$r_s$ & not fixed & $11.4\sim14$,0.2(kpc) & 14(kpc) \\
\bottomrule
\end{tabularwithnotes}
\end{table}

The starcounts resulting from the best fit model that includes the simple wave are shown in Figure 19.  Note that now the north near region near the anticenter is very well fit.  The rest of the panels are for the most part not better, but at least they are not worse.

One can see from the figure that the reason the model cannot be made to exactly match the data is that the model is axisymmetric, but the data is clearly different on either side of the Galactic anticenter.  For example look at the row where $b=-12.5^\circ$.  All of the data to the right of $l=178^\circ$ (high Galactic longitude) has a fairly sharp peak while all of the data to the left of $l=178^\circ$ (low Galactic longitude) is quite rounded.  While the data and the model are a good match at $l=130^\circ$ and $l=150^\circ$, the model and the data do not match at $l=203^\circ$ and $l=229^\circ$.  It is clear from the southern data that there is an asymmetry around the anticenter in the shape and position of the peak of the apparent magnitude distribution.  Since our model is axisymmetric, there was no way for the model to include this feature of the data.

 The wave amplitudes and positions we measure are reasonably similar to the simulations of \citet{2013MNRAS.429..159G}, which predicts disk oscillations due to infall of the Sagittarius dwarf galaxy.  Our measurements are more similar to the light ($M_{vir} = 10^{10.5} M_\odot$) Sagittarius dwarf.  However, it is unclear whether this is the only possible explanation, or in particular whether this is the only satellite that could be causing the observed disk oscillations.

\section{Relationship of rings to spiral arms and vertical waves}

\citet{2013ApJ...777...91Y} and \citet{2012ApJ...750L..41W} found an asymmetry in the number of stars above and below the Galactic plane, but the sign of the asymmetry oscillates with height above and below the disk.  Closer than 0.5 kpc, there are more stars in the south; at $0.5<|z|<1$ kpc there are more stars in the north.  At $1<|z|<2$ kpc, there are more stars in the south again.  Our near north structure is about 2 kpc from the Sun, and we probe $0.35<|z|<1.15$ kpc, so our finding that there are more stars in the north is reasonably consistent with their observation.  If it is true that there are also density oscillations with height, then we are possibly observing the combination of radial oscillations with vertical oscillations.   Because the vertical oscillations are apparently due to an oscillation of the disk midplane, the observations seem more consistent with a bending mode due to satellite infall \citep{2014MNRAS.440.1971W} rather than a spiral-induced perturbation \citep{2014MNRAS.440.2564F}.

One thing we noticed about Figure 12 is that the rings appear to be slightly farther from the Galactic center in the third quadrant than they are in the second quadrant, opening in the direction of the Milky Way's spiral arms.  We therefore asked ourselves whether these structures could be related to spiral arms.  Figure 16 of \citet{2014A&A...569A.125H} shows the locations of HII regions, giant molecular clouds (GMCs), and 6.7-GHz methanol masers that are used to trace the Milky Way's spiral arms.  This figure shows the Galactic warp; the spiral arms in quadrants 1 and 2 are north of the plane, and quadrant three is primarily south of the plane.  We note that the Perseus arm is about two kpc from the Sun in the direction of the anticenter; roughly at the same place as our near north structure.  However, most of the tracers are below the plane while our near north structure is denser above the plane.

Spiral arm tracers that could be related to the Outer Arm in Figure 16 of \citet{2014A&A...569A.125H}, in the region from $110^\circ<l<230^\circ$, appear to be 4-6 kpc from the Sun.  This location is similar to our south middle structure, but while the spiral arm tracers are predominantly above the plane, our south middle structure is below the plane.  We see the south middle structure at $10^\circ<|b|<20^\circ$, which corresponds to $-2.2<z<-1.5$~kpc, while the Outer Arm observations are typically 0.2 to 0.4 kpc  above the plane.

We are struck by the fact that previous authors have noted that in the solar neighborhood the asymmetry close to the Galactic plane is in the opposite direction to the asymmetry about one kpc above the plane.  We are seeing similar behavior at other Galactocentric radii, and that the waves may be related to the Milky Way's spiral structure.
 Note, however, that the structures could also be related to perturbations from a satellite galaxy; \citet{2013MNRAS.429..159G} have shown (see Figure 6 of that paper) that the vertical perturbations from satellite galaxies are not expected to be perfect rings, but more like spirals.  It is unclear how spiral structure in gas should be related to oscillations of stars induced by dwarf galaxy infall, but we note here that their structures are similar.

\section{The Monoceros and TriAnd Rings}

The identity of the Monoceros Ring in the north is fairly straightforward and consistent between different authors.  The only ambiguity arises from how much of the substructure at that distance is assigned to or associated with the ring and how much has been pulled out as separate structures.

The TriAnd Ring and the southern structures have been labeled differently by different authors and at different Galactic longitudes.  In \citet{2002ApJ...569..245N}, the structure S200-24-19.8, at $(l,b)=(200^\circ,-24^\circ)$, was tentatively associated with the Monoceros stream in the north (see Figure 26 of that paper), even though it was 0.4 magnitudes fainter than the northern structure.  When \citet{2003MNRAS.340L..21I} wrote the paper identifying the ``one ring to encompass them all,'' which is their description of the Monoceros Ring, they identified structure that we recognize as the Monoceros Ring in the north, but in their southern fields they identified stars that we call the south middle structure with the ``one ring."  \citet{2004ApJ...615..732R} originally discovered the overdensity covering the Triangulum and Andromeda constellations, with $100^\circ < l < 150^\circ$ and $-40^\circ<b<-20^\circ$, and by their estimation about 30 kpc from the Galactic center.  In the Rocha-Pinto paper, the nearer debris in the south, that we are calling the south middle structure, is labeled as the Galactic Anticenter Stellar Stream (GASS), which is their name for the Monoceros Ring.  \citet{2014ApJ...793...62S} showed that the structure studied by \citet{2004ApJ...615..732R} actually corresponds to the so-called ``TriAnd2'' overdensity, identified by \citet{2007ApJ...668L.123M} at a distance of $\sim28$~kpc. Sheffield et al. identify the more nearby ``TriAnd1'' at a distance of $\sim15-21$~kpc. \citet{2014MNRAS.444.3975D} independently found a heliocentric distance to the Triangulum-Andromeda overdensity of 20~kpc (in agreement with the \citealt{2007ApJ...668L.123M} distance to TriAnd1), which translates to 23 kpc from the Galactic center.  Our calculation of just over 20 kpc from the Sun agrees well with previous determinations of the TriAnd distance.

Our findings might suggest that the Monoceros and TriAnd Rings, along with the north near and south middle structures, are part of the disk.  These rings appear to fit into the pattern of spiral arms and density waves.  One can imagine that if the disk continues to decline in metallicity and increase in scale height past 14 kpc from the Galactic center, our observations would be in reasonable agreement with this finding.  In this scenario, the reason the disk appears to end at around 14 kpc is due to an oscillation in density, and the reason it appears again in a more distant ring is also due to an oscillation.  The spiral density wave pattern, though possibly not the spiral arms themselves (which require gas and star formation), would extend out to 20 kpc or more from the Galactic center.  The fact that stars in these structures appear to be rotating in approximately circular orbits in the same direction as the disk would then be explained.  It is also possible that the apparently narrow line-of-sight depth of the ring structures (especially the Monoceros Ring) could be explained by an oscillation of the disk up into our line of sight and then down out of our observation window.  The conflicting argument for these structures to be tidal features then rests on the observation of apparent filamentary streams in these structures, that might arise from the capture of groups of subhalos.

It remains to be determined whether: (1) the disk really extends as a single entity past 20 kpc from the Galactic center, and (2) disk oscillations, warps, flares, and spiral density waves can produce the highly substructured overdensities that are observed in for example the Pan-STARRS1 picture of anticenter substructure \citep{2014ApJ...791....9S}.  A possible solution to the controversy could be that the outer disk is built up from the accretion of small satellites, as proposed in \citet{2003ApJ...597...21A}.  If the individual accreted satellites have not yet fully mixed with the others, then we could simultaneously be seeing both satellite accretion and disk oscillations. 

\section{Conclusion}

In this paper we show that there is an oscillation in the number counts of stars in color-magnitude diagrams generated for patches of sky above and below the Galactic plane, in the Galactic latitude range $110^\circ<l<229^\circ.$  This number count oscillation corresponds to an oscillation of the stellar density asymmetry as a function of distance from the Sun, in the direction of the Galactic anticenter.

We identify four ``substructures" that represent the locations of peaks in the oscillations of the disk midplane, observed at about $\pm 15^\circ$ Galactic latitude, towards the Galactic anticenter.  Assuming a distance of 8 kpc from the Sun to the Galactic center, the North Near structure is 10.5 kpc from the Galactic center, the South Middle structure is 12-14 kpc from the Galactic center, the Monoceros Ring is 16.5 kpc from the Galactic center, and the TriAnd Ring is 21 kpc from the Galactic center.  The distances are determined from isochrone fits to main sequence stars.  In three of four substructures (South Middle, Monoceros, and TriAnd), the distance from the Galactic center appears to increase slightly with Galactic longitude, in the direction of Milky Way spiral arms.  We are unable to observe a significant extent of the North Near structure.

The Monoceros Ring has been studied by many authors, but here we associate only the identifications north of the Galactic plane.  Since the Monoceros Ring in the north is in between the South Middle and TriAnd Ring, both of these have, at different locations and by different authors, been previously associated with the Monoceros Ring in the north.  In this paper, we suggest that this association is incorrect.  The southern structures also cover a large range of Galactic longitudes and are on either side of the Monoceros Ring by more than four kpc.  The part of the TriAnd Ring that is in the longitude range $110^\circ<l<150^\circ$ has previously been identified as the Triangulum Andromeda Stream.

We fit the two nearer oscillations with a toy model in which the disk plane is offset by 70 pc up at 10.5 kpc from the Galactic center and 170 pc down at 14 kpc from the Galactic center in a somewhat sinusoidal pattern.  This model can fit the star counts in the anticenter, but because the disk is demonstrably not symmetric around the Galactic anticenter, this axisymmetric model is not sufficient to describe the data.  More complex and physically motivated models are needed to match the observed data.

We can tentatively connect the North Near structure with the position of the Perseus spiral arm and the South Middle structure with the position of the Outer spiral arm.  Maps by \citet{2014A&A...569A.125H} suggest that the Perseus arm is more obvious below the plane, while the North Near structure shows more stars above the plane.  Likewise, the Outer arm is more obvious above the plane while the South Middle structure shows more stars below the plane.  This can be reconciled if we also compare with Widrow et al. (2013) and Yanny \& Gardner (2014), who show that there are vertical oscillations in the stellar density in the solar neighborhood; close to the plane the density of stars is higher in the south while at $0.5<|z|<1.0$ kpc the density of stars is higher in the north.  If similar vertical density waves appear throughout the disk, then we can reconcile the opposite density patterns of the Outer arm and South Middle structure as well.  One imagines in this case that the vertical and radial oscillations are connected to spiral density waves in the Galaxy.

Several recent papers have discussed the origin of the Monoceros and TriAnd Rings.  TriAnd in particular is thought to be the result of satellite accretion, while the debate of the disk vs. satellite origin for the Monoceros Ring has lasted more than a decade.  In both of these structures, a number of smaller substructures that look like dwarf galaxies and tidal streams have been observed.  We suggest that the TriAnd and Monoceros Rings could look like both satellite accretion and like the disk if they in fact consist of accreted satellites that form the outer disk.   If the Monoceros and TriAnd rings are in the outer disk, then the stellar disk extends to at least 25 kpc from the Galactic center.  Previous measurements of the disk scale length and the observation of a sharp cutoff in stellar density about 15 kpc from the Galactic center \citep{1992ApJ...400L..25R} should be re-examined in light of the observed oscillations of the disk midplane.

 Our results roughly resemble the radial and vertical oscillations expected from the infall of the Sagittarius dwarf galaxy \citep{2013MNRAS.429..159G}.  As this is not the only effect that can result in such oscillations and as other satellites may have contributions, future work with stringent quantitative comparisons to our measurements is warranted.

\section{Acknowledgments} 
This work is supported by NSFC grant Nos. 11203030 and the National Key Basic Research Program of China 2014CB845703, as well as the US National Science Foundation under grant AST 09-37523 and AST 14-09421.   We thank the anonymous referee for comments that led to a clearer presentation of our results.
Funding for SDSS-III has been provided by the Alfred P. Sloan Foundation, the Participating Institutions, the National Science Foundation, and the U.S. Department of Energy Office of Science. The SDSS-III web site is http://www.sdss3.org/.
SDSS-III is managed by the Astrophysical Research Consortium for the Participating Institutions of the SDSS-III Collaboration including the University of Arizona, the Brazilian Participation Group, Brookhaven National Laboratory, Carnegie Mellon University, University of Florida, the French Participation Group, the German Participation Group, Harvard University, the Instituto de Astrofisica de Canarias, the Michigan State/Notre Dame/JINA Participation Group, Johns Hopkins University, Lawrence Berkeley National Laboratory, Max Planck Institute for Astrophysics, Max Planck Institute for Extraterrestrial Physics, New Mexico State University, New York University, Ohio State University, Pennsylvania State University, University of Portsmouth, Princeton University, the Spanish Participation Group, University of Tokyo, University of Utah, Vanderbilt University, University of Virginia, University of Washington, and Yale University.

\newpage

\begin{figure}
\includegraphics[scale=0.75]{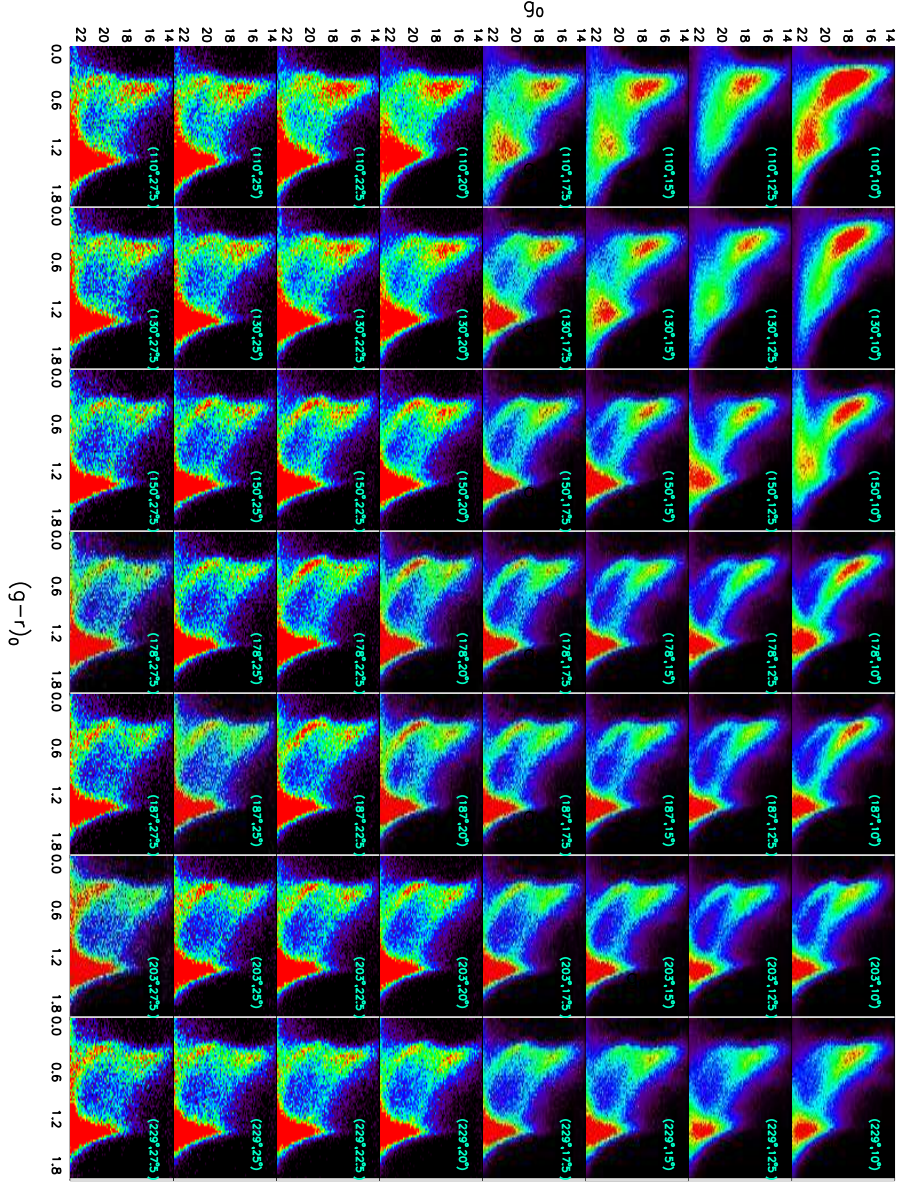}
\caption[hessnorth] {
\footnotesize
$g_0$ vs. $(g-r)_0$ Hess diagrams of SDSS data in $2.5^\circ \times 2.5^\circ$ patches of sky north of the Galactic plane.  Each panel is labeled with the central longitude of the data in that panel, and the lowest latitude.  For example, the top left panel includes SDSS photometric data with $128.75^\circ<l<131.25^\circ$, and $10^\circ<b<12.5^\circ$.  There are possibly two main sequences apparent in each of the patches: a brighter main sequence with a turnoff near $g_0=16.5$, and a fainter main sequence with a turnoff near $g_0=19.5$.  The fainter main sequence is associated with the Monoceros Ring.  The brighter main sequence (the ``near north" structure) is possibly related to the thin and thick disks, with the fraction of stars in the thin disk increasing at lower latitudes.  While the structure of the fainter main sequence is fairly similar in all panels, the brighter structure is narrower and bluer at low latitude, and is wider and brighter at higher latitude.  The Hess diagram bins are 0.1 wide in magnitude and 0.03 wide in color.  The color scale saturates with red at a maximum value of 140, 120, 80, 60, 40, 30, 25, and 20 at latitudes $10^\circ, 12.5^\circ, 15^\circ, 17.5^\circ, 20^\circ, 22.5^\circ, 25^\circ,$ and $27.5^\circ,$ respectively.
}\label{hessnorth}
\end{figure}

\begin{figure}
\includegraphics[scale=0.75]{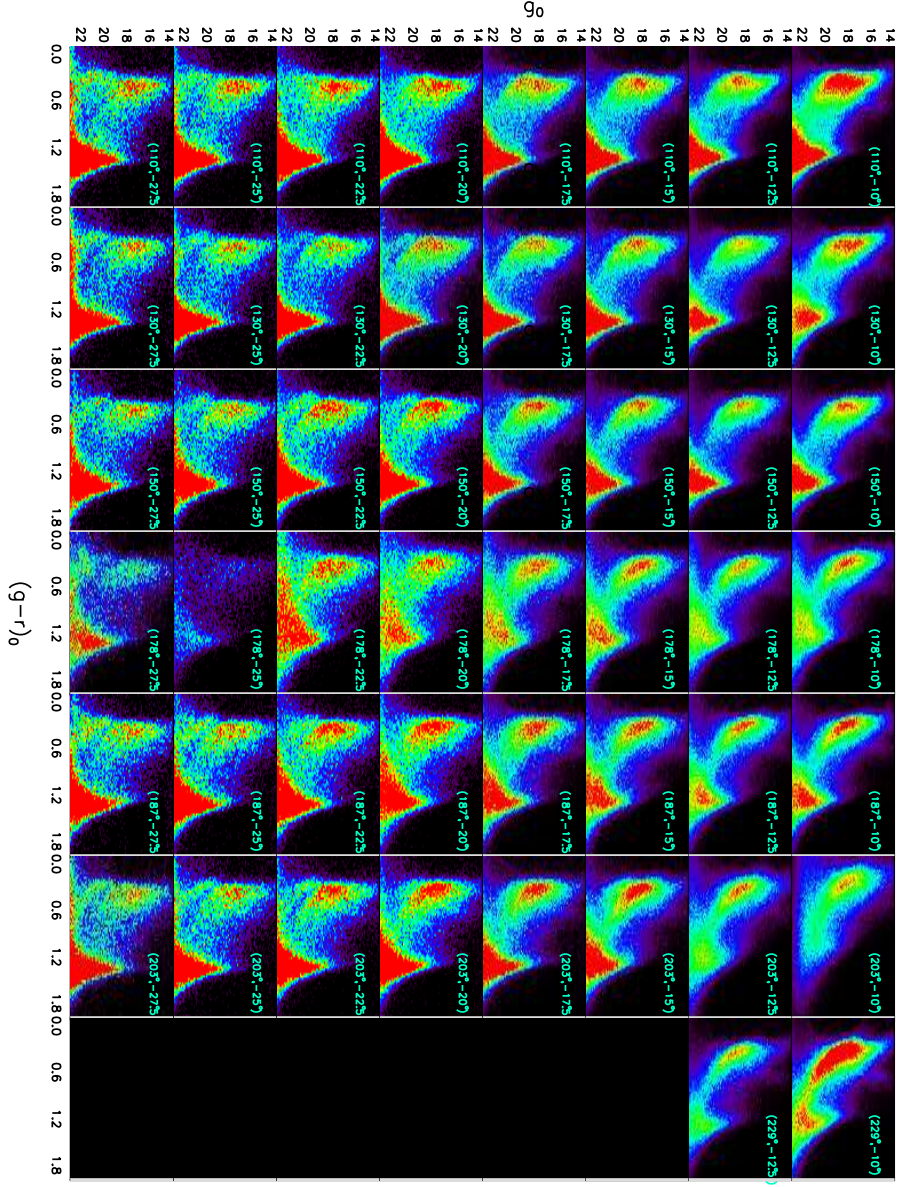}
\caption[hesssouth] {
\footnotesize
$g_0$ vs. $(g-r)_0$ Hess diagrams of SDSS data in $2.5^\circ \times 2.5^\circ$ patches of sky south of the Galactic plane.  Each panel is labeled with the central longitude of the data in that panel, and the highest latitude.  For example, the top left panel includes SDSS photometric data with $128.75^\circ<l<131.25^\circ$, and $-12.5^\circ<b<-10^\circ$.  Unlike the north, most of the panels in this figure include only one, broad main sequence with a turnoff near $g_0=18$ (the ``south middle" structure), which is half way between the magnitudes of the turnoffs in the north.  The panels further from the Galactic plane include a faint, narrow main sequence with a turnoff at $g_0 \sim 20$ that is slightly fainter than the turnoff of the fainter, narrow main sequence in the north.  The Hess diagram bin sizes and color scales are the same as those for the northern panels in Figure 2.
}\label{hesssouth}
\end{figure}

\begin{figure}
\includegraphics[scale=0.75]{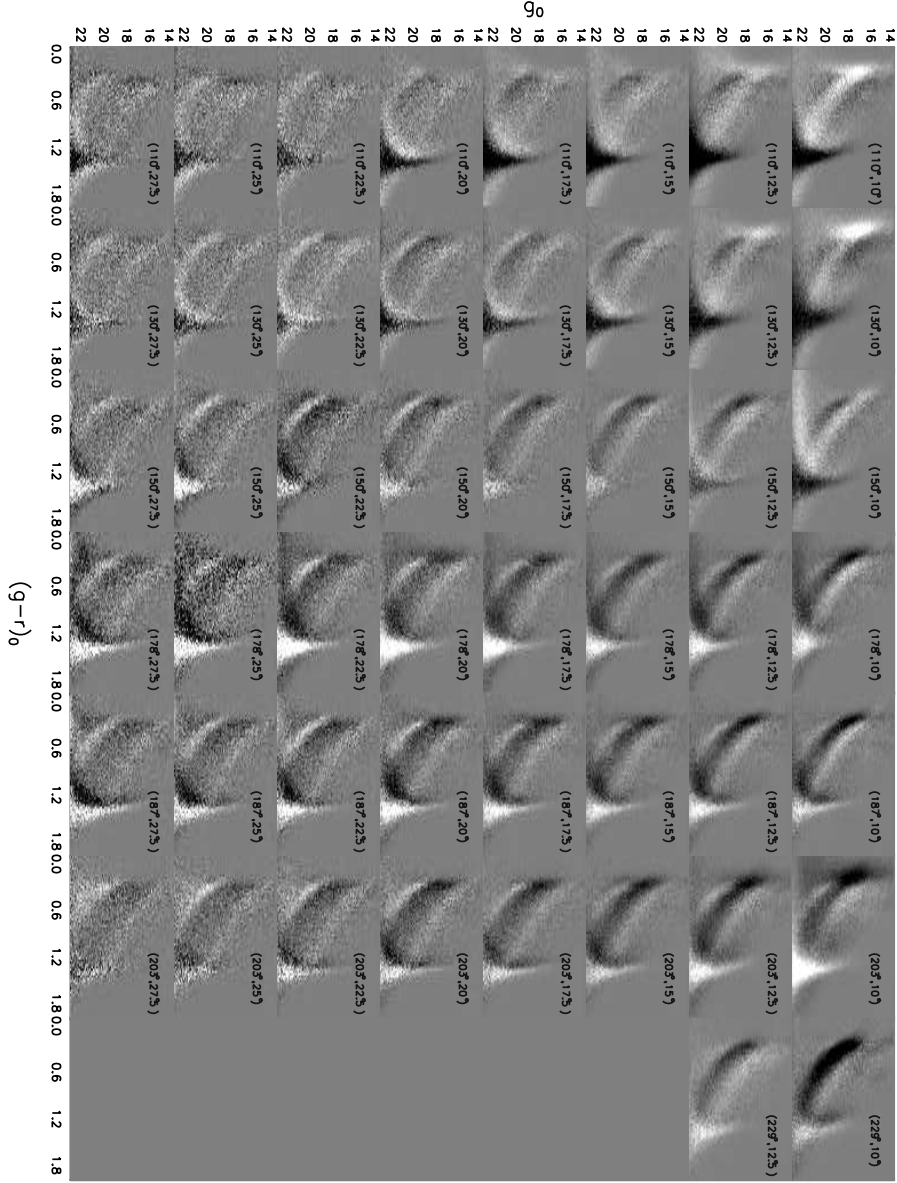}
\caption[residualhess] {
\footnotesize
Hess diagrams showing the asymmetry in star counts north and south of the Galactic plane.  Each panel is the difference of the corresponding panel in Figure 2 minus the same panel in Figure 3.  The grey scale varies with latitude; for panels with Galactic latitude of $10^\circ, 12.5^\circ, 15^\circ, 17.5^\circ, 20^\circ, 22.5^\circ, 25^\circ$ and $27.5^\circ$, the dynamic range of star counts of each pixel is $\pm 100, \pm 90, \pm 80, \pm 60, \pm 40, \pm 30, \pm 25,$ and $\pm 20$, respectively.  Notice the alternating pattern of black and white stripes as a function of apparent magnitude.  One explanation for this pattern is that the midpoint of the disk could oscillate up and down through the Galactic plane as a function of distance from the Galactic center.  Some of the panels show an excess of M dwarf stars ($(g-r)_0>1.2$) in the south or in the north, or a color offset between the north and south M dwarfs.  These differences are likely due to nearby, low latitude extinction that removes many fainter stars from the sample in some directions; for example the panels $(l,b)=(130^\circ,10^\circ)$ and $(l,b)=(203^\circ, -10^\circ$) are regions of high extinction. 
}\label{residualhess}
\end{figure}

\begin{figure}
\includegraphics[scale=0.65]{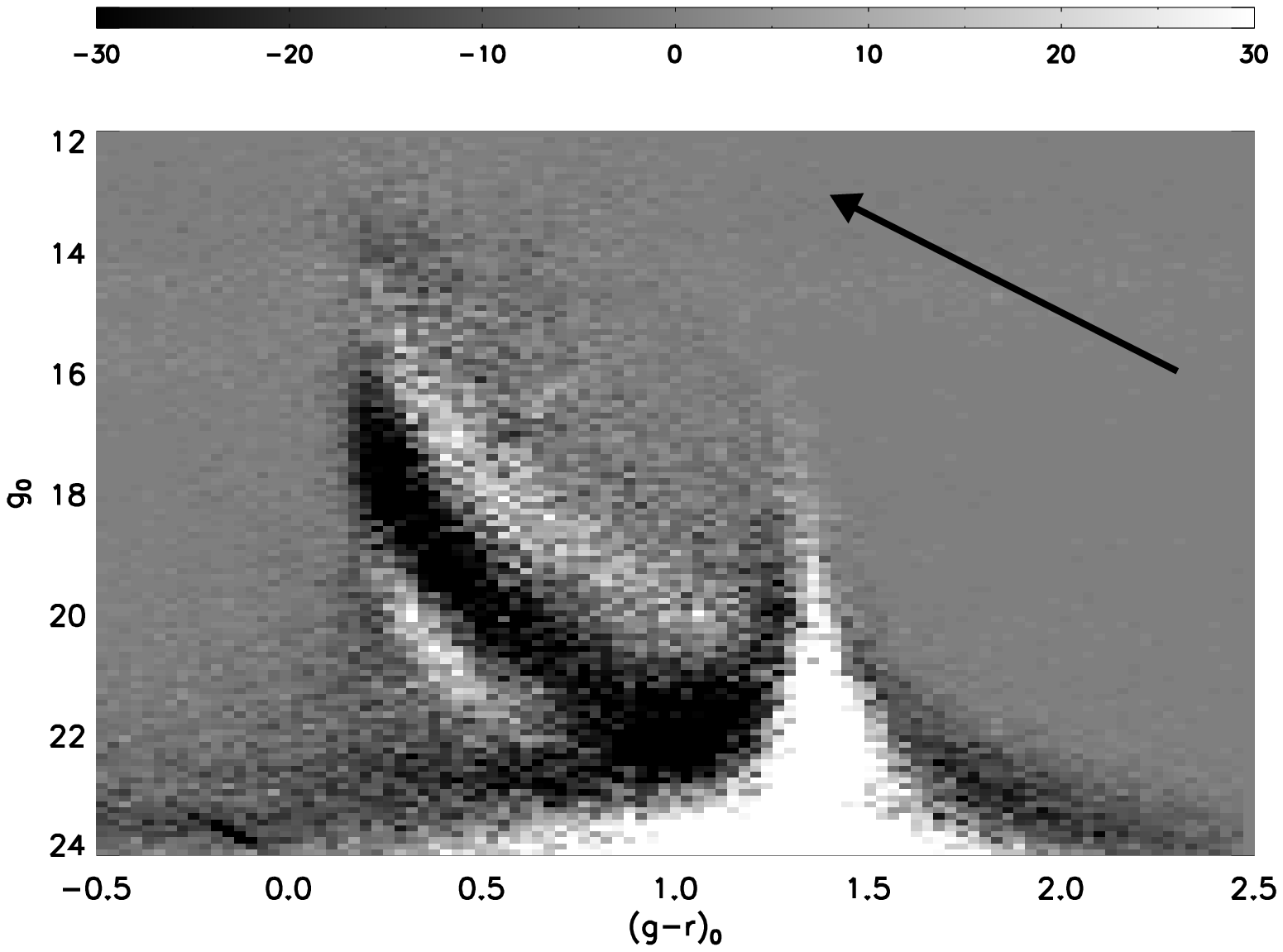}
\includegraphics[scale=0.65]{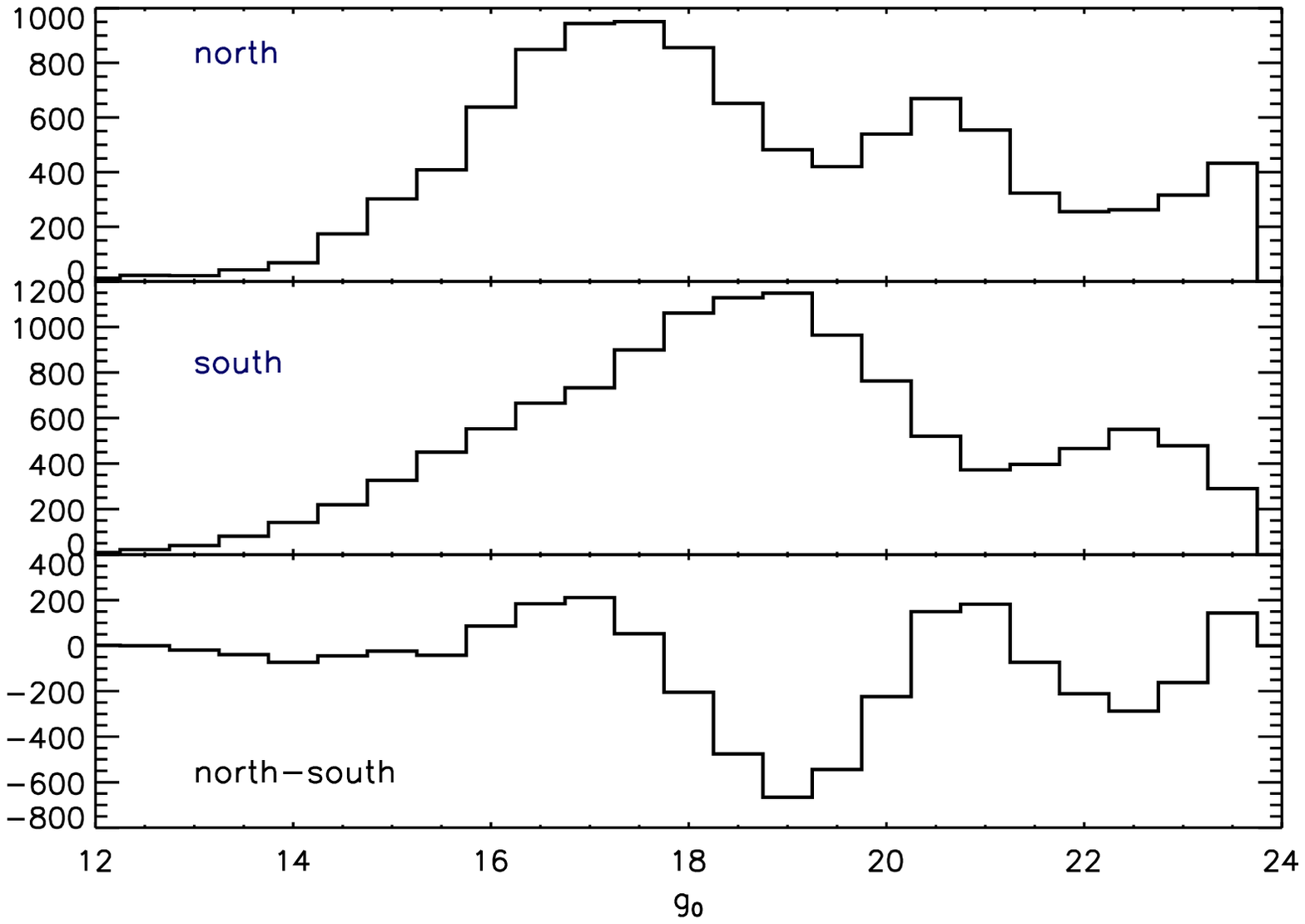}
\caption[residualhess178.15] {
\footnotesize
Sample Hess diagram showing the alternating black and white striping that indicates an asymmetry in the disk with a sign that alternates in distance from the Galactic center.  The top panel is the result of subtracting a Hess diagram of the southern sky, $176.75^\circ<l<179.25^\circ$ and $-17.5^\circ<b<-15^\circ$, from a Hess diagram of a symmetric section of the northern sky, $176.75^\circ<l<179.25^\circ$ and $15^\circ<b<17.5^\circ$.  The arrow shows the direction of the reddening vector.  Clearly, a poor correction for reddening will not reduce the measured asymmetry.  The lower panel shows the star counts ($0.4<(g-r)_0<0.5$) as a function of magnitude for the northern patch, the southern patch, and the subtraction of the two.  The apparent magnitudes of the peaks differ by more than a magnitude (a 60\% implied distance error).  The star counts can differ by as much as a factor of two between the north and south at a given apparent magnitude.
}\label{residualhess178.15}
\end{figure}

\begin{figure}
\includegraphics[scale=0.8]{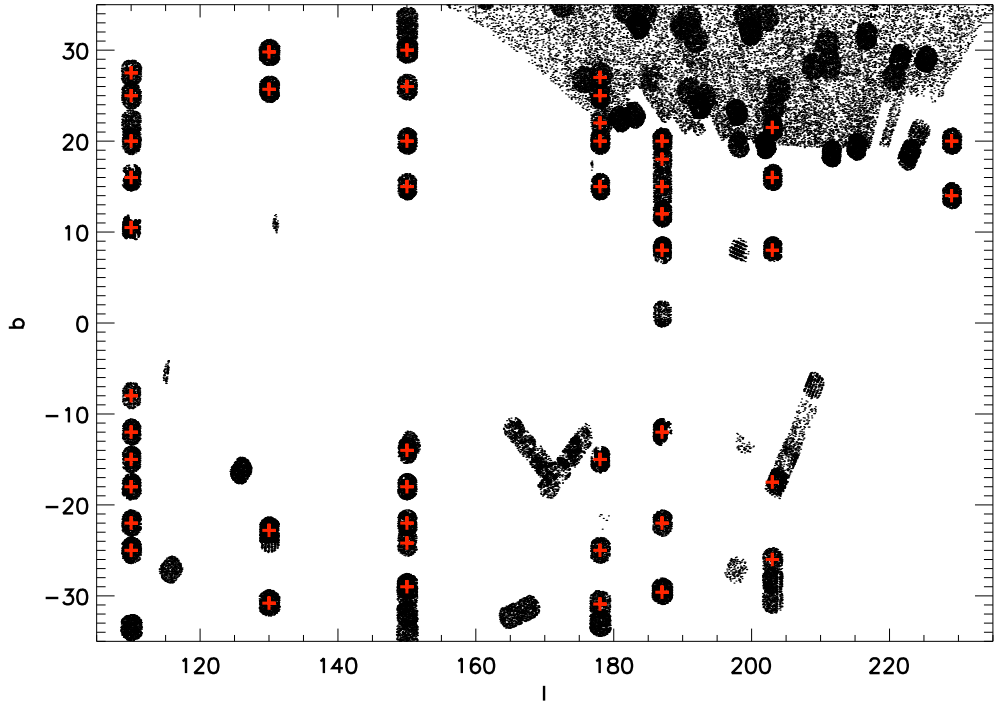}
\caption[footprint] {
\footnotesize
SDSS Stellar Spectra.  Black dots show the locations of stars with spectra in the SDSS, in the region near the Galactic anticenter.  Red crosses show the positions of the 46 plates (26 in the north and 20 in the south) that were selected to study the metallicities of the nearer ridgelines.
}\label{footprint}
\end{figure}

\begin{figure}
\includegraphics[scale=0.65]{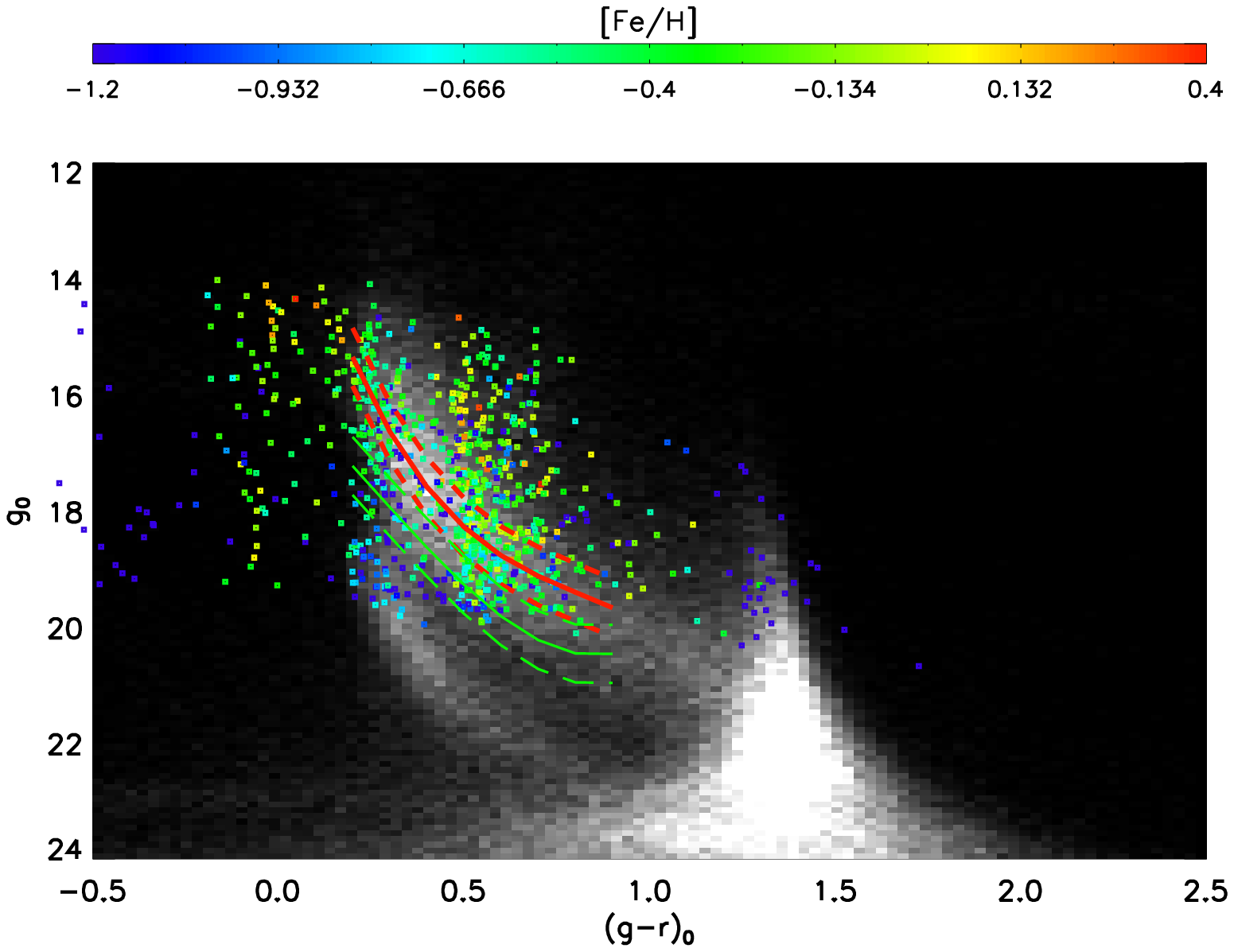}
\includegraphics[scale=0.65]{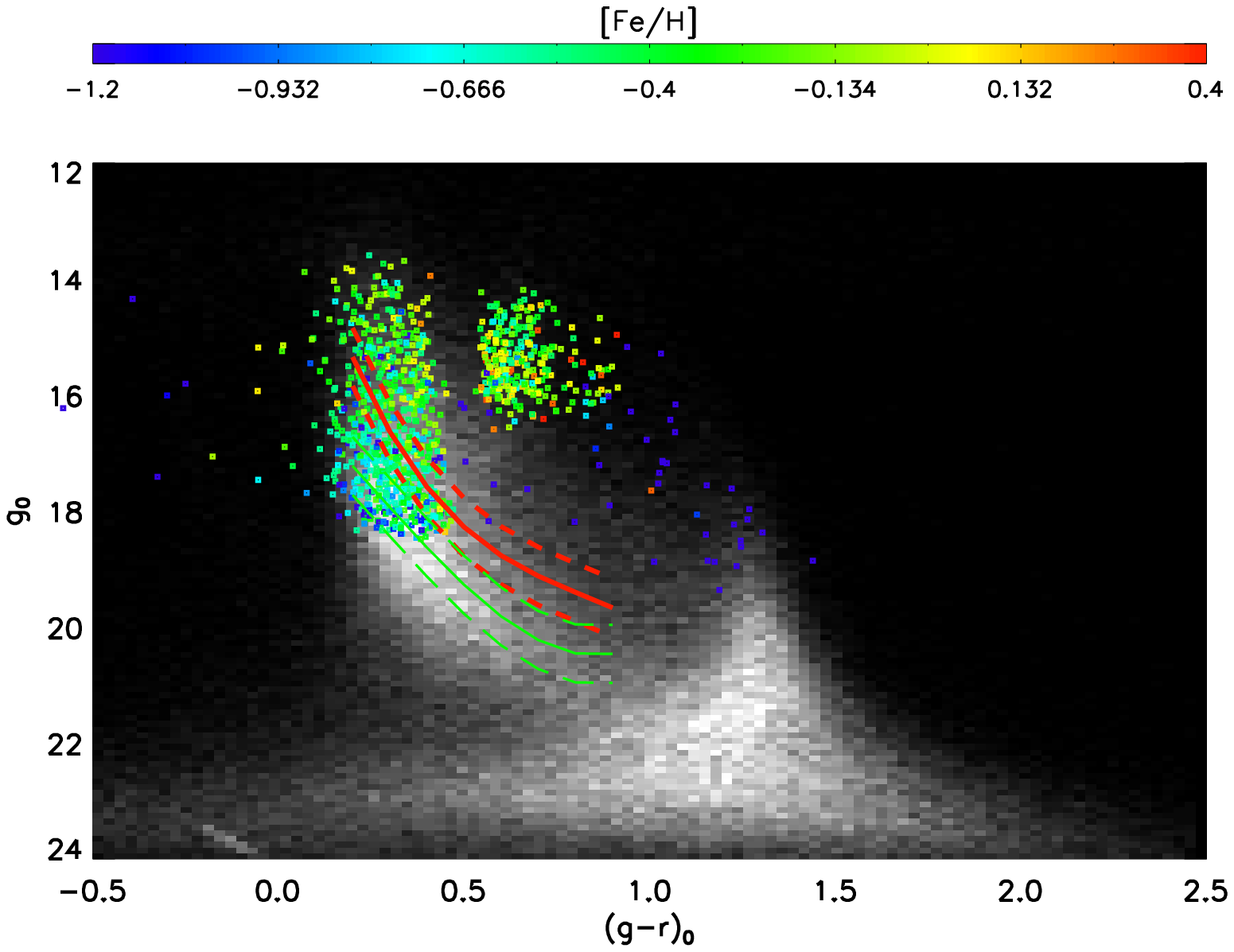}
\caption[selectspecn] {
\footnotesize
Selection of spectra in north near and south middle structures.  The top panel shows a Hess diagram of the stars with photometric measurements in SDSS, in the region of the sky near $(l,b)=(178^\circ, 15^\circ)$.  The lower panel shows a Hess diagram of the corresponding patch of sky south of the Galactic plane.  The positions of SDSS spectra in this region of the sky are overlaid on the CMD; the colors indicate the measured metallicity of each point as shown in the color bar.  A polynomial was fit to the apparent center of the brighter main sequence of stars in the north, as shown by the solid red line. The dashed lines show the region of the CMD in which north near spectra were selected.  The red lines are also shown on the southern Hess diagram, to show the selection of the north near comparison spectra.  The green line in the lower panel shows a polynomial fit to the ridgeline of south middle structure. The dashed green lines show the region of the CMD in which spectra were selected for south middle structure.  The green lines are also shown in the upper panel, indicating the region from which south middle comparison spectra were selected.  
}\label{selectspecn}
\end{figure}

\begin{figure}
\includegraphics[scale=0.5]{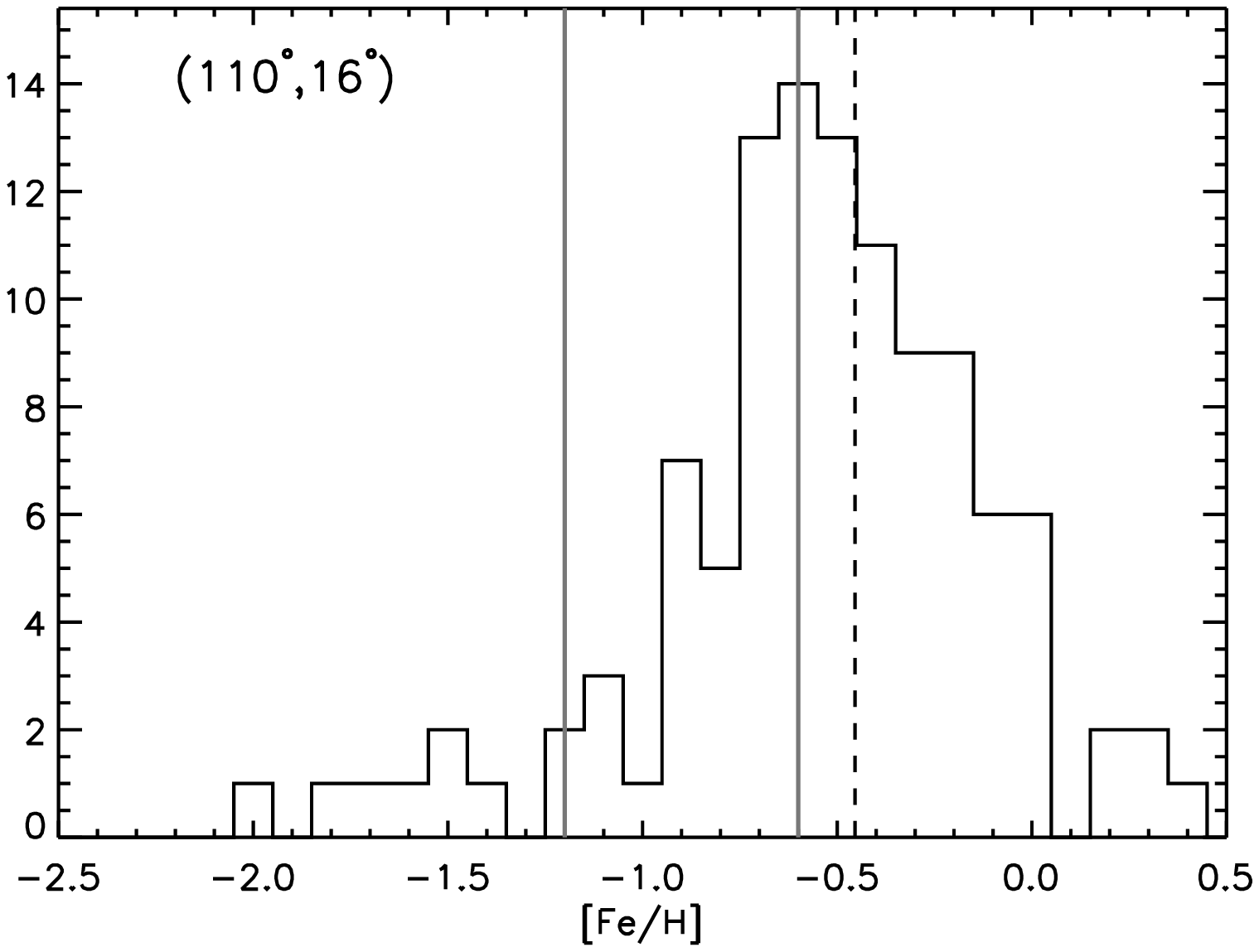}
\includegraphics[scale=0.5]{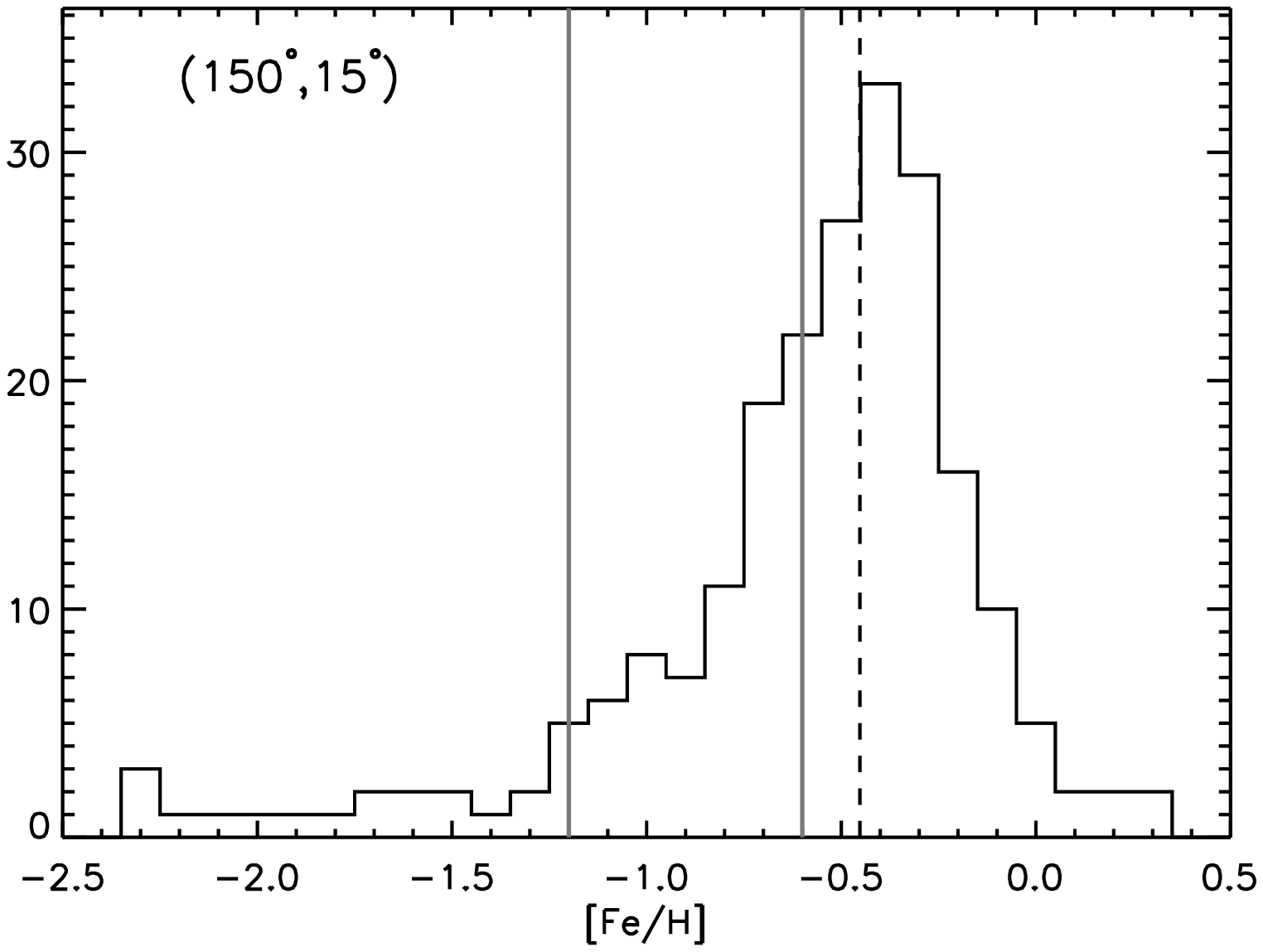}
\includegraphics[scale=0.5]{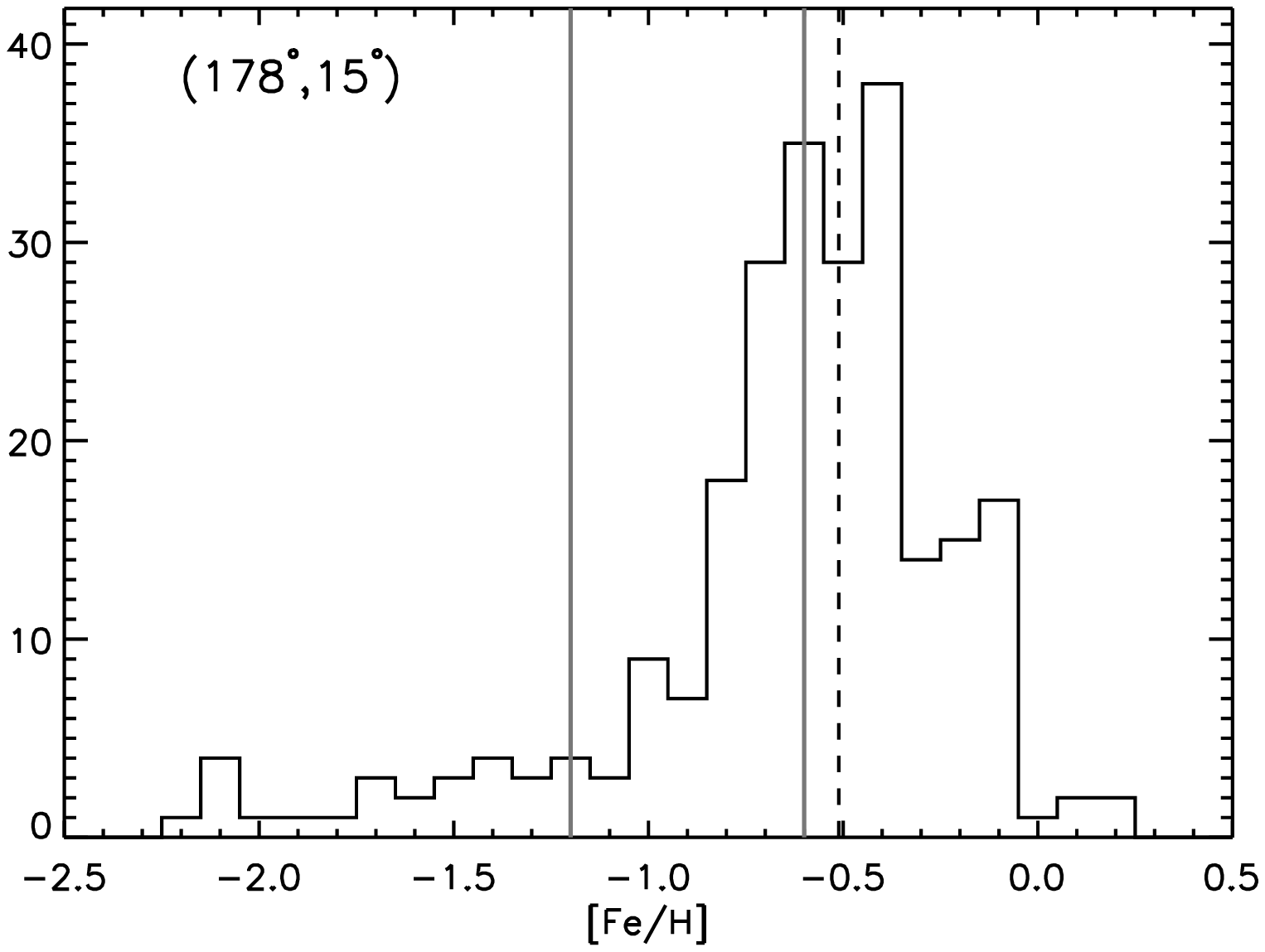}
\includegraphics[scale=0.5]{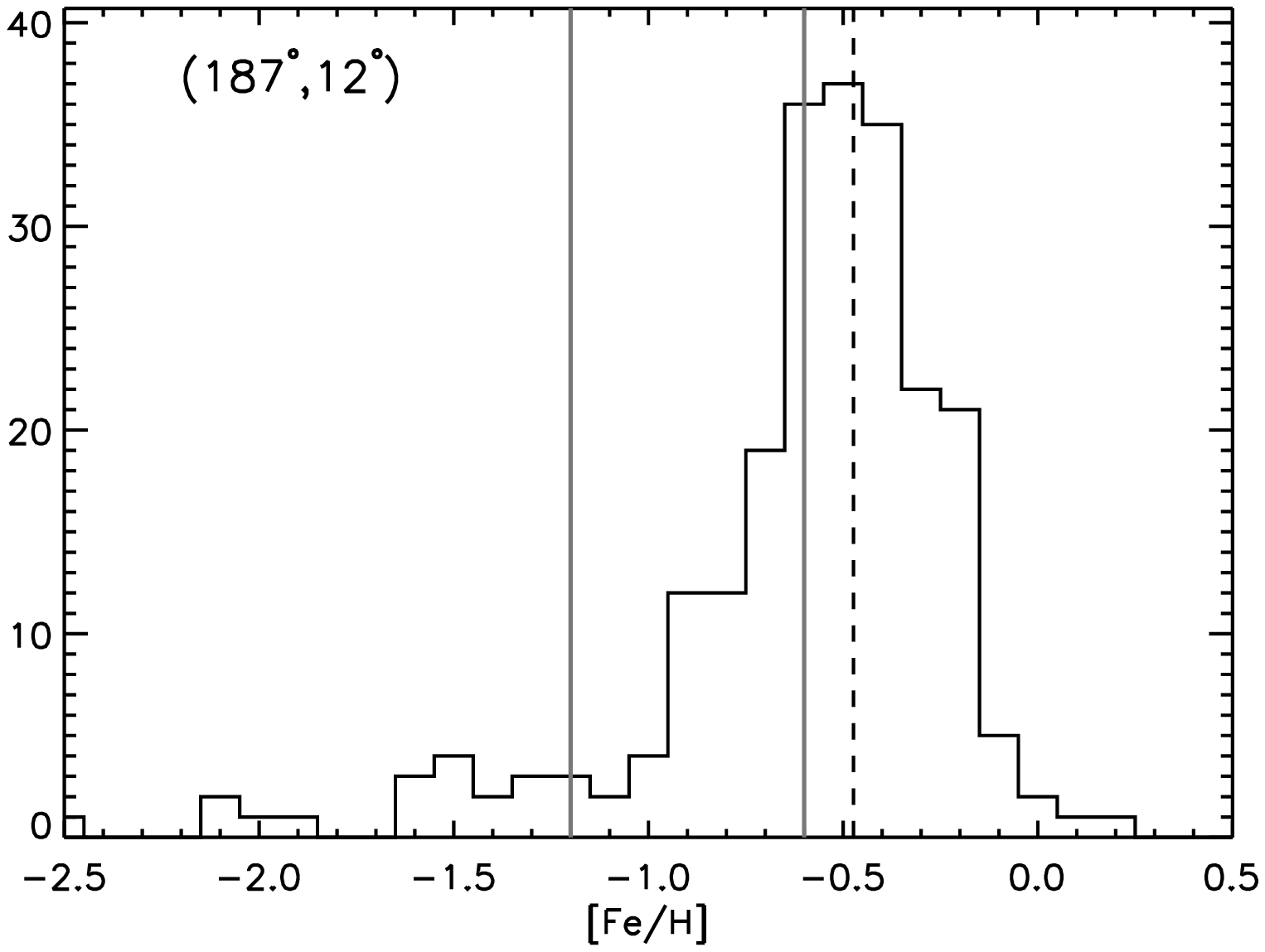}
\includegraphics[scale=0.5]{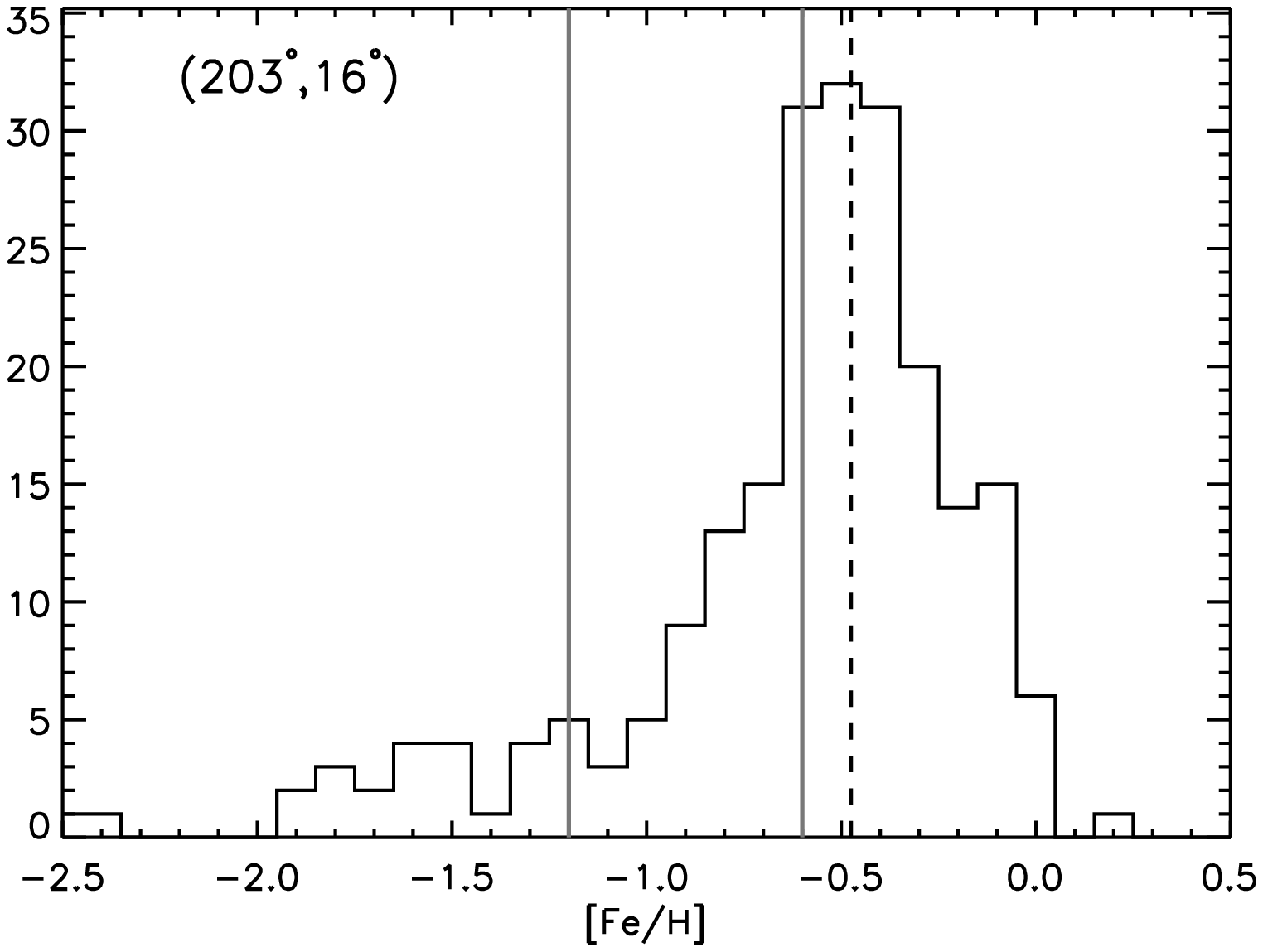}
\includegraphics[scale=0.5]{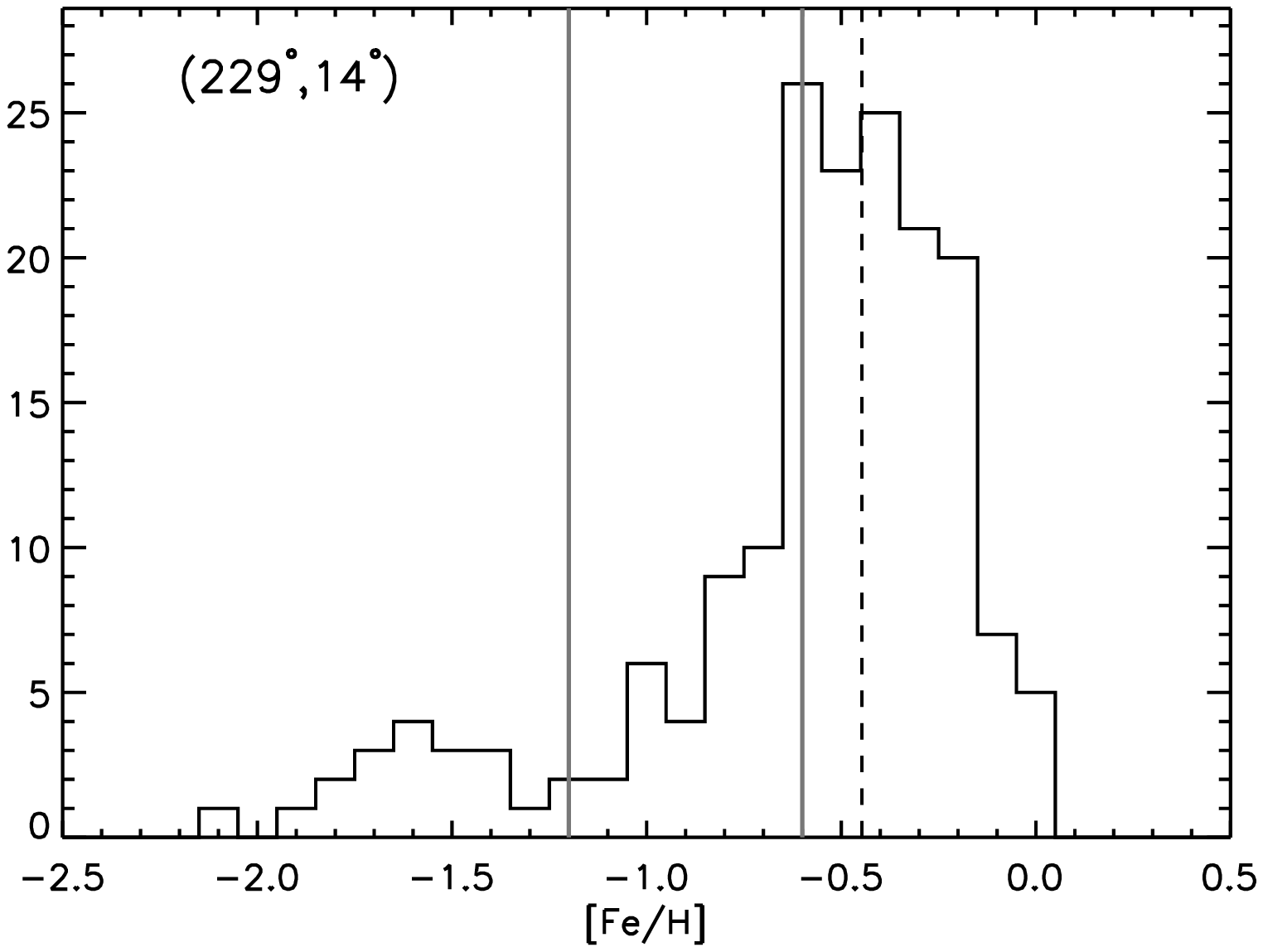}
\caption[metdist] {
\footnotesize
 Metallicity distribution in the north near structure, at six different longitudes.  We present the spectra that coincide with the position in the Hess diagrams of the north near structure near $b\sim 15^\circ$. The vertical lines at [Fe/H]=-0.6,-1.2 roughly separate stars with the metallicities of the thin disk, thick disk, and halo populations (Li et al. 2012).  The majority of the stars in the north near structure have metallicities consistent with our expectations for the thin disk. The dashed line shows the median value of [Fe/H] in each panel. 
}\label{metdist}
\end{figure}

\begin{figure}
\includegraphics[scale=0.5]{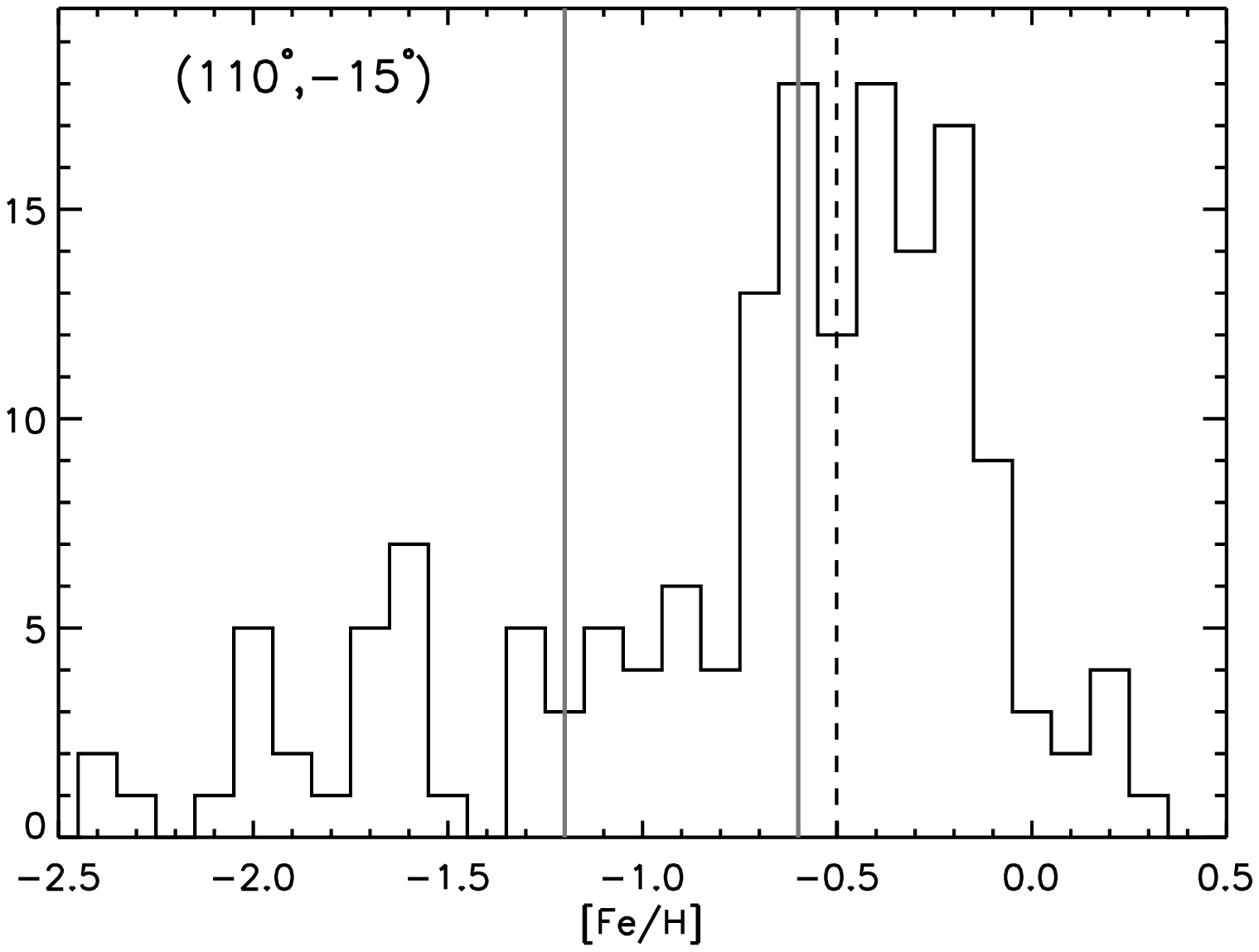}
\includegraphics[scale=0.5]{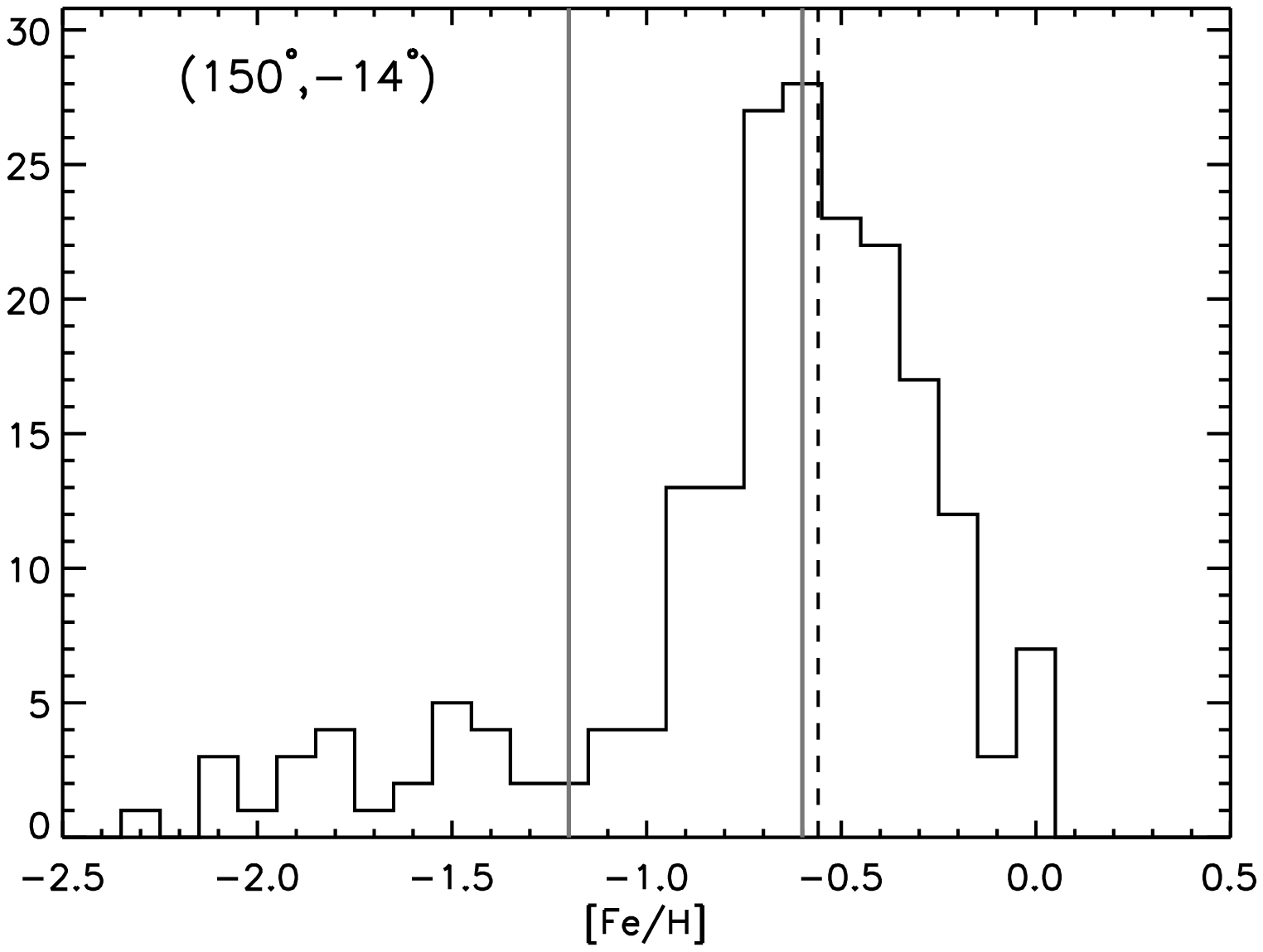}
\includegraphics[scale=0.5]{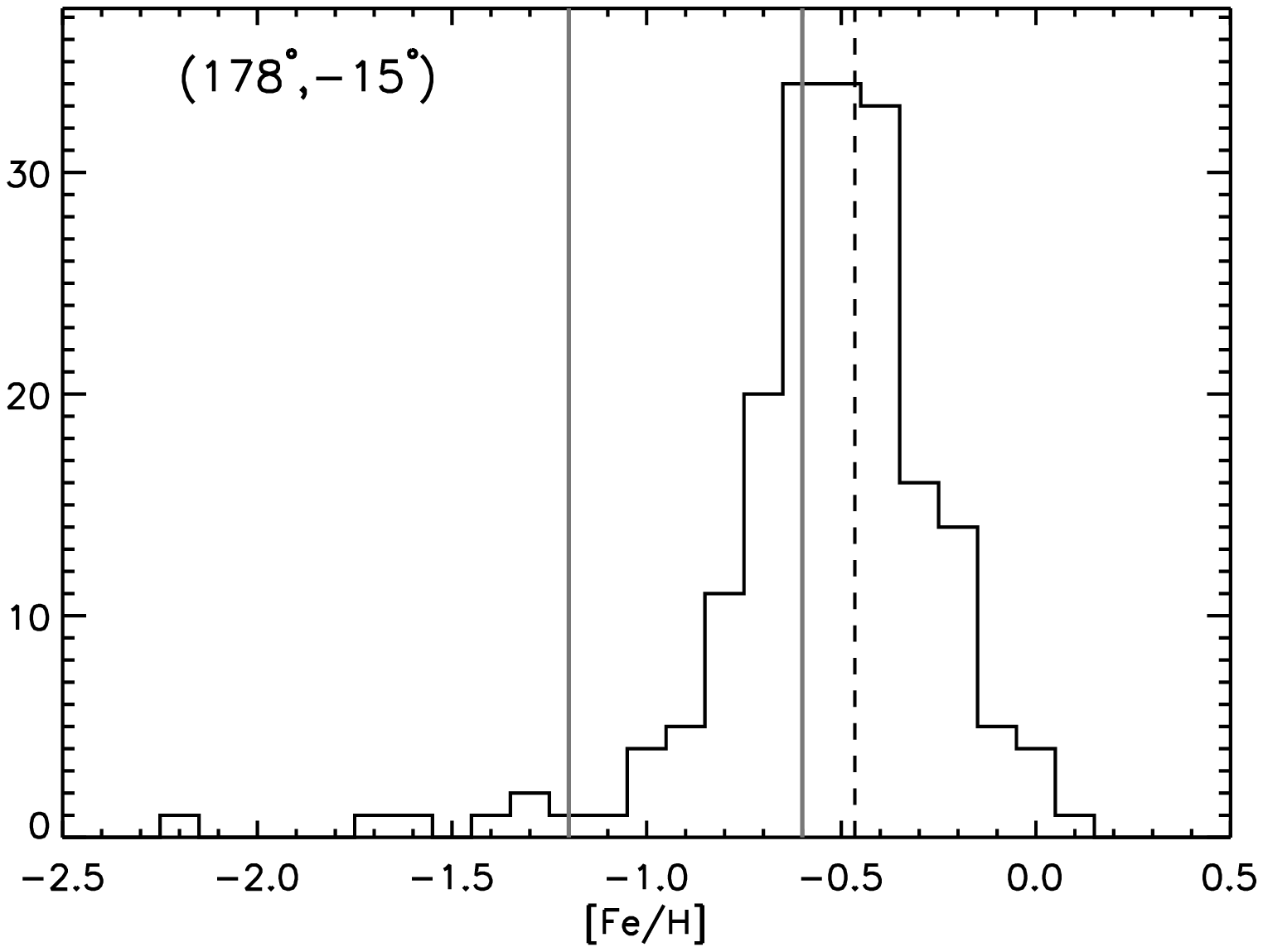}
\includegraphics[scale=0.5]{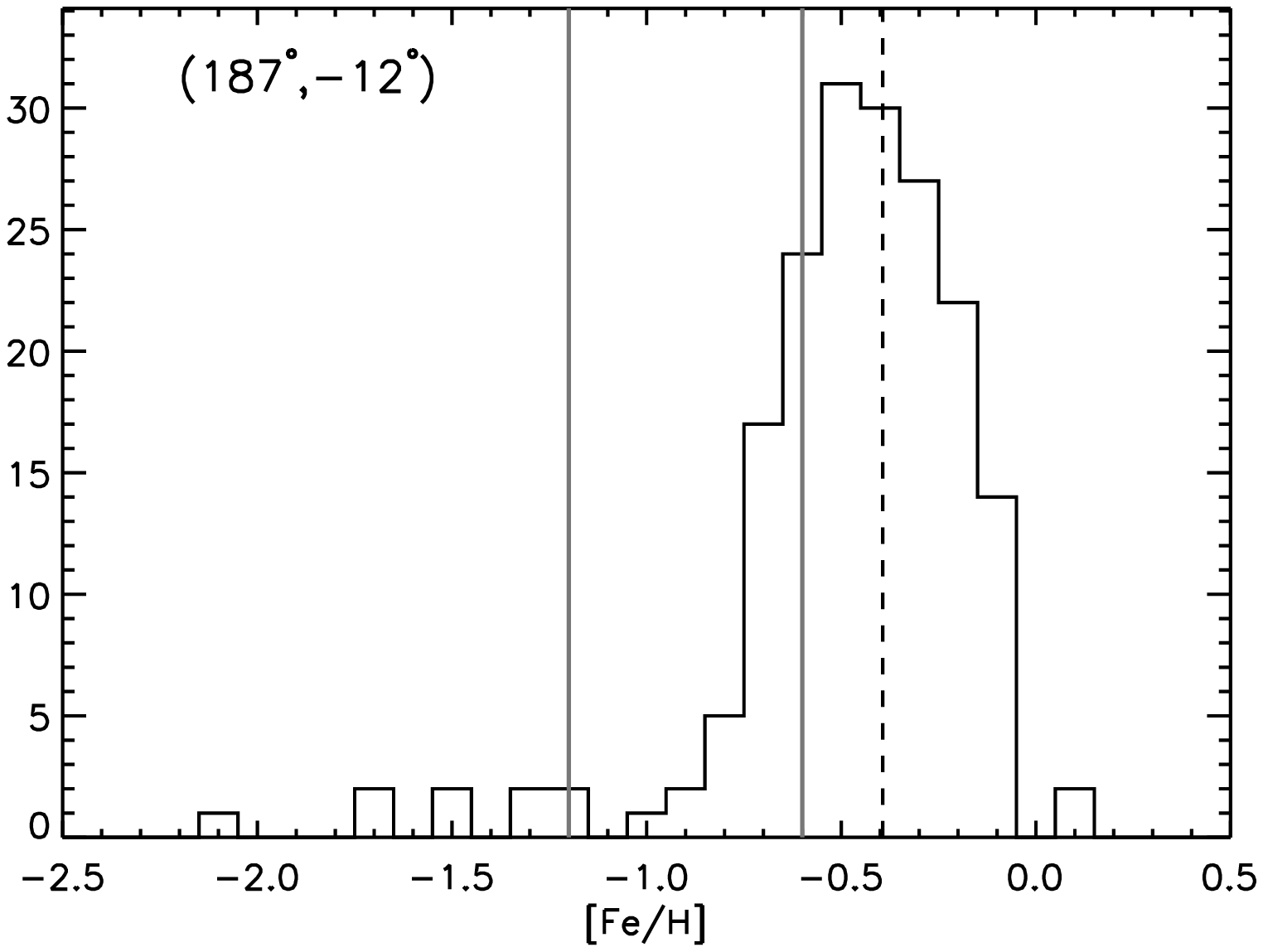}
\includegraphics[scale=0.5]{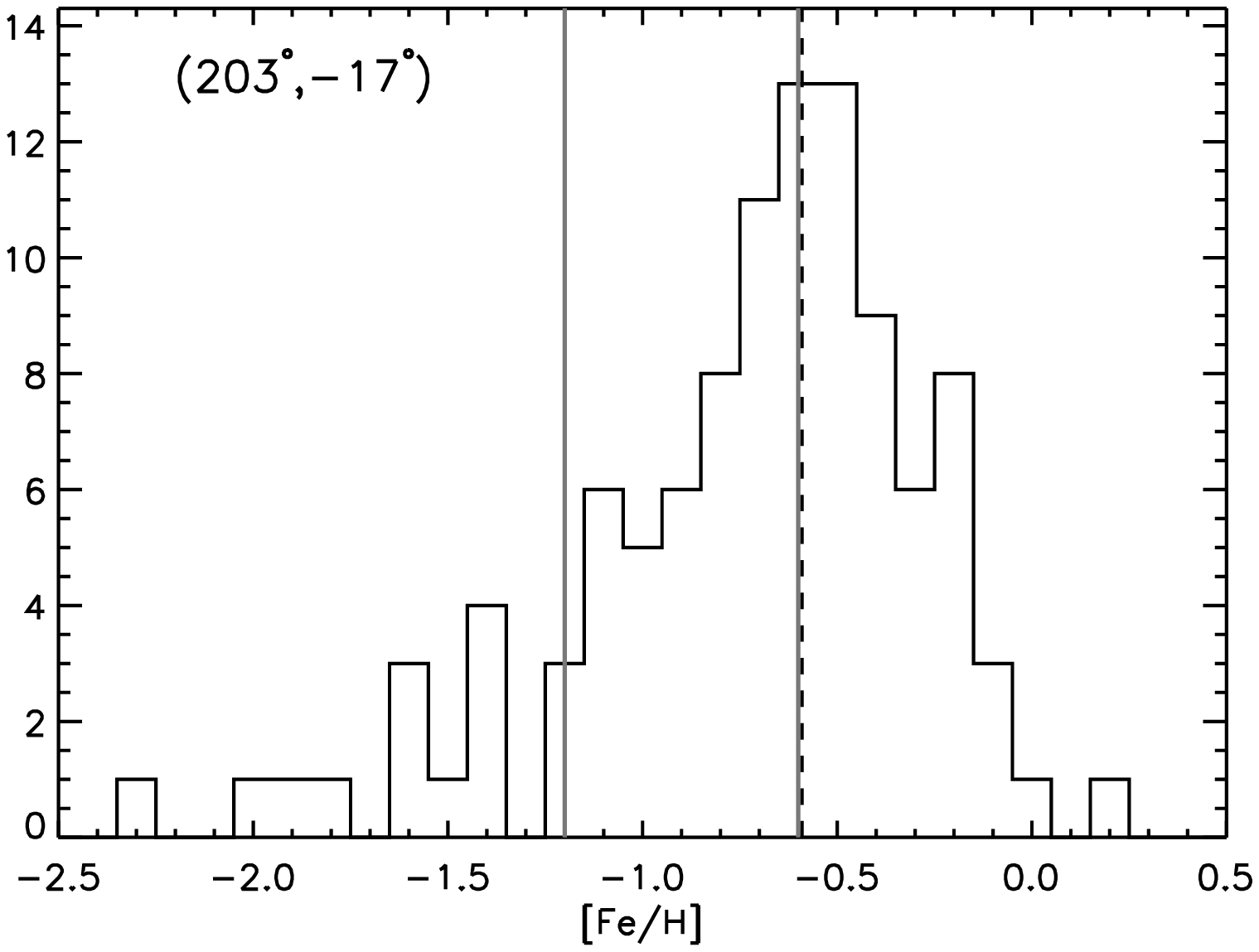}
\caption[velmet] {
\footnotesize
 Metallicity distribution in the south comparison fields for the north near structure.  The dashed line shows the median value of [Fe/H] in each panel.  We present spectra that coincide with the position in the Hess diagrams of the north near structure at five different Galactic longitudes, but look in the south near $b \sim -15^\circ$ where we don't expect to see the north near structure.  We see that the stars at a symmetric distance below the Galactic plane have metallicities that are similar to the north near structure, and match our expectations for disk stars.
}\label{velmet}
\end{figure}

\begin{figure}
\includegraphics[scale=0.5]{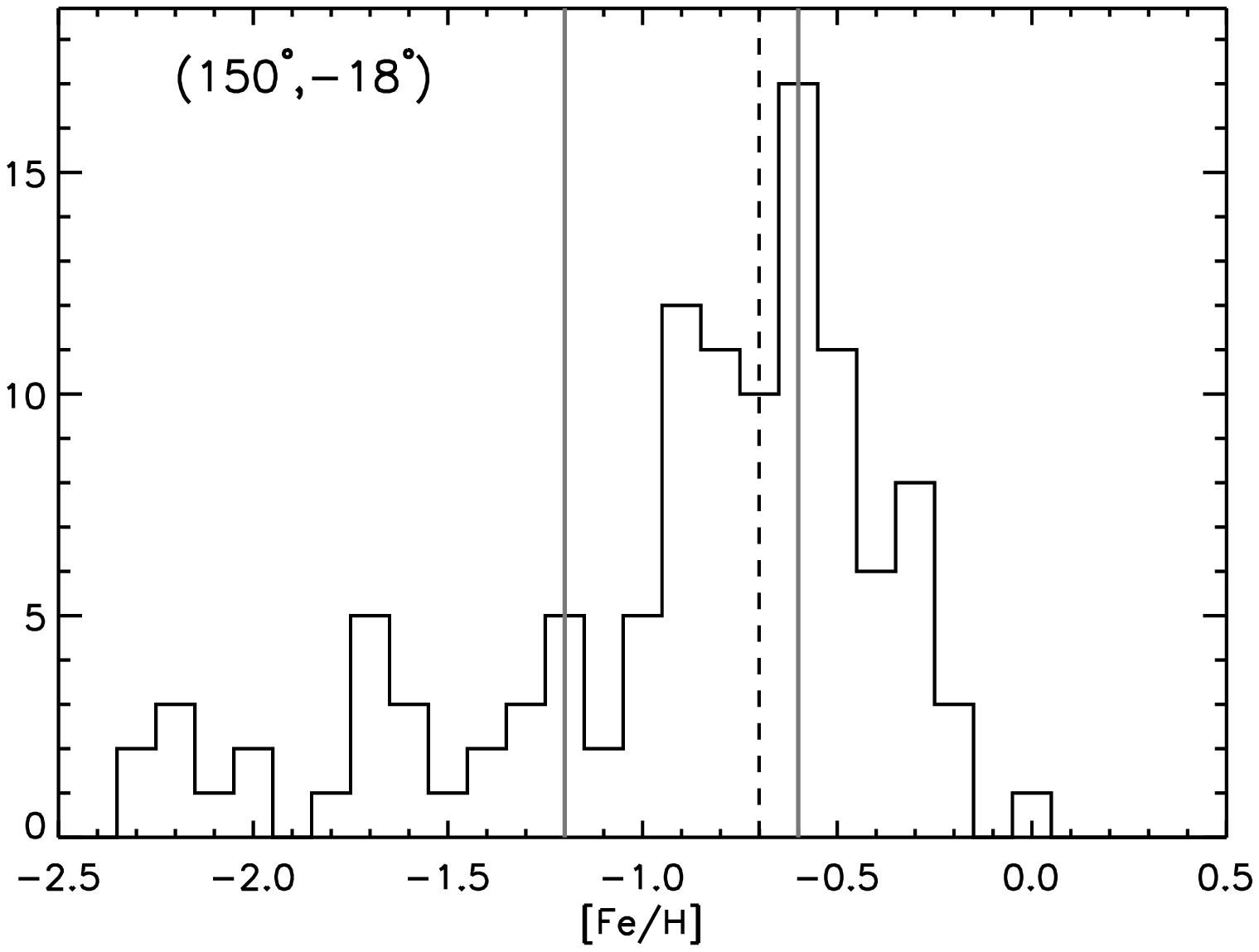}
\includegraphics[scale=0.5]{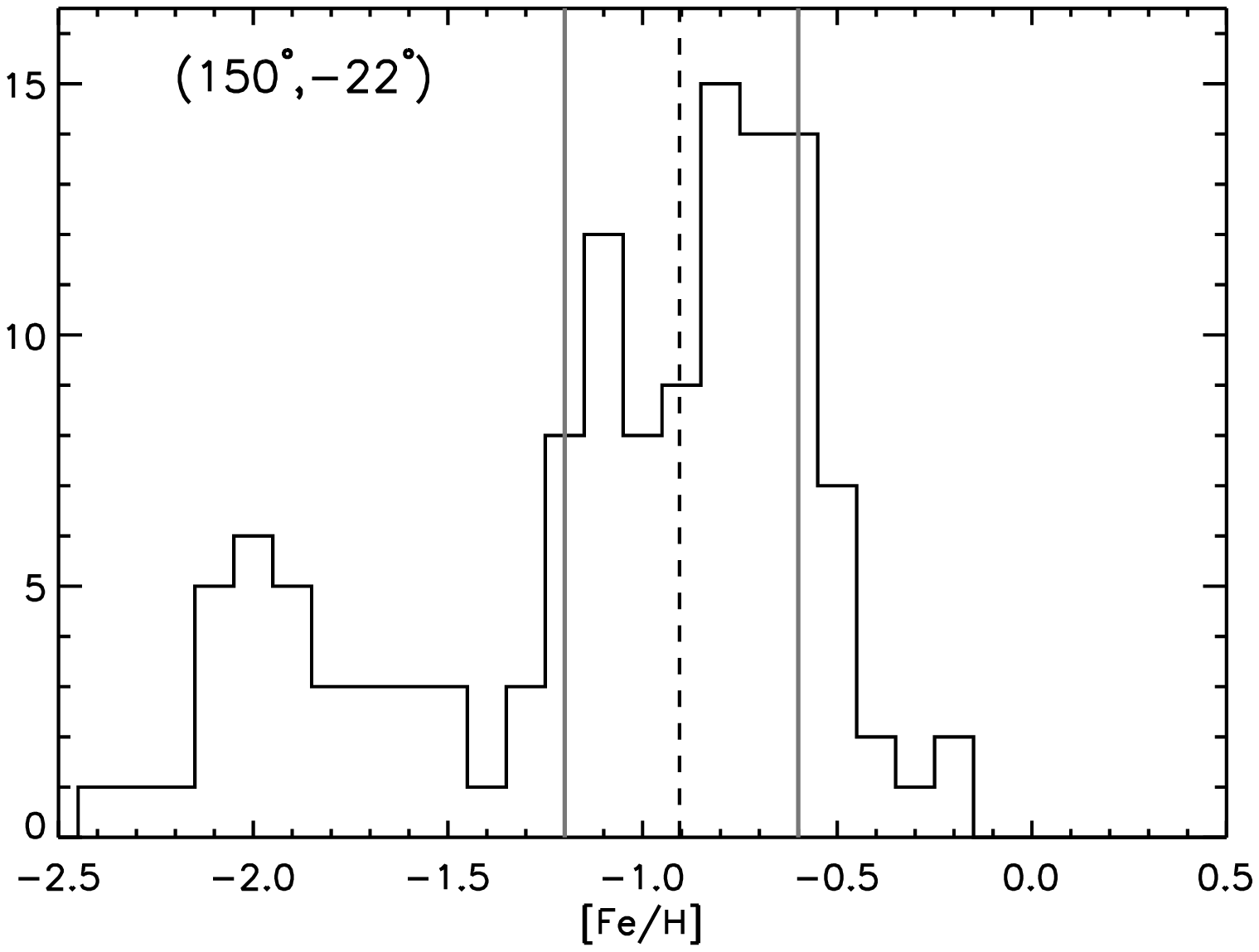}
\includegraphics[scale=0.5]{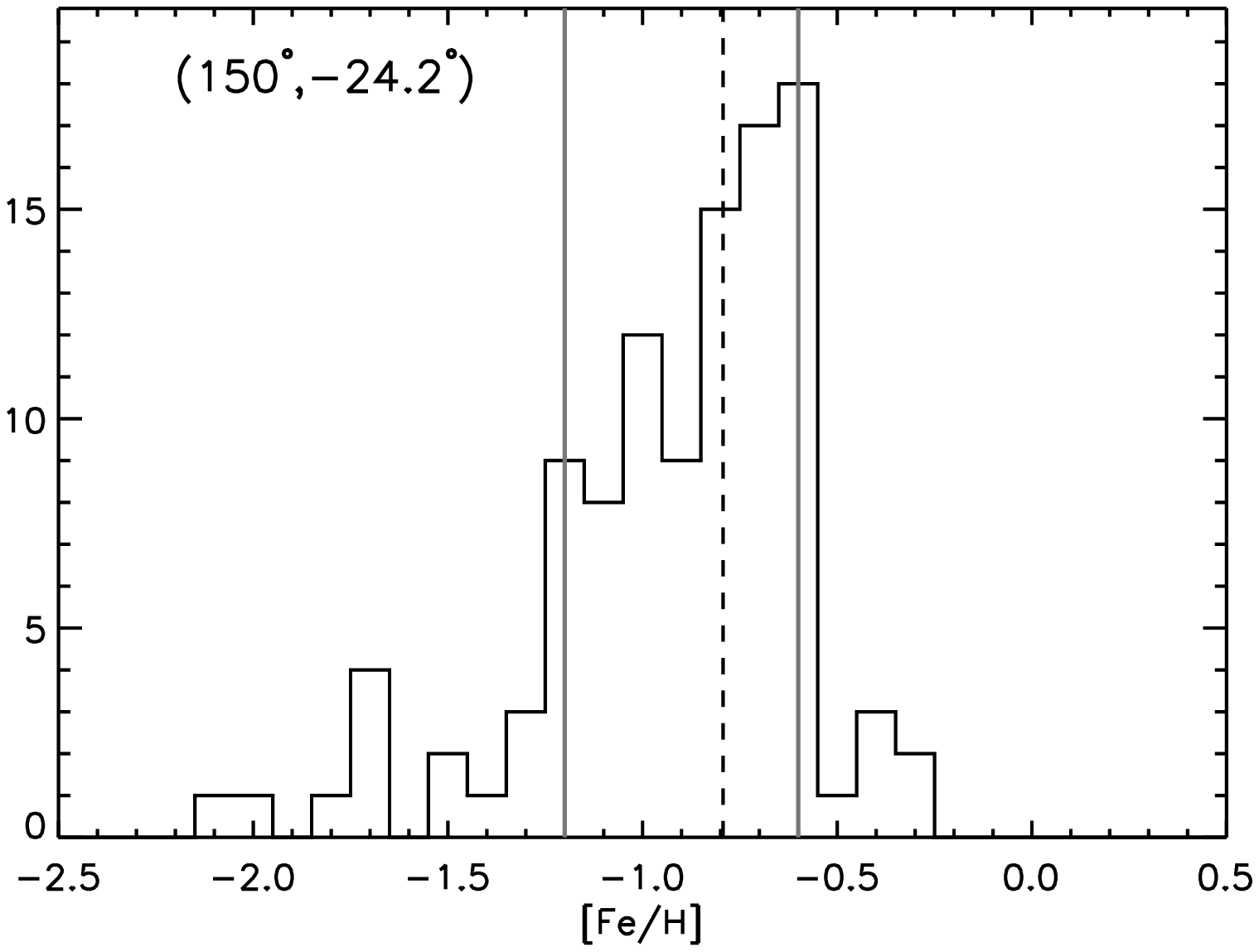}
\includegraphics[scale=0.5]{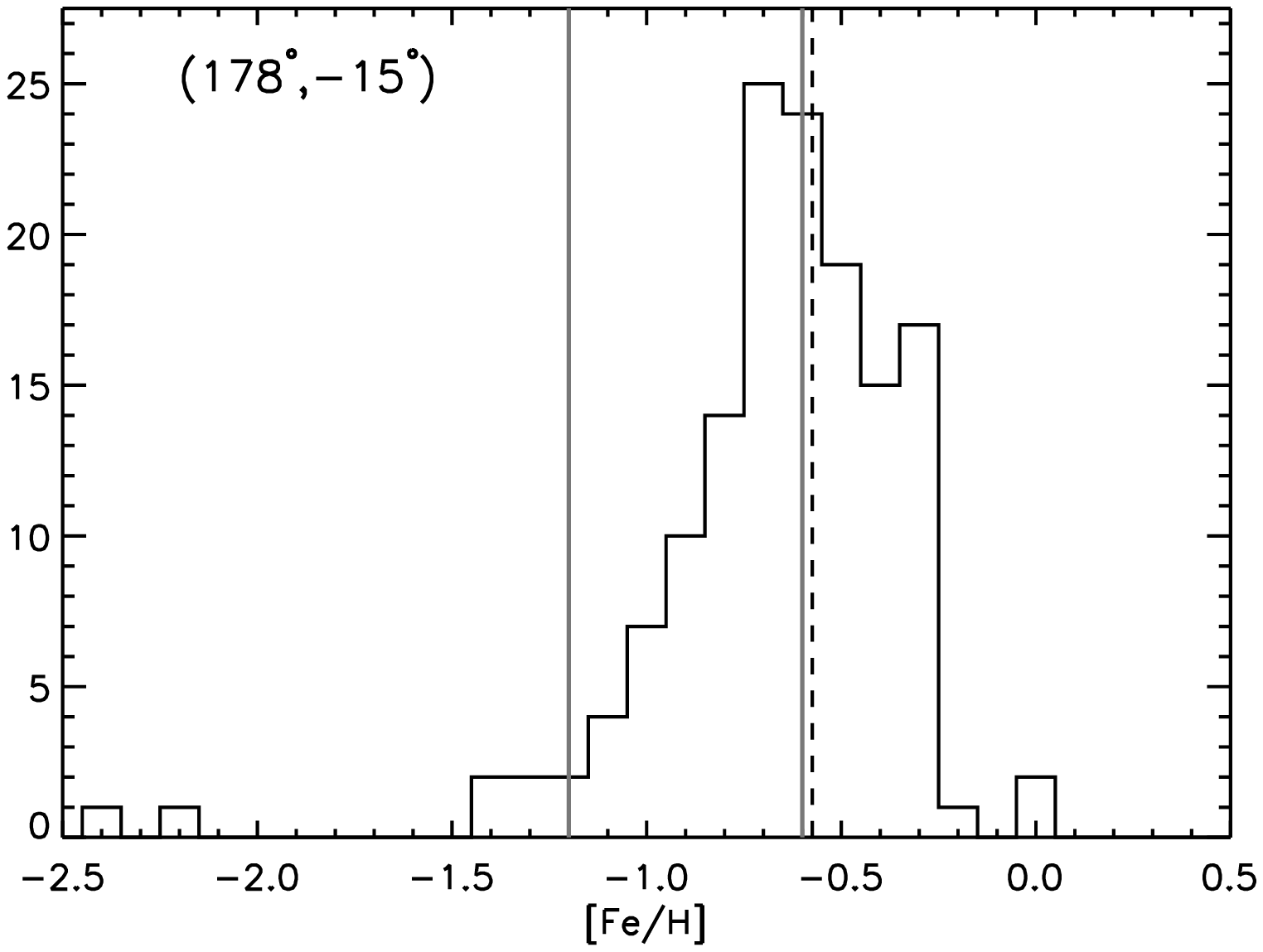}
\includegraphics[scale=0.5]{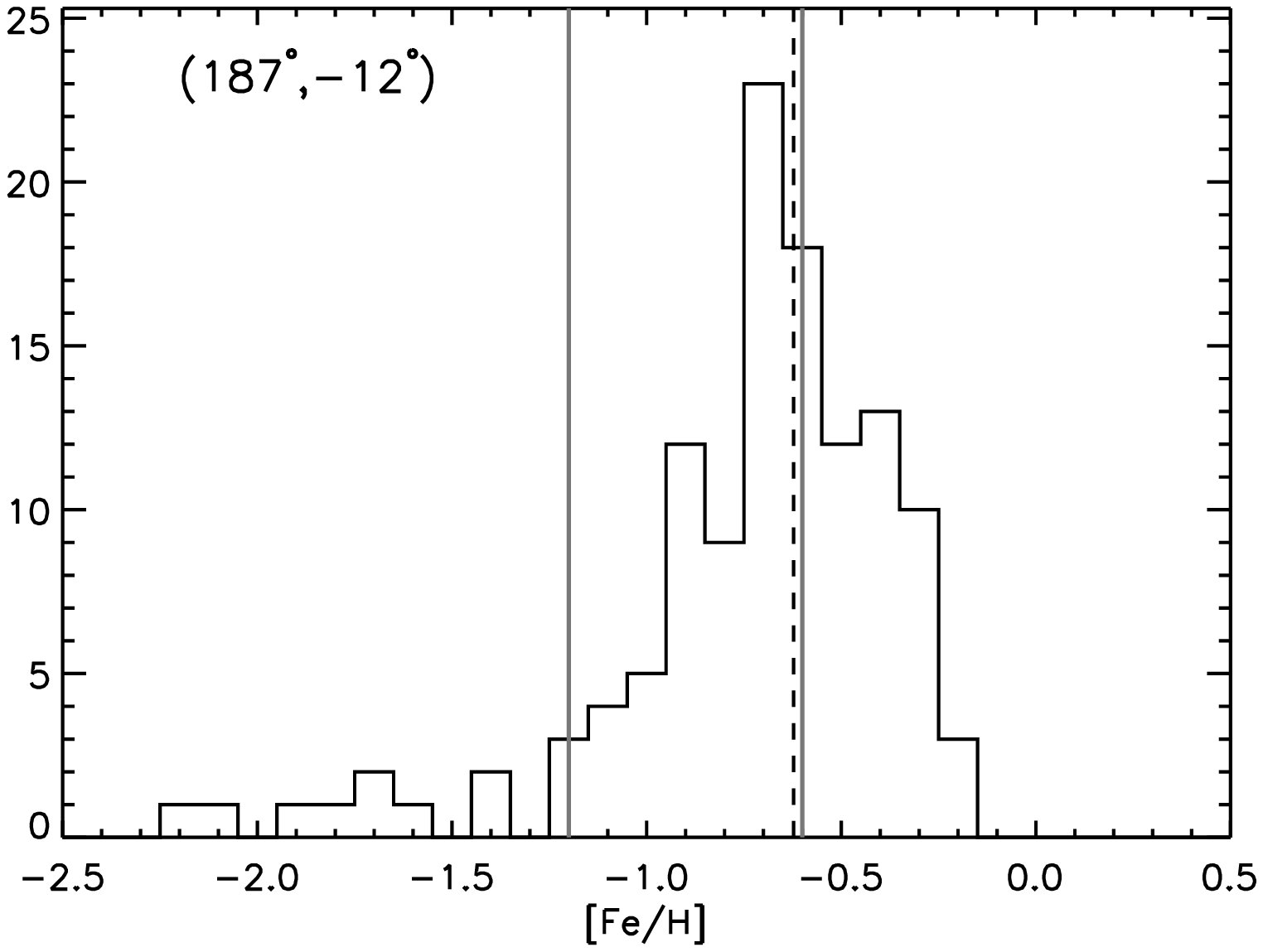}
\includegraphics[scale=0.5]{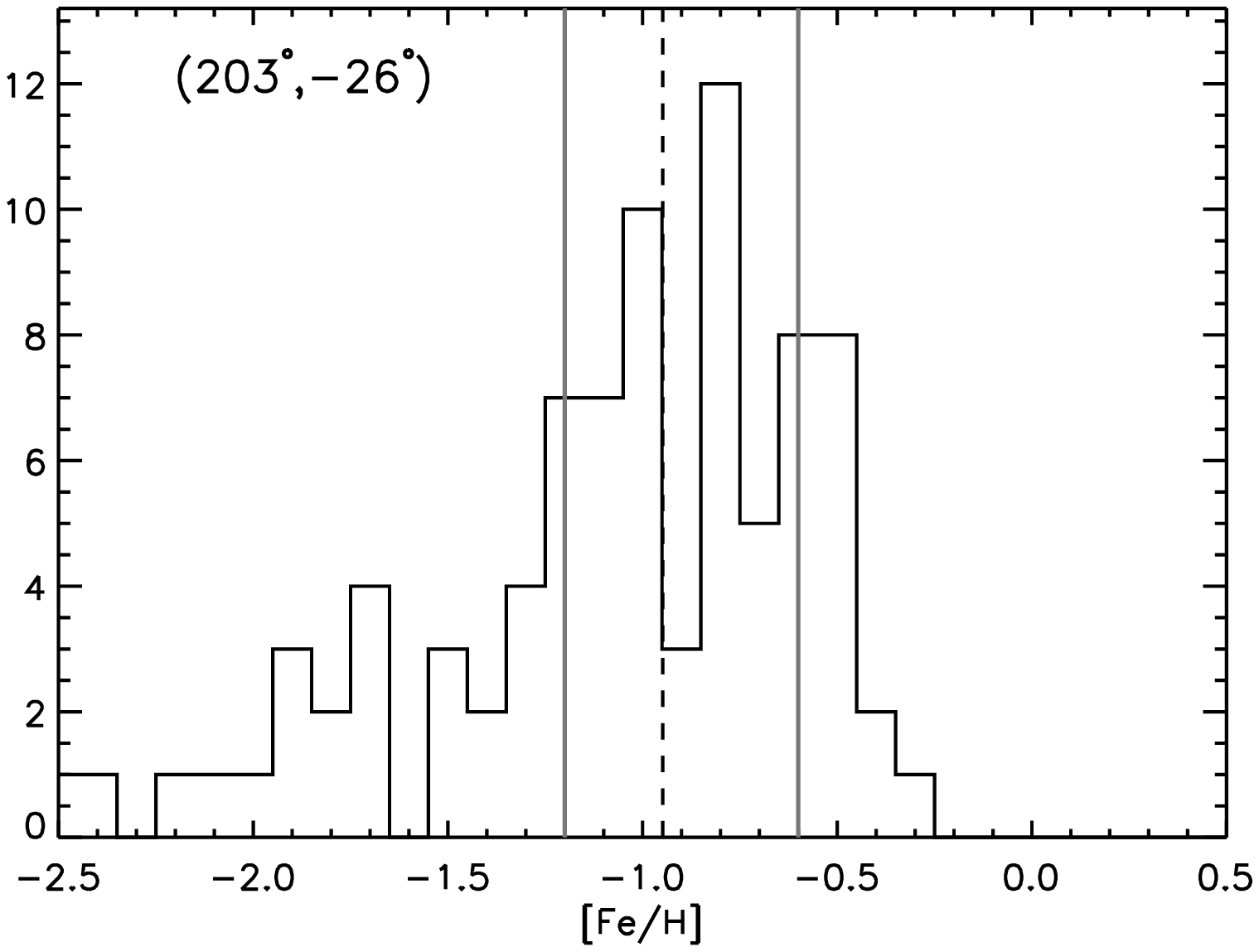}
\caption[velmet2] {
\footnotesize
 Metallicity distribution in the south middle structure. We present the metallicities for spectra that coincide with the position in the Hess diagrams of the south middle structure at five different positions in the Galaxy, with varying latitude and longitude.  The dashed line shows the median value of [Fe/H] in each panel.  Note that stars in the south middle structure that are closer to the Galactic plane have a higher fraction of stars with metallicities of thin disk stars.  Just as with the north near structure, the stars with spectra in the south middle structure are indistinguishable from disk stars, though a larger fraction have metallicities of thick disk stars.
}\label{velmet2}
\end{figure}

\begin{figure}
\includegraphics[scale=0.65]{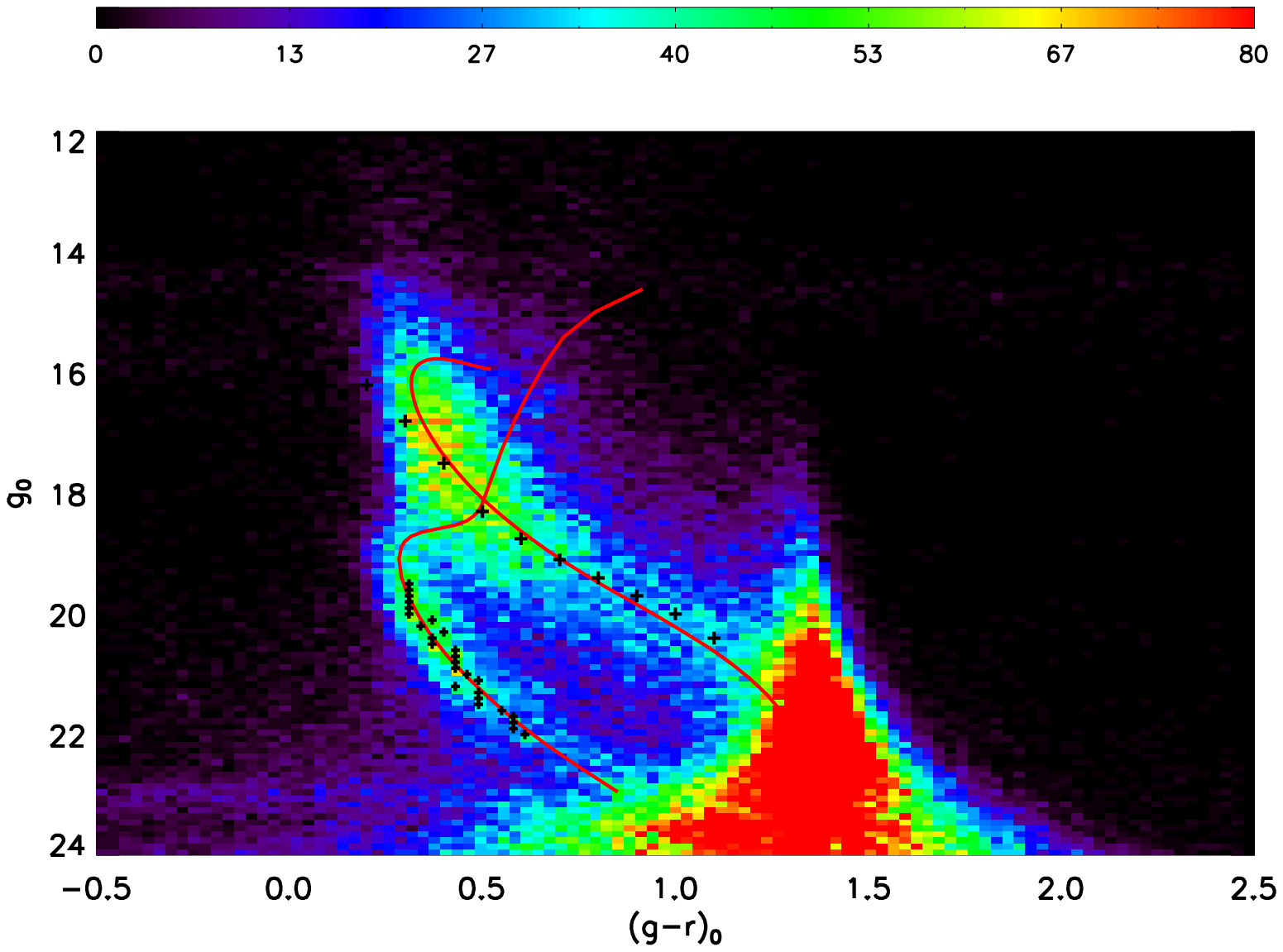}
\includegraphics[scale=0.65]{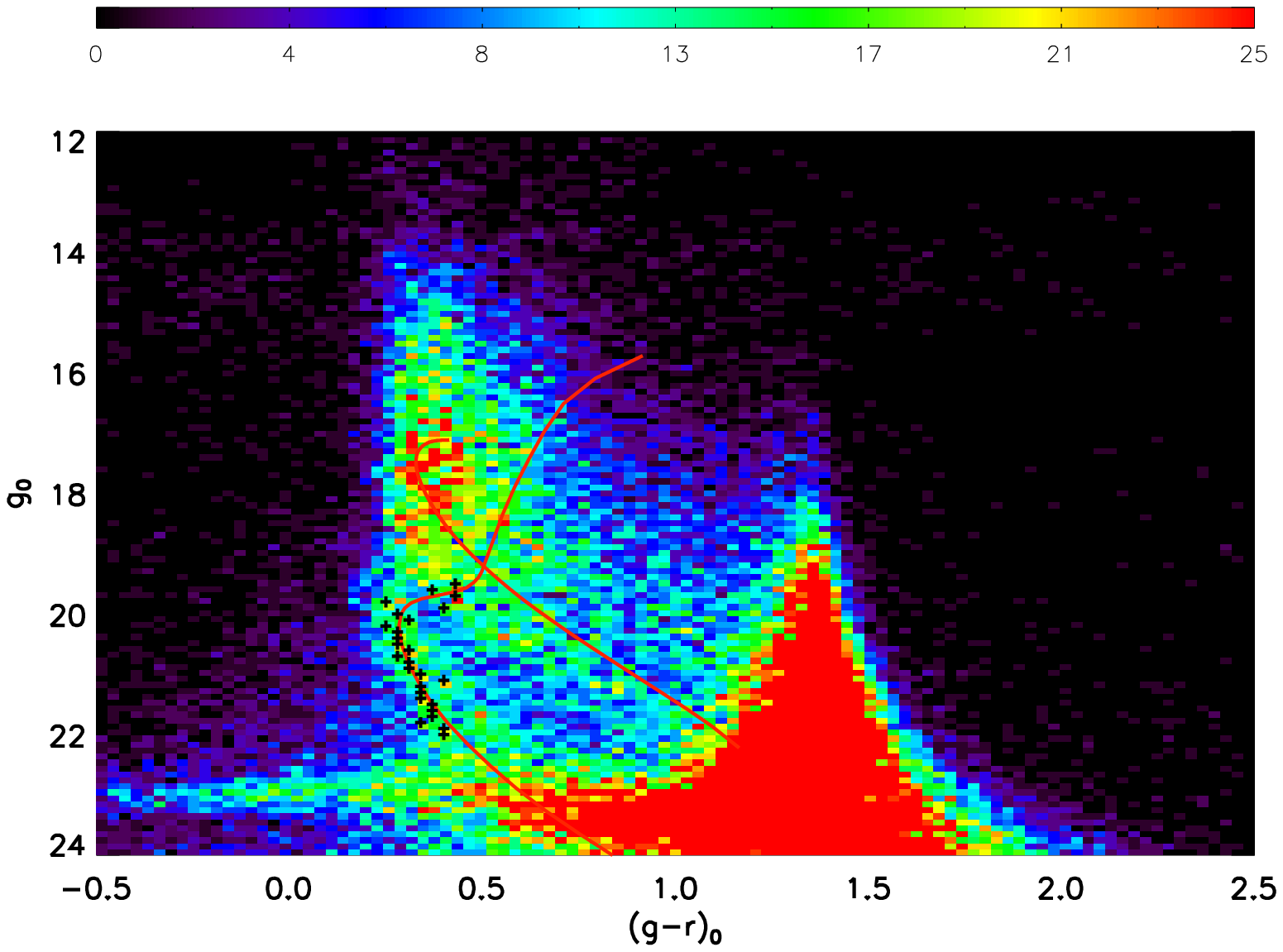}
\caption[isochronefit] {
\footnotesize
Hess diagrams of sky areas (l,b)=$(178^\circ,15^\circ)$, upper panel; and (l,b)=$(203^\circ,-25^\circ)$, lower panel. The plus signs in the upper panel indicate the ridge lines of the north near structure and the Monoceros Ring. The red curve that fits the ridge line of north near structure is an isochrone with [Fe/H]=-0.5, which is the mean metallicity of the stars in the spectral sample for this strucure.  The red curve that fits the Monoceros Ring is an isochrone matching the globular cluster M5. 
The plus signs in the lower panel follow the ridges of the south middle structure and the TriAnd Ring.  An isochrone with [Fe/H]=-0.88 is adopted to fit the peak of the blue stars ($(g-r)_0<0.4$) in the south middle structure in this sky area.  The M5 isochrone is adopted to fit the TriAnd Ring.
The isochrones with [Fe/H]=-0.5 and -0.88 are interpolated from the empirical isochrons of An et al. (2009).
}\label{isochronefit}
\end{figure}

\begin{figure}
\includegraphics[scale=0.65]{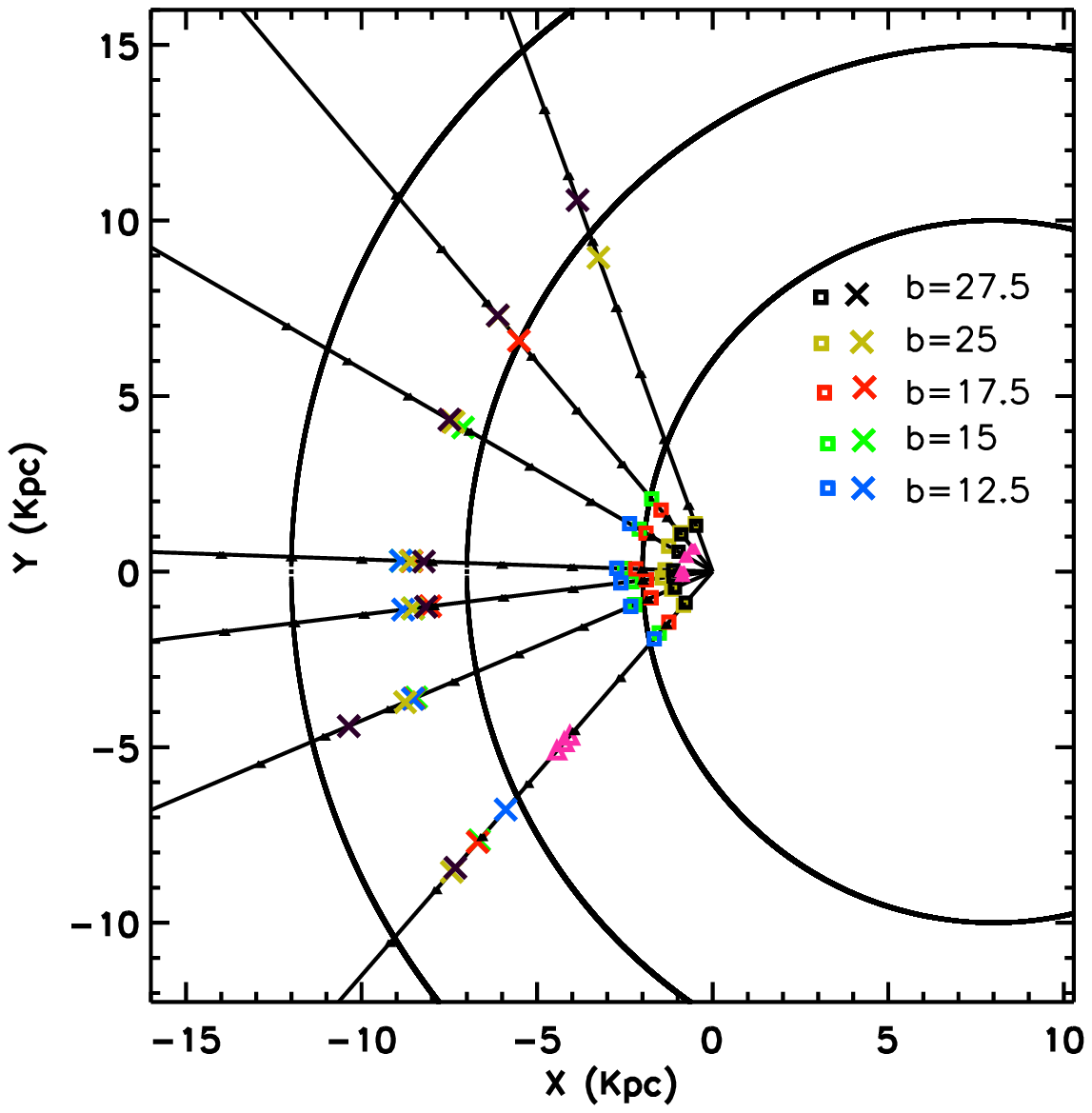}
\includegraphics[scale=0.65]{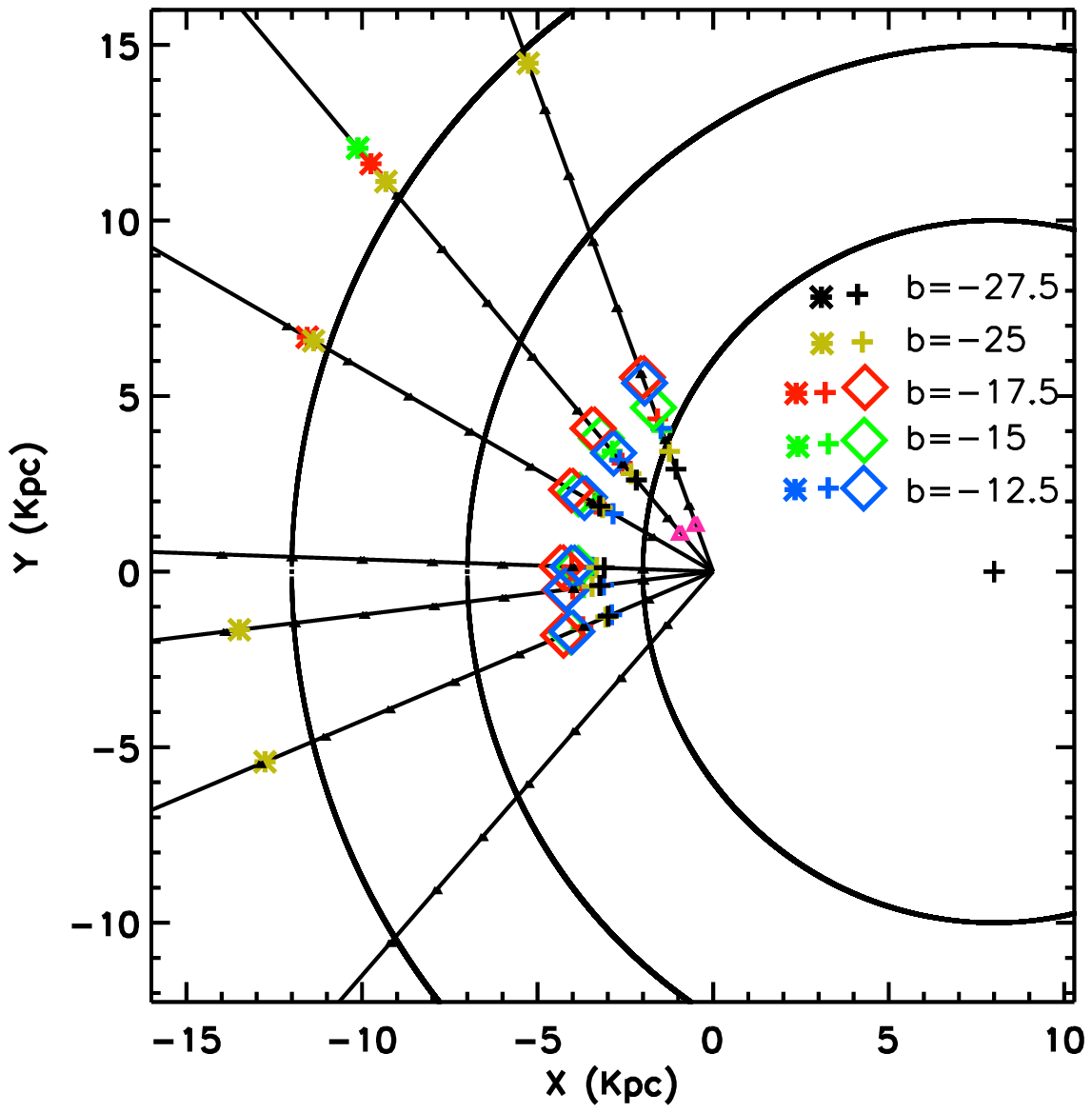}
\caption[rings] {
\footnotesize
We show here the positions of the density peaks projected on the Galactic plane, centered on the Sun.  The upper panel shows peaks above the Galactic plane ($b>0^\circ$), and the lower panel shows peaks below the Galactic plane ($b<0^\circ$)  A different symbol is used for each ringlike structure we have identified: Monoceros Ring - crosses; north near - squares, south middle - diamonds; TriAndromeda - asterisks; unidentified structure - pink triangles.  The color of the symbol  (see legend) encodes the Galactic latitude range used to determine the distance: $12.5^\circ<|b|<15^\circ$ - blue; $15^\circ<|b|<17.5^\circ$ - green; $17.5^\circ<|b|<20^\circ$ - red; $28^\circ<b<30^\circ$ - black.  The Galactic center is at $(X,Y)=(0,8)$ kpc.  Lines fanning out radially from the Sun indicate the directions for which there is low latitude data in SDSS, at Galactic longtudes of $70^\circ, 94^\circ, 110^\circ, 130^\circ, 150^\circ, 178^\circ, 187^\circ, 203^\circ,$ and $229^\circ$.  The nodes along the lines are spaced at 2 kpc intervals.  The circles are concentric around the Galactic center, at 10 kpc, 15 kpc, 20 kpc. 
}\label{rings}
\end{figure}

\begin{figure}
\includegraphics[scale=0.5]{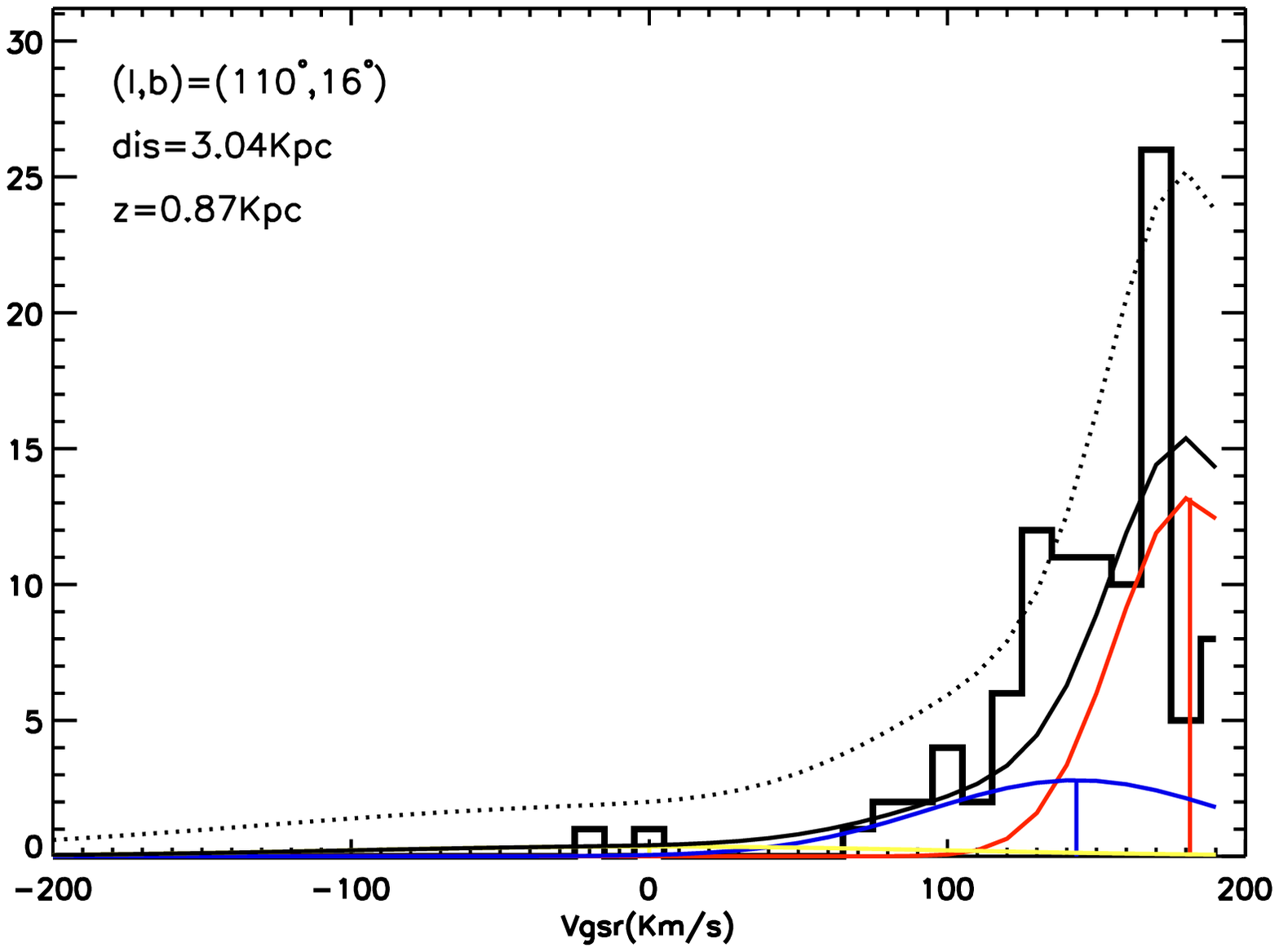}
\includegraphics[scale=0.5]{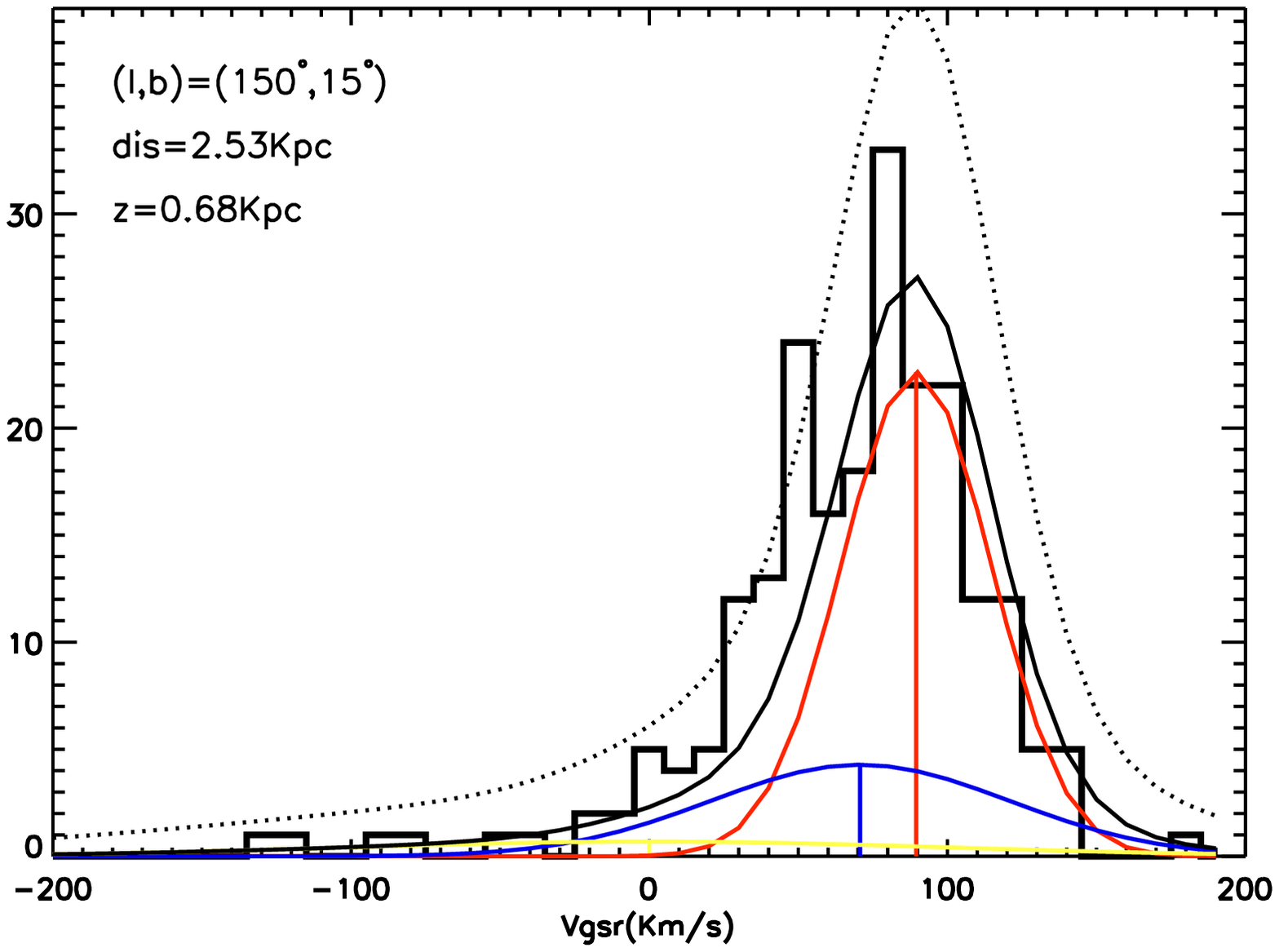}
\includegraphics[scale=0.5]{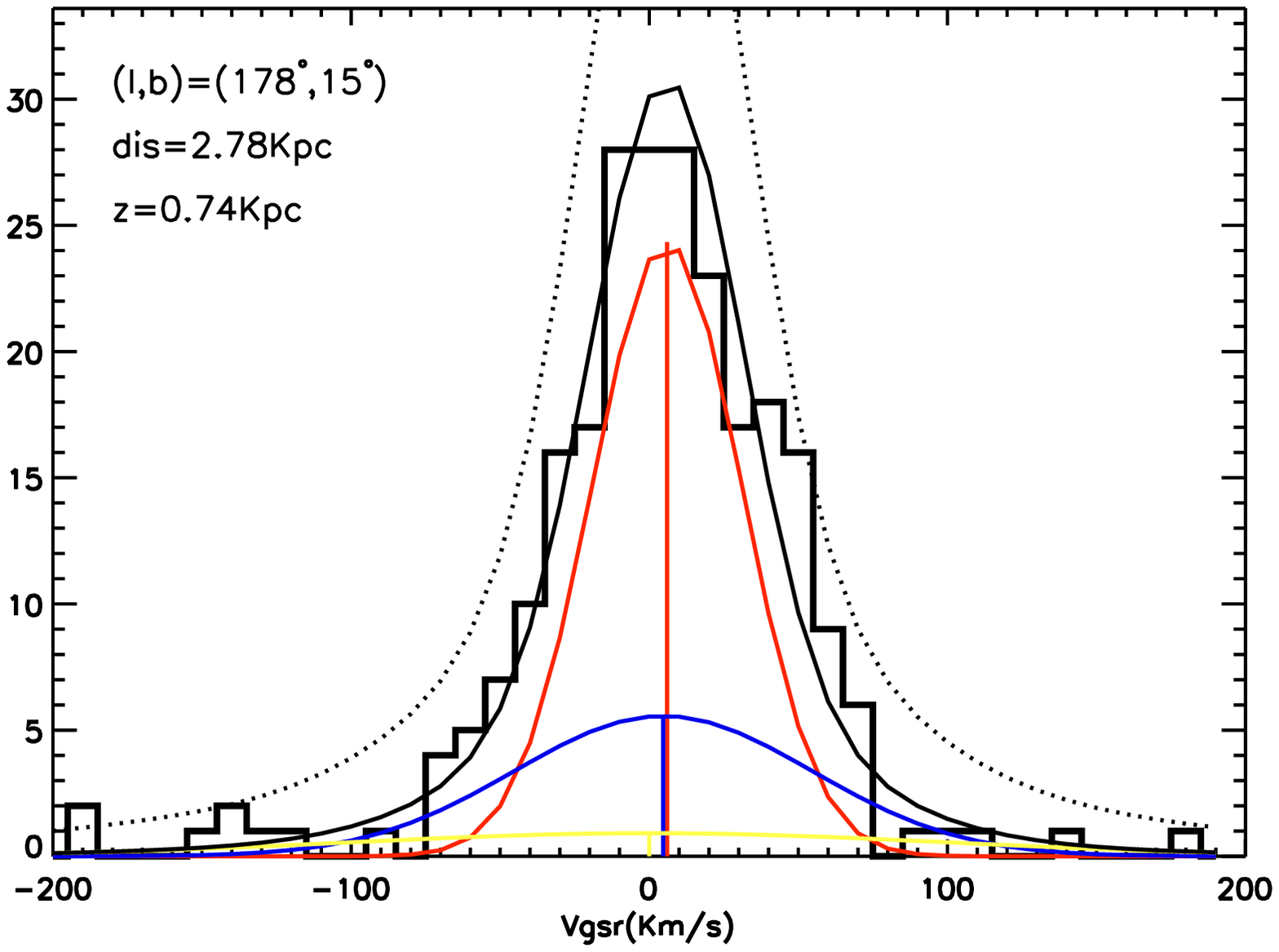}
\includegraphics[scale=0.5]{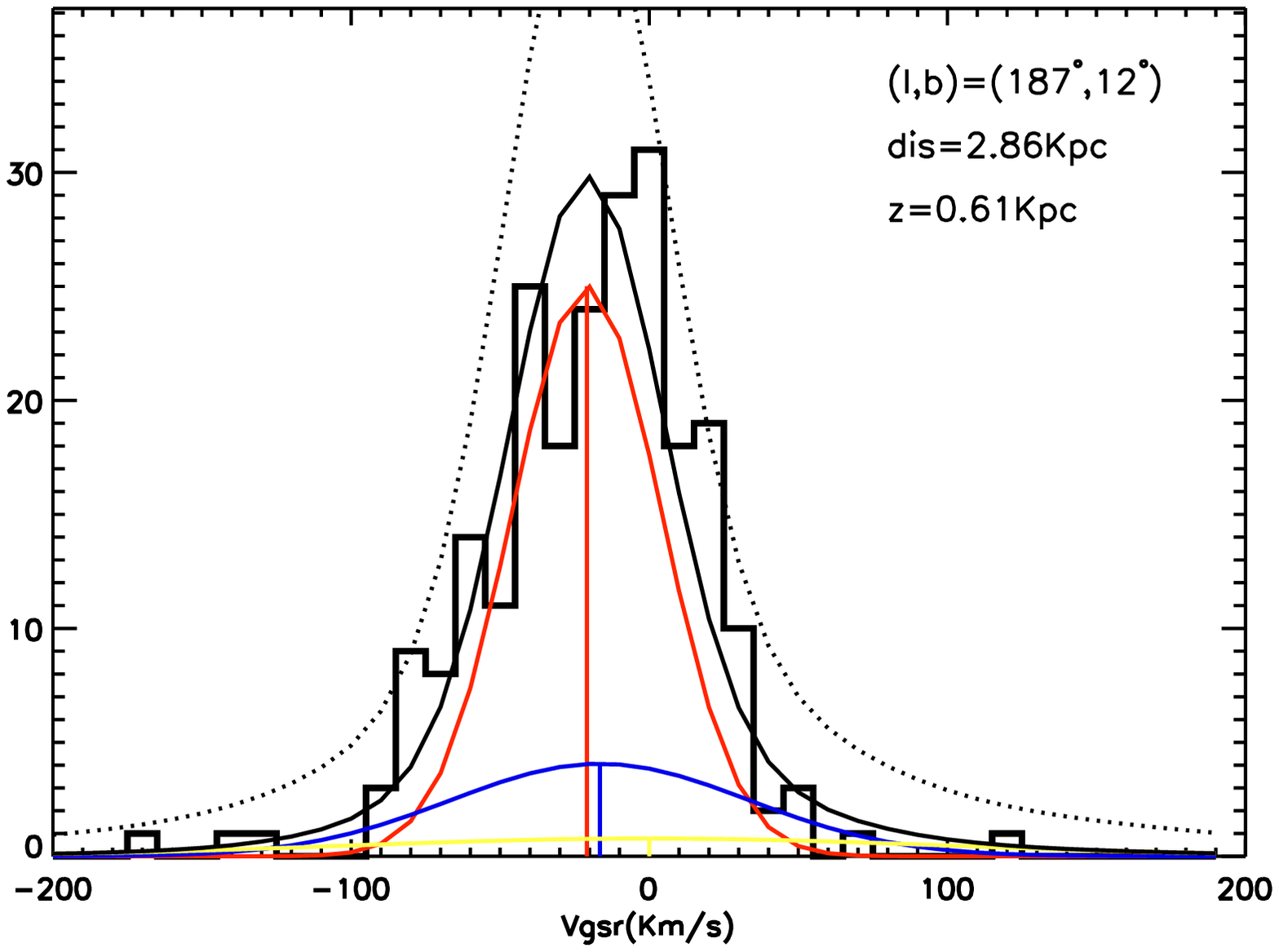}
\includegraphics[scale=0.5]{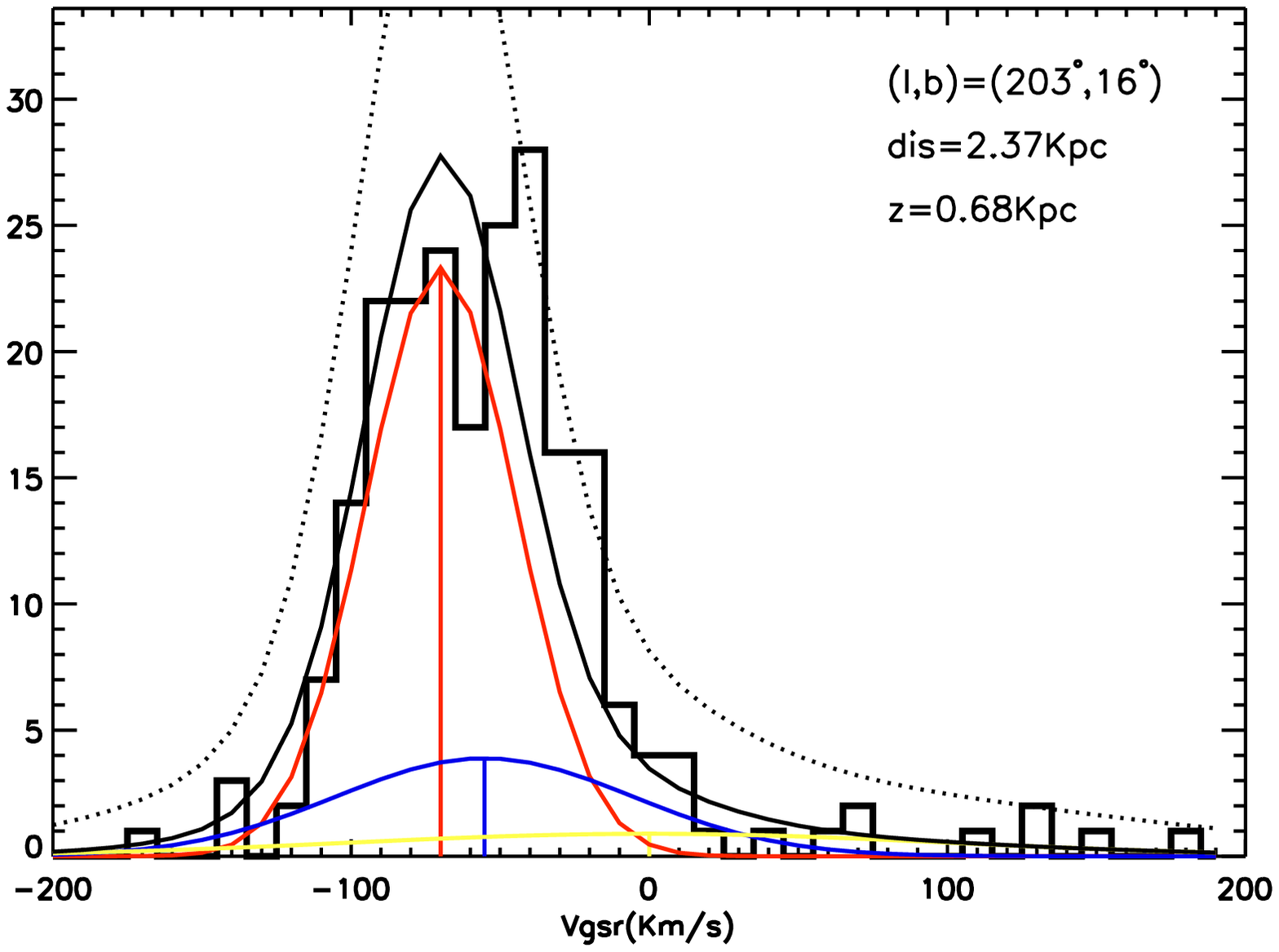}
\includegraphics[scale=0.5]{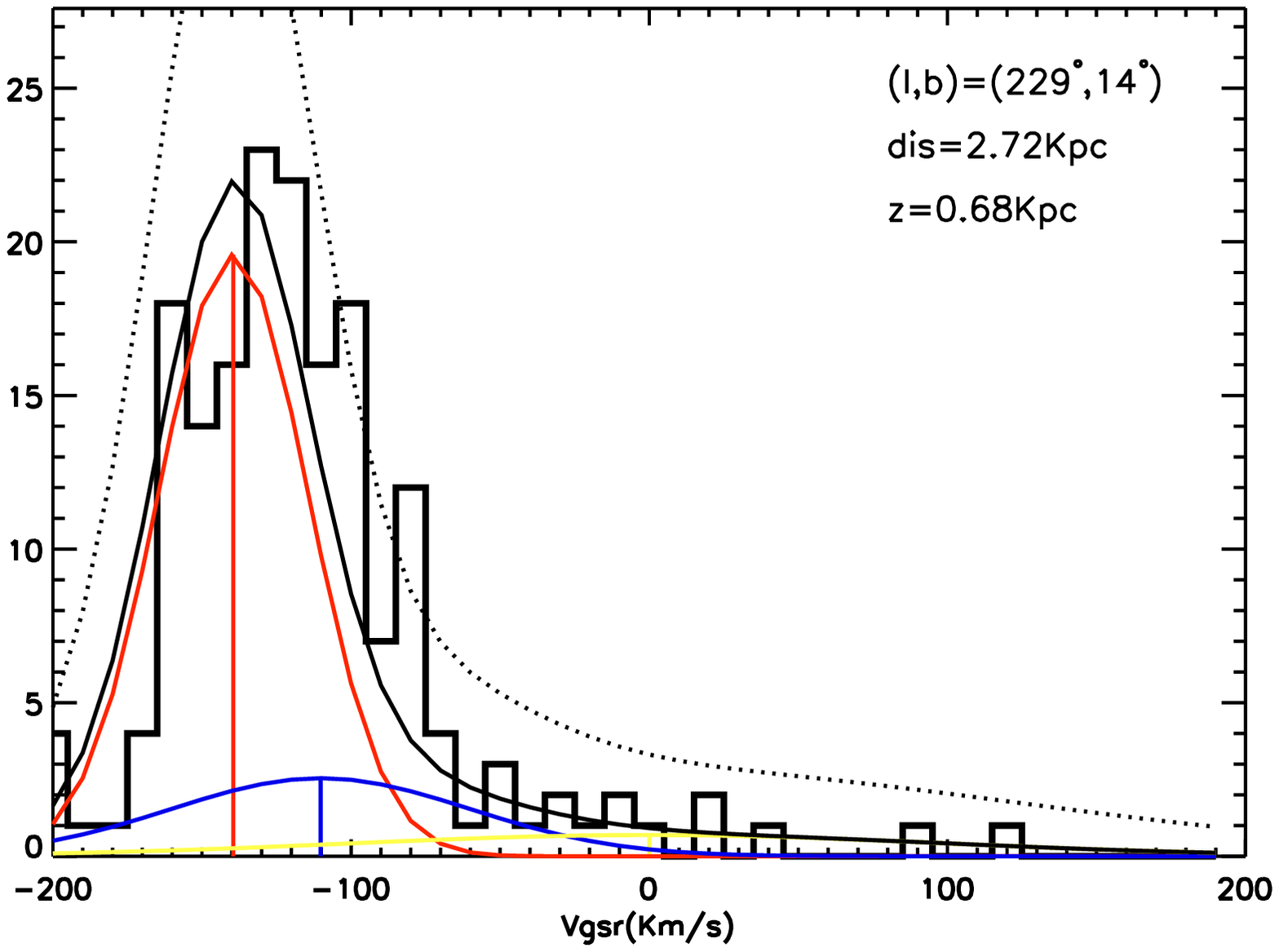}
\caption[vgsrnorthnear] {
\footnotesize
$V_{\rm gsr}$ distribution of stars which sample the north near structure at $b \sim 15^\circ.$  These are the same stars as are shown in Figure 8.  The black histogram shows the $V_{\rm gsr}$ distribution of the data. The theoretical $V_{\rm gsr}$ distribution of thin disk, thick disk, and halo are indicated by the red curve, blue curve, and yellow curve, respectively.  The black curve shows the sum of the three, weighted relative to the fractions of stars in each metallicity range in Figure 8. The dotted line indicates a 2.5 sigma excess from the theoretical prediction. The distance and Galactic height are estimated by fitting the isochrone with [Fe/H]=-0.5,
which is the mean metalicity of the spectral sample along the strucure.
}\label{vgsrnorthnear}
\end{figure}

\begin{figure}
\includegraphics[scale=0.5]{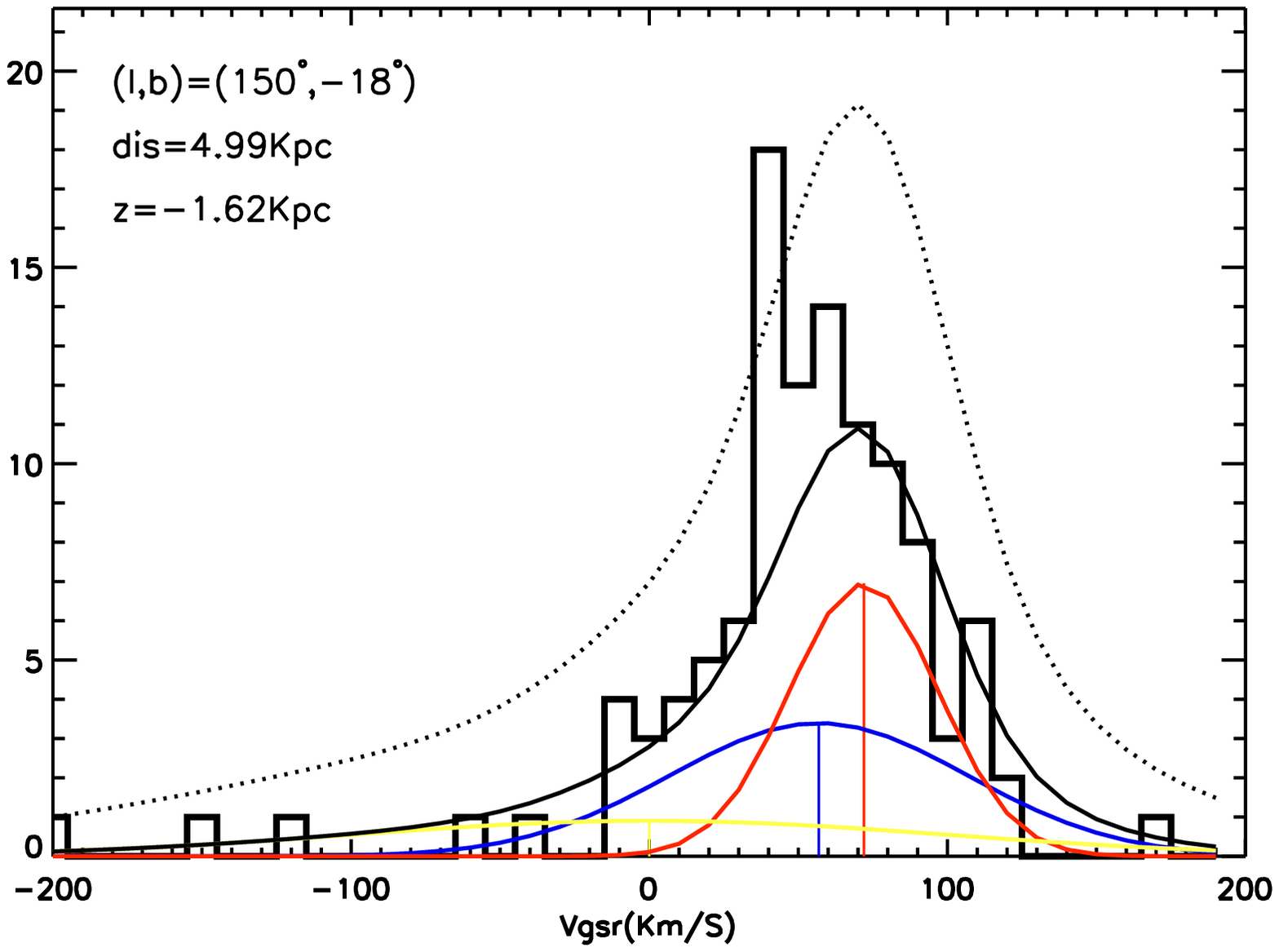}
\includegraphics[scale=0.5]{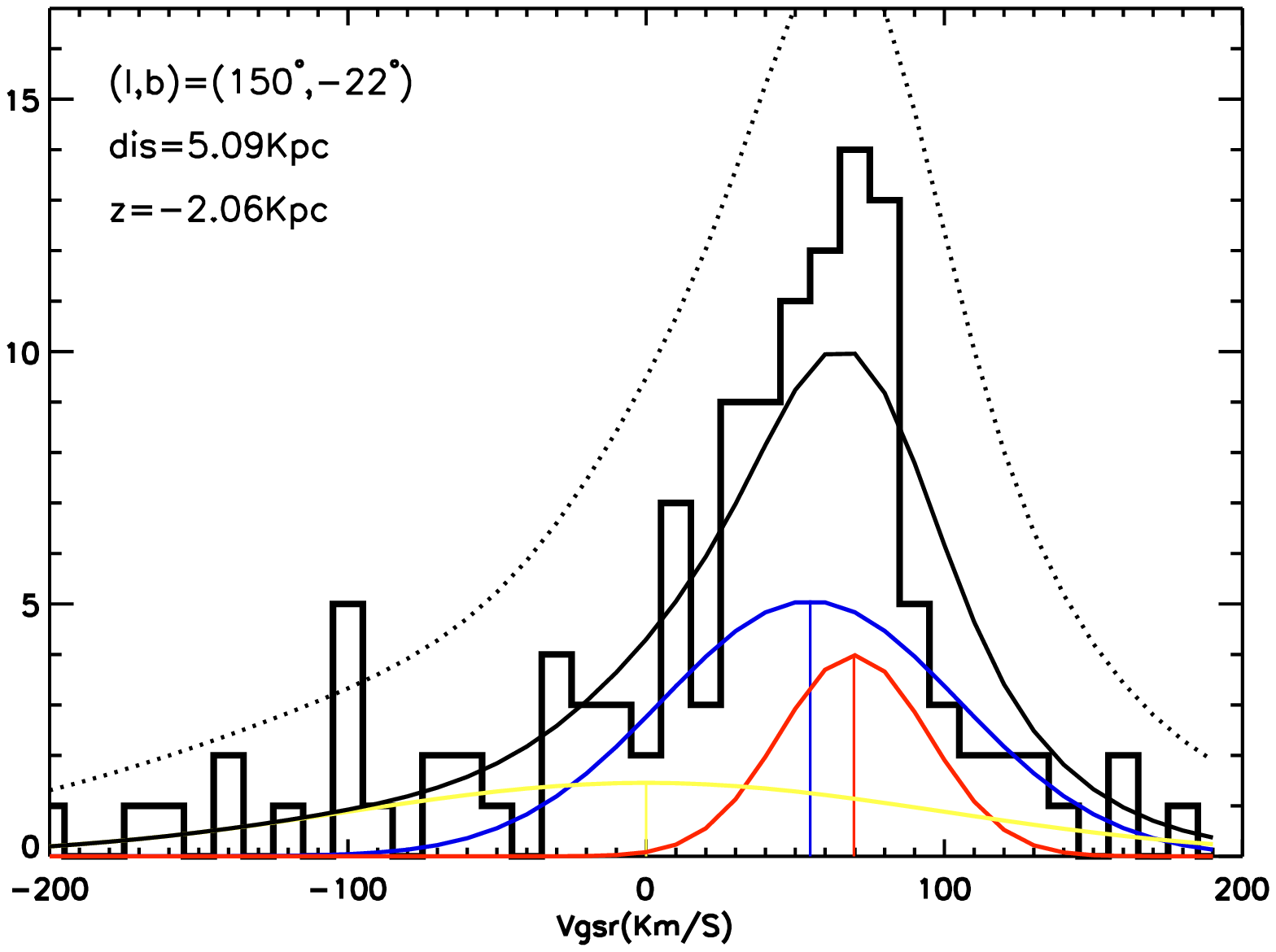}
\includegraphics[scale=0.5]{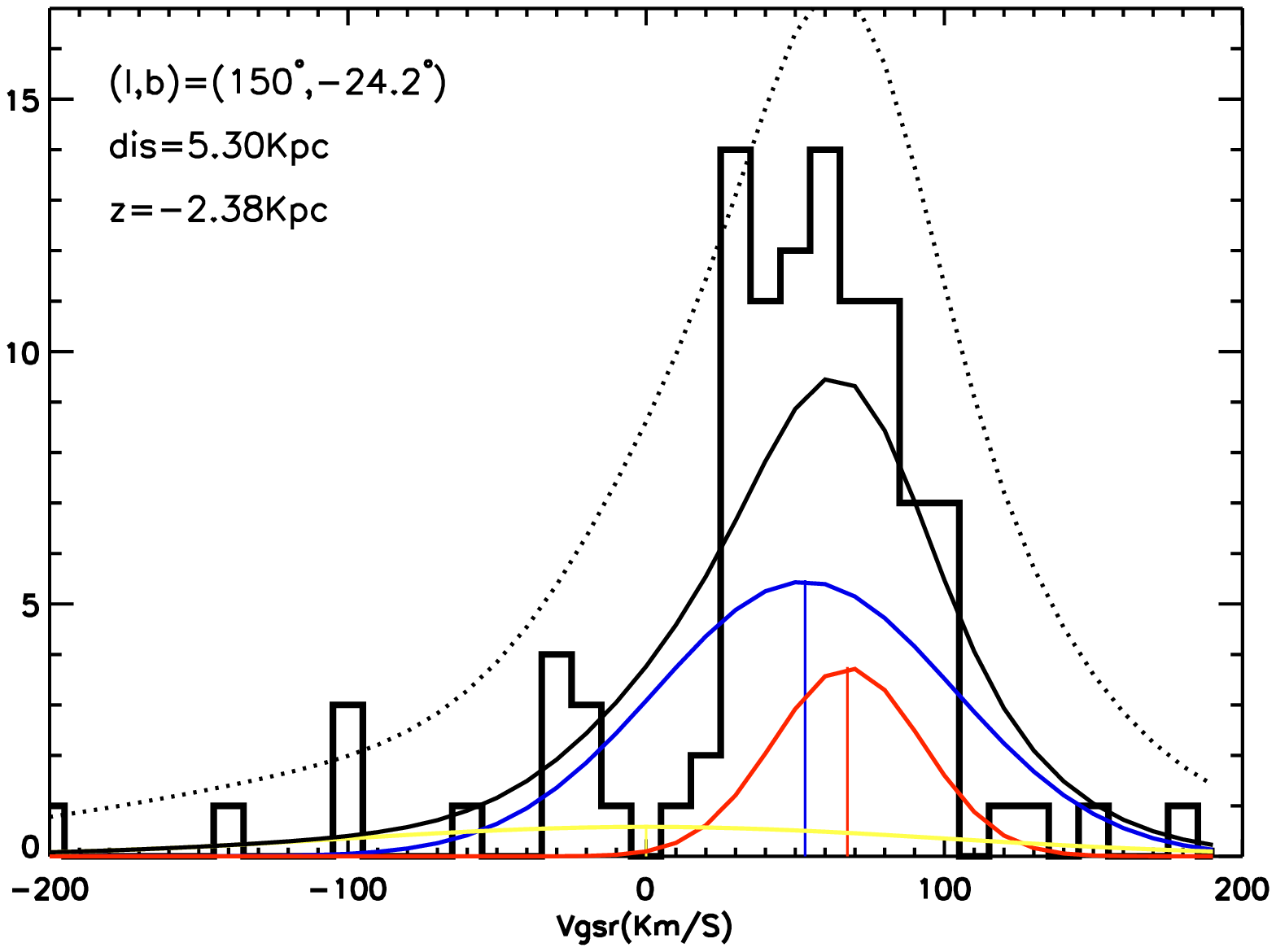}
\includegraphics[scale=0.5]{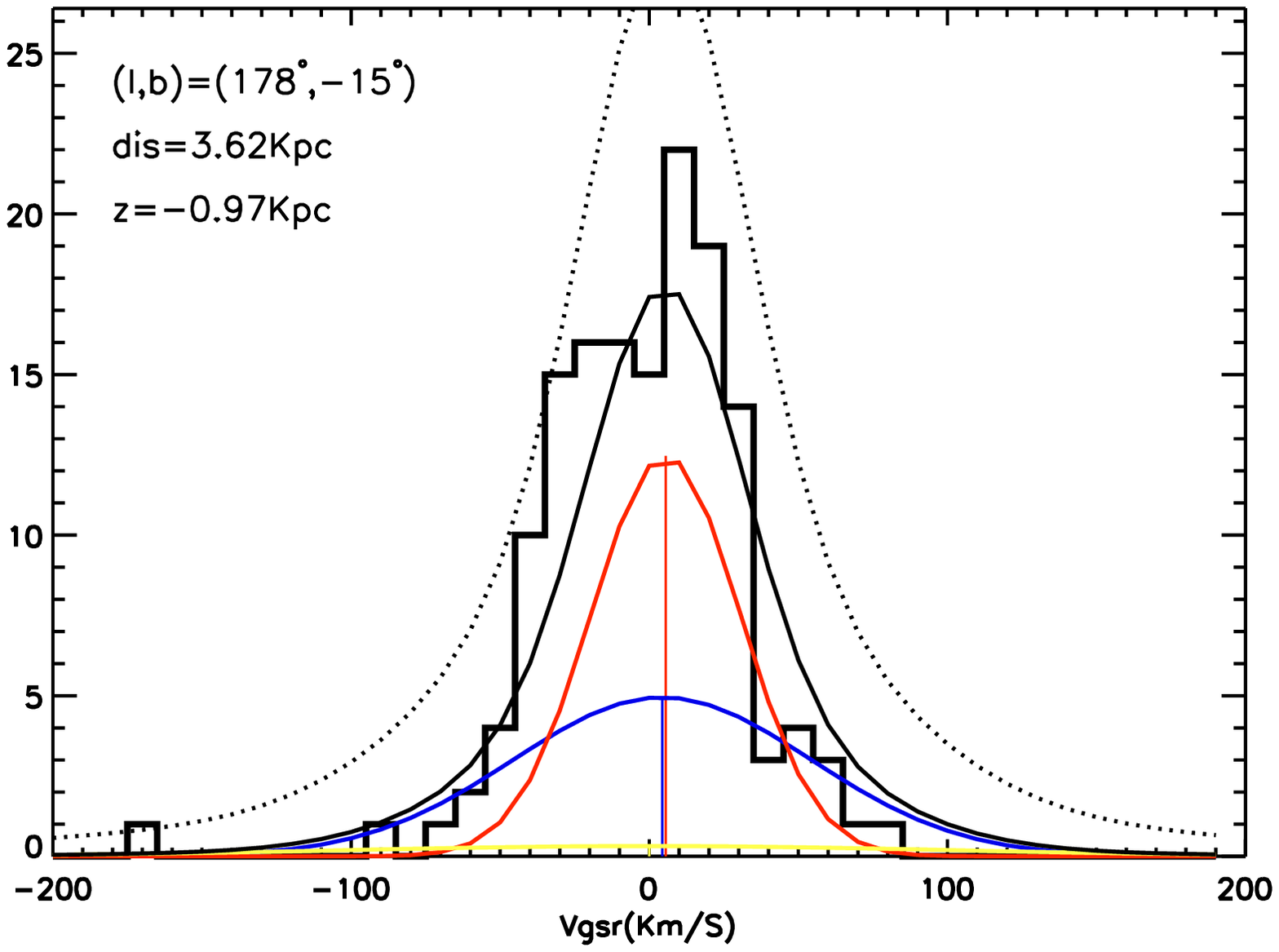}
\includegraphics[scale=0.5]{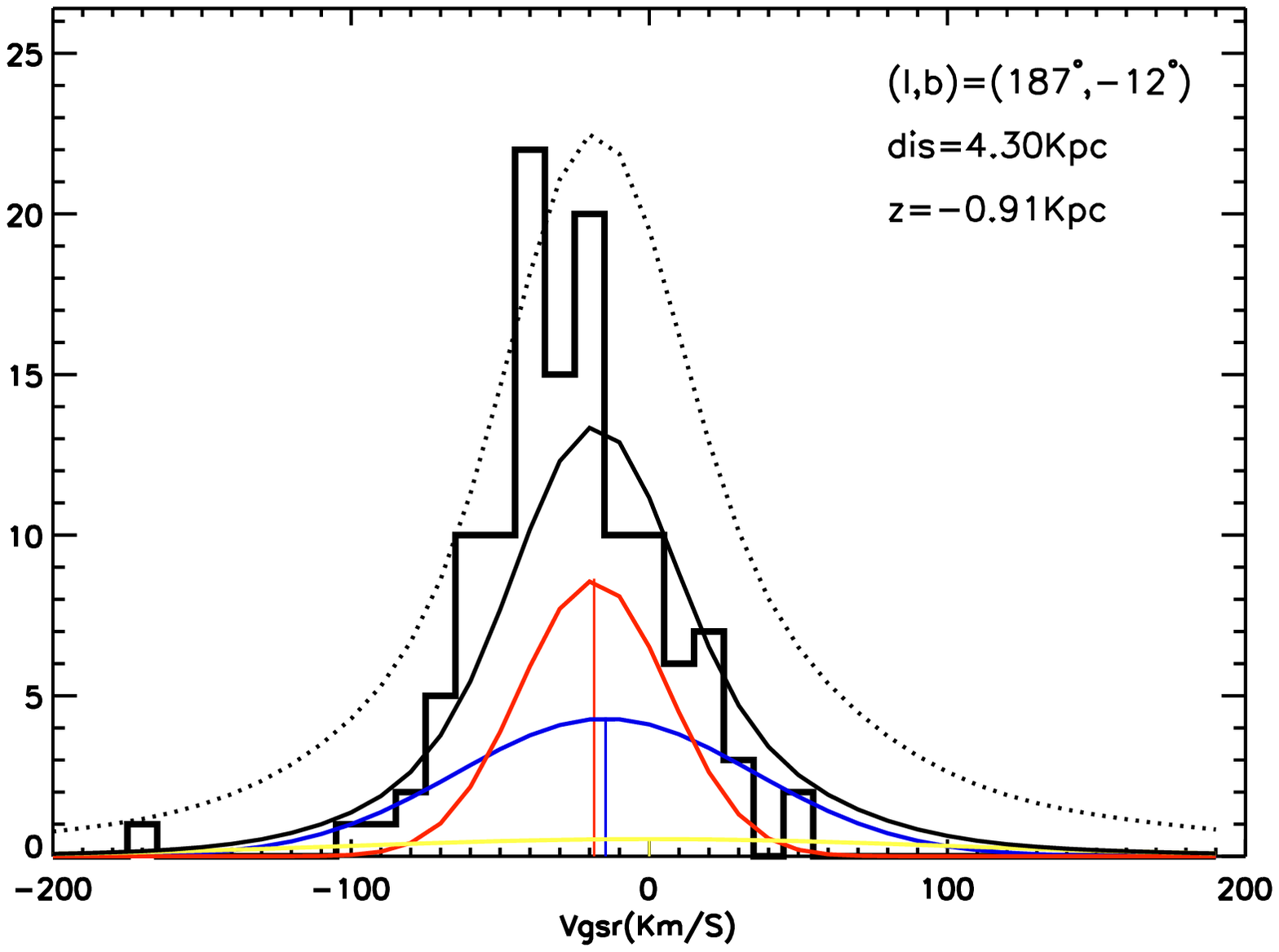}
\includegraphics[scale=0.5]{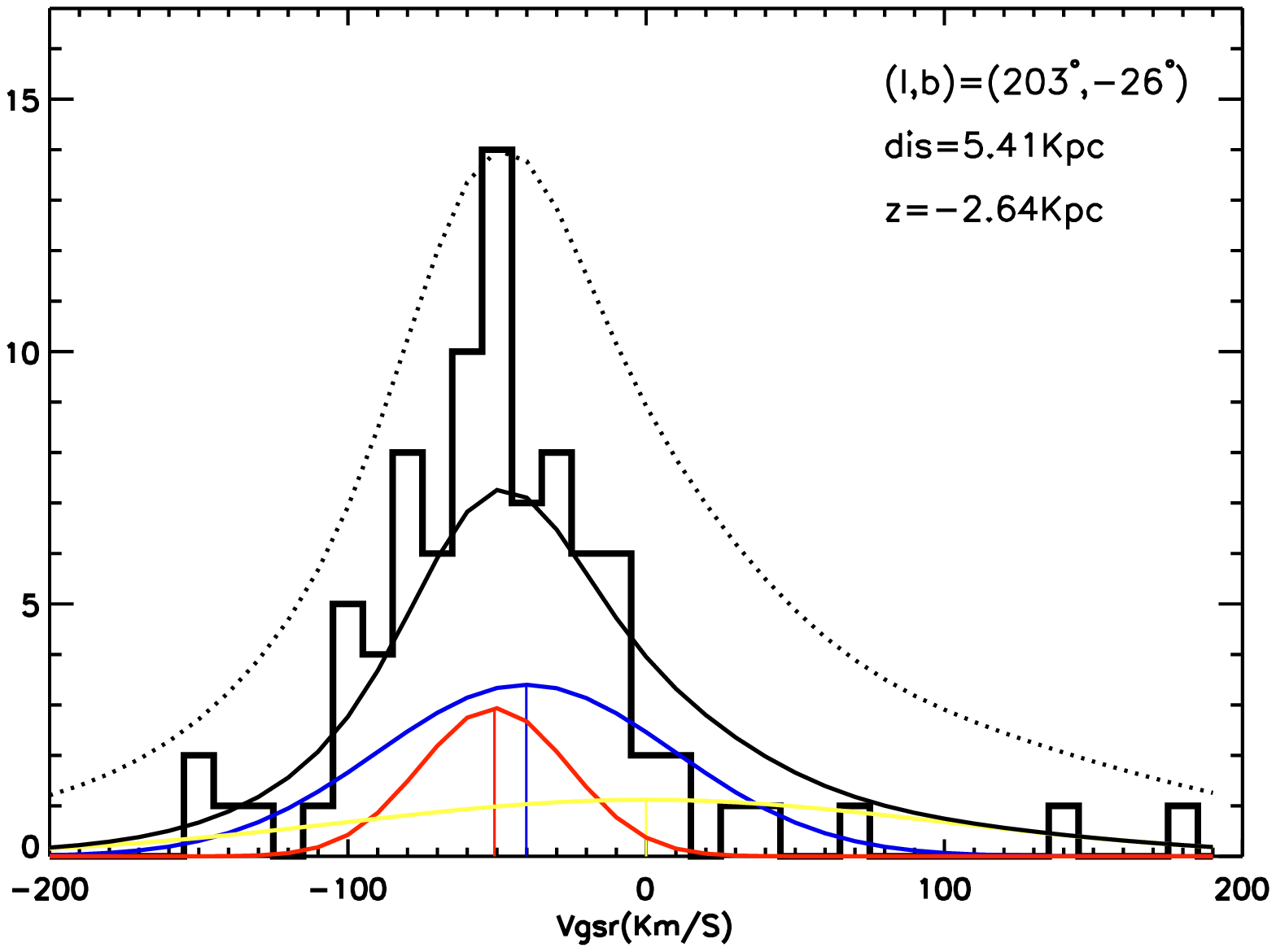}
\caption[vgsrnorthnear] {
\footnotesize
$V_{\rm gsr}$ distribution of stars in the south middle structure. The model curves areconstructed similarly to those in Figure 13. The fraction of theoretical thin disk, thick disk, halo are determined from the fraction of stars in each metallicity region in Figure 10.
}\label{vgsrnorthnear}
\end{figure}

\begin{figure}
\includegraphics[scale=0.5]{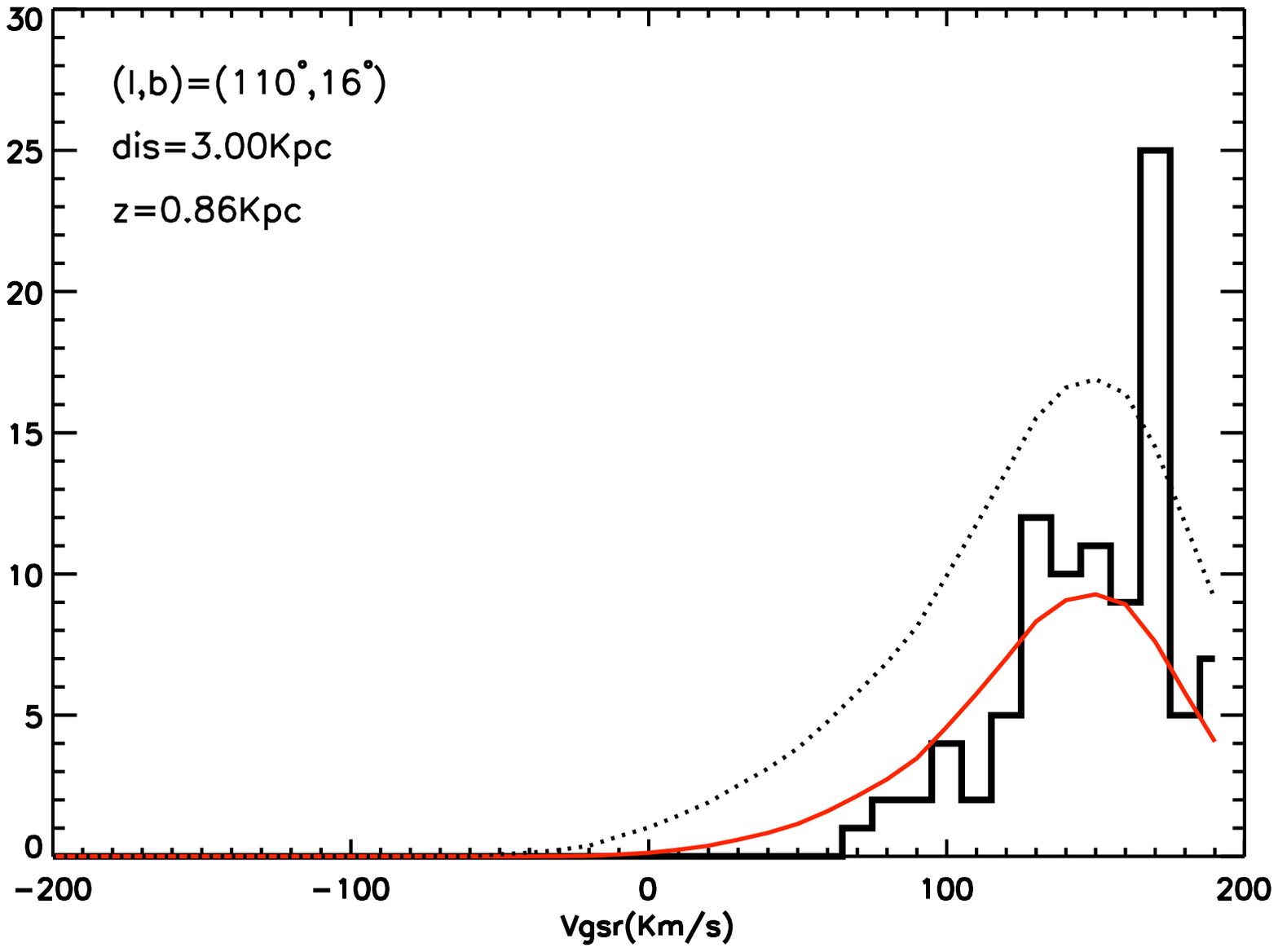}
\includegraphics[scale=0.5]{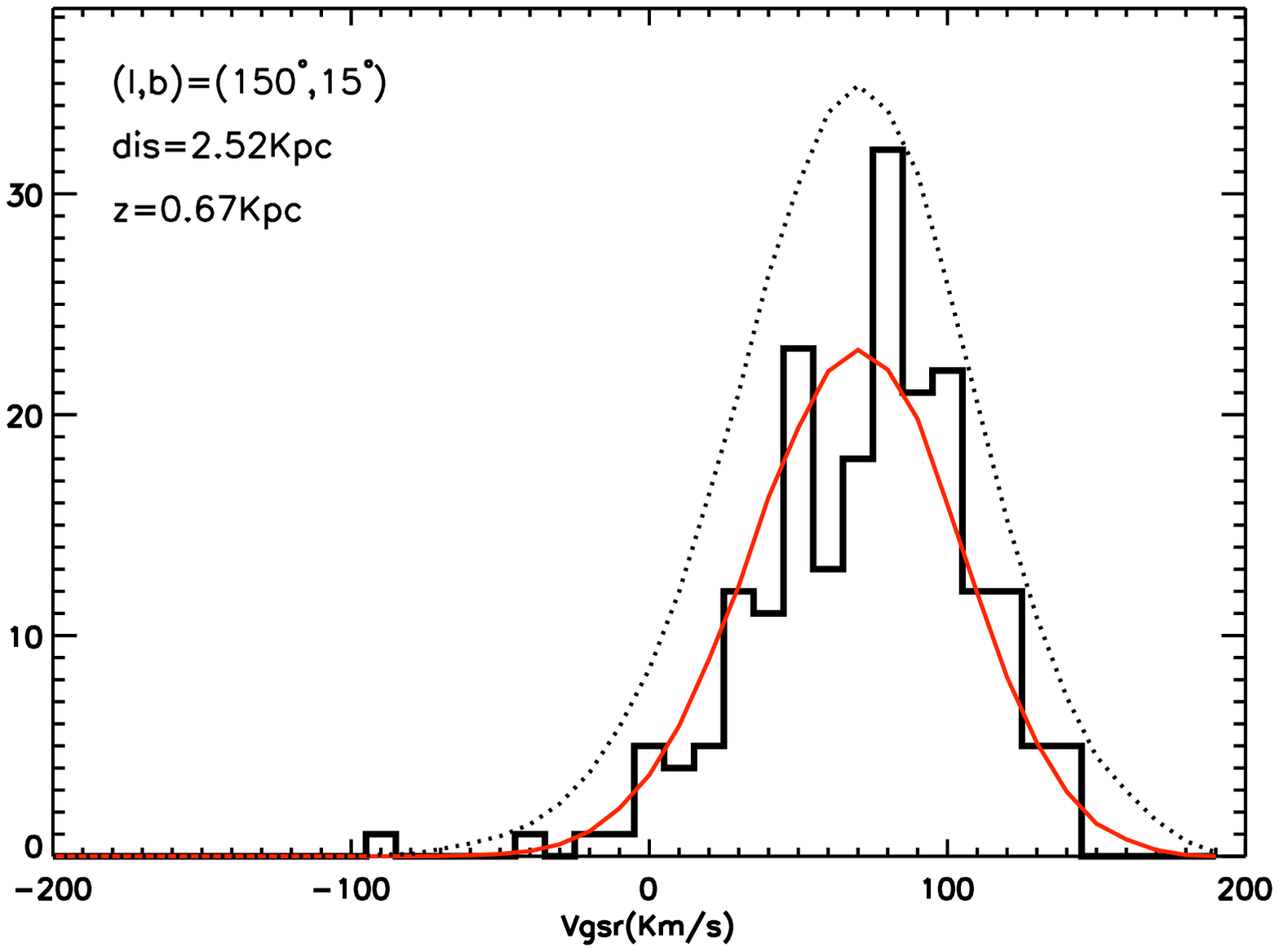}
\includegraphics[scale=0.5]{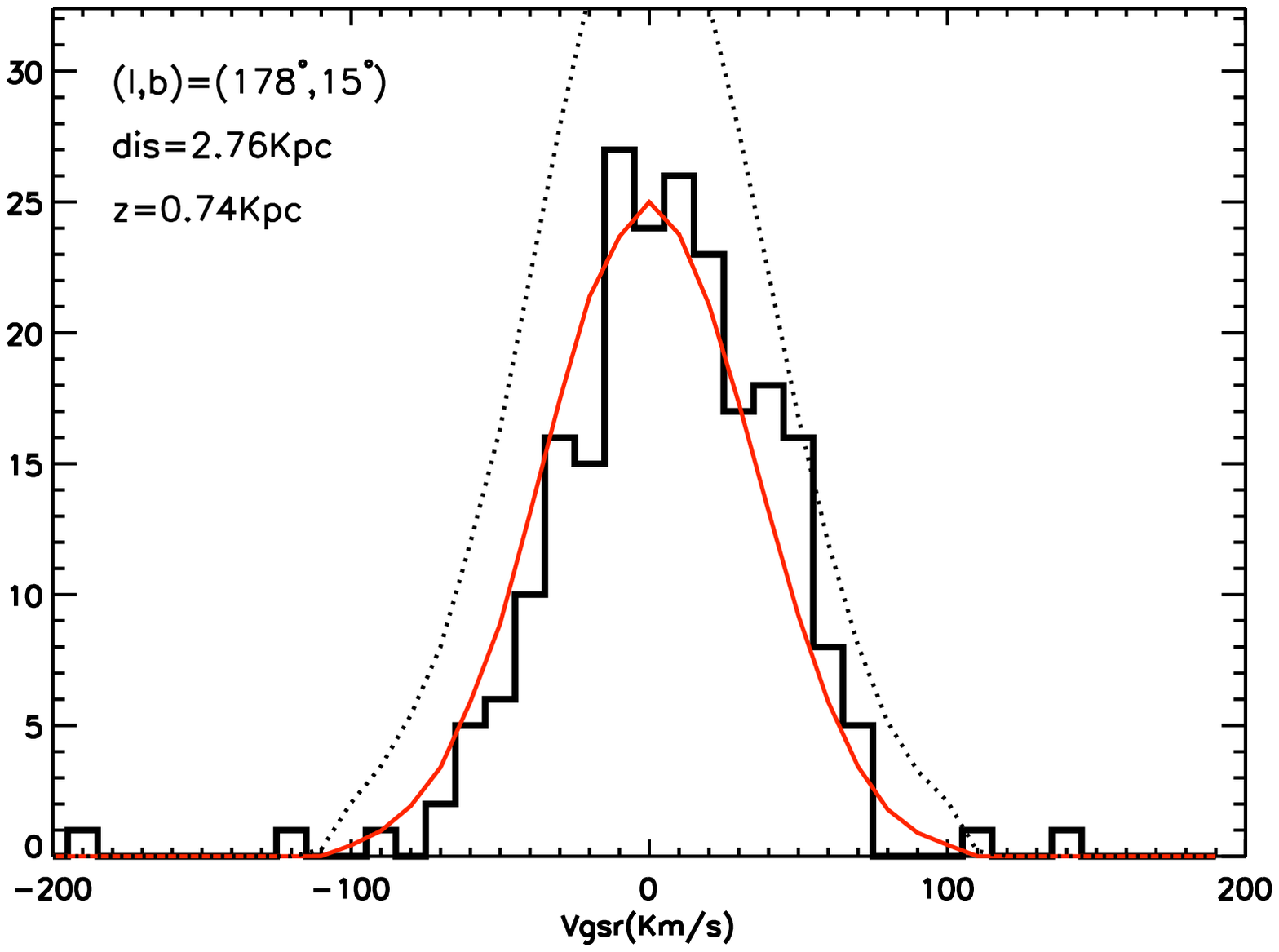}
\includegraphics[scale=0.5]{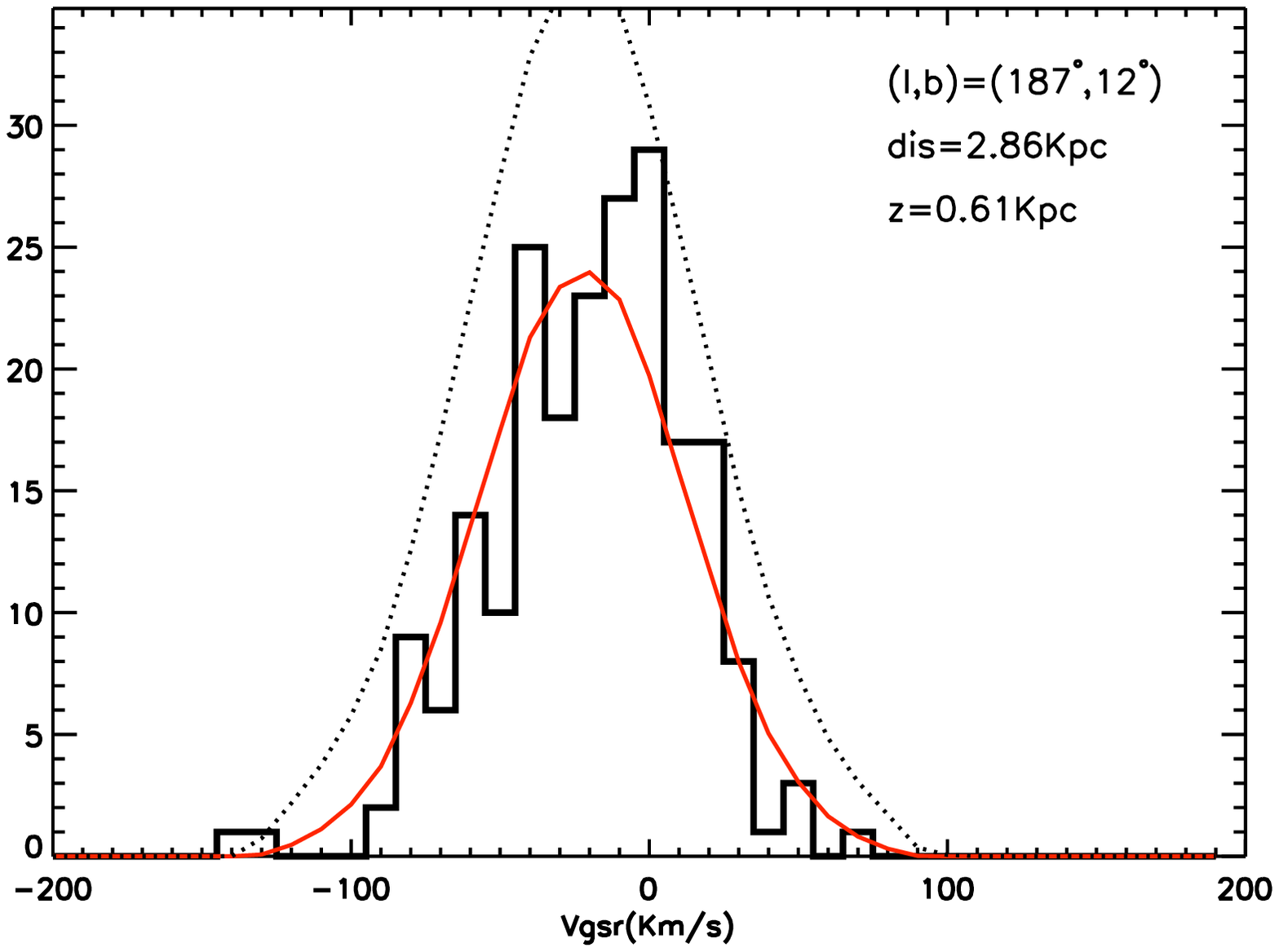}
\includegraphics[scale=0.5]{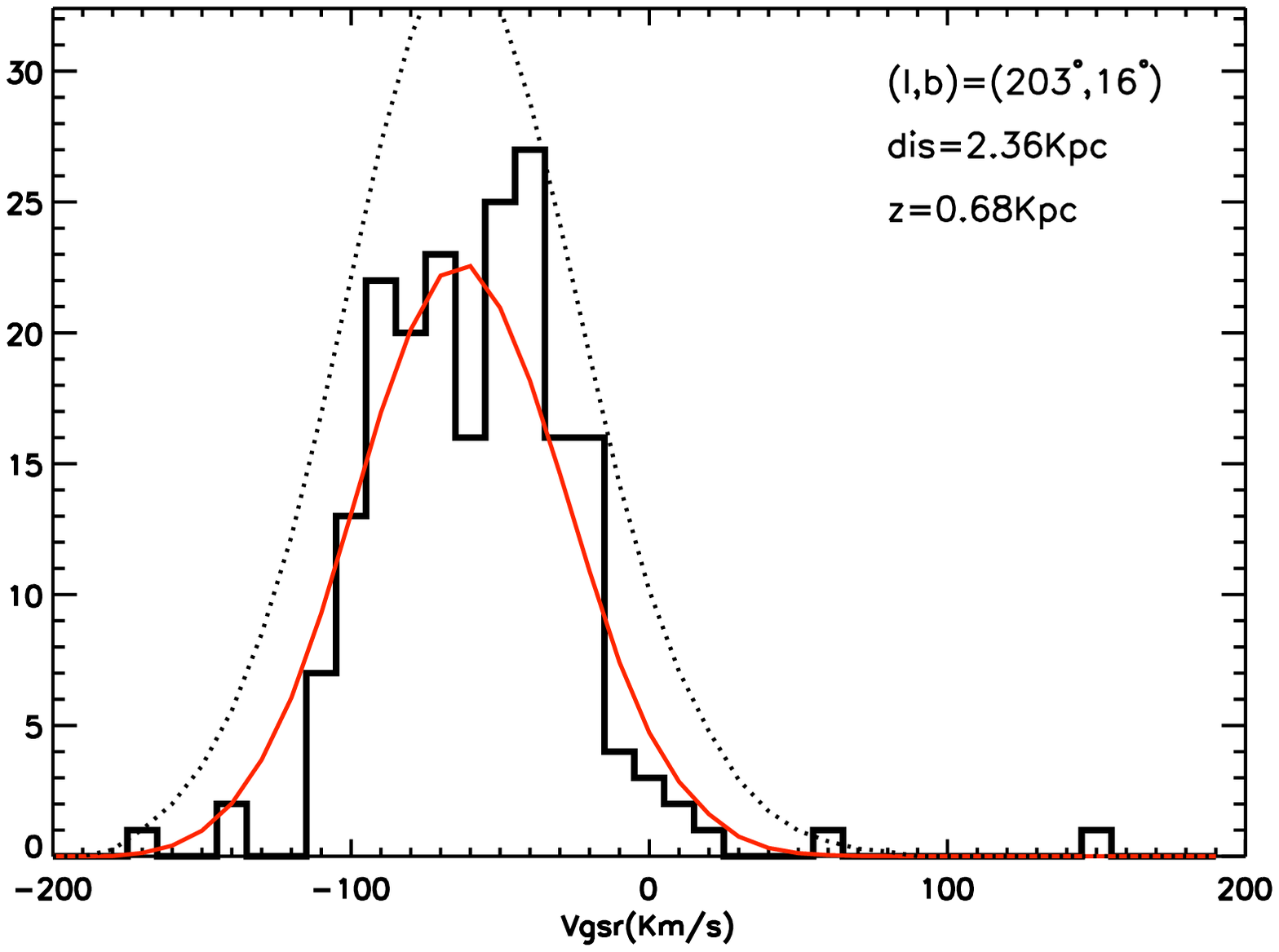}
\includegraphics[scale=0.5]{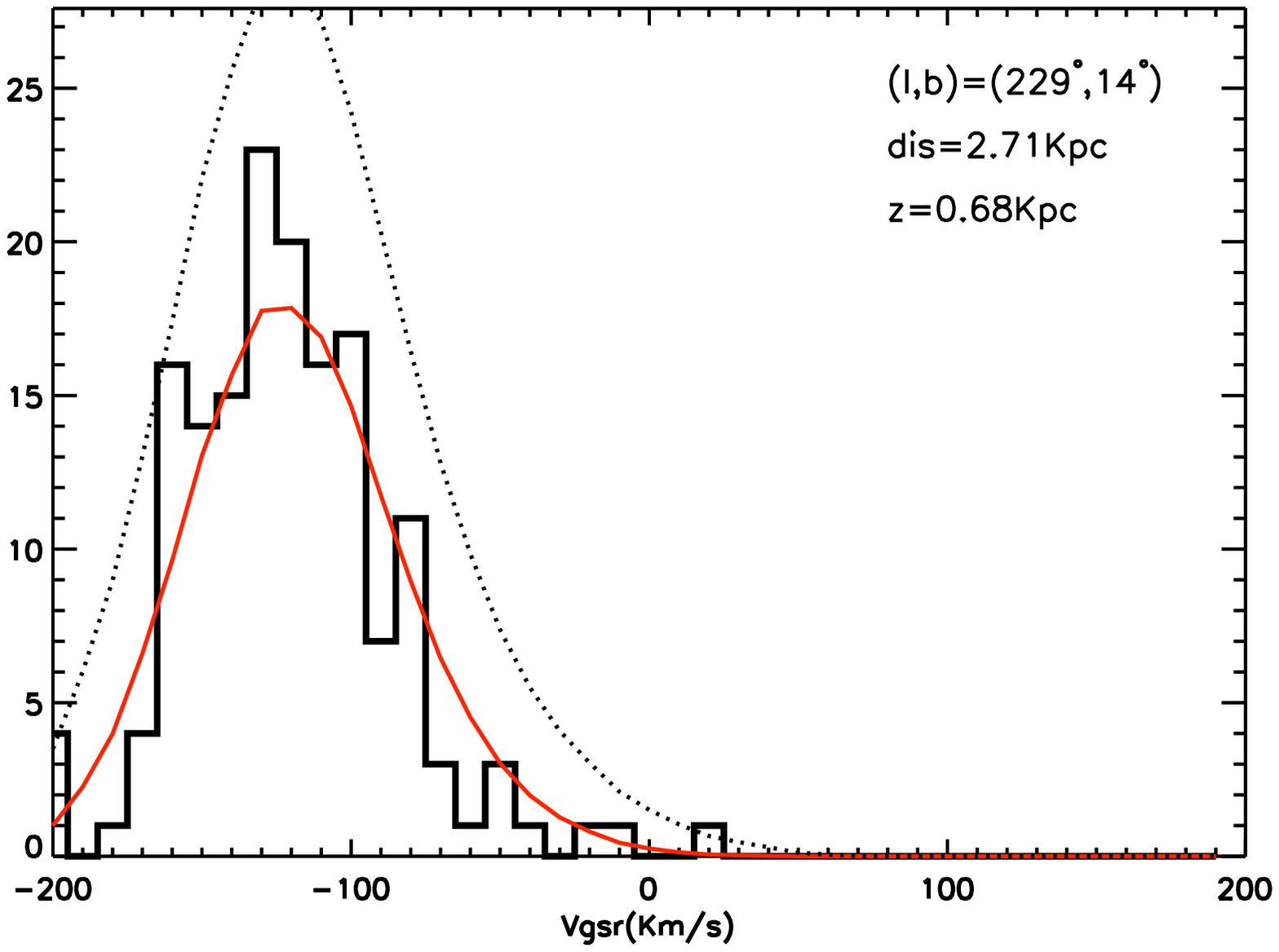}
\caption[vgsrdistralph] {
\footnotesize
$V_{\rm gsr}$ distribution of stars with metallicities of disk stars in the north near structure, fit  with the result of a dynamics equation from Sch{\"o}nrich \& Binney (2012), which includes the asymmetric drift.  The black histogram is the $V_{\rm gsr}$ distribution of data higher metallicity stars that are likely to be thin and thick disk members, while the red curve is the expected $V_{\rm gsr}$ distribution.  The black dotted curve indicates where a 2.5 sigma overdensity would lie.  Note that the fit to the velocity distribution is significantly improved when asymmetric drift is included in the model.
}\label{vgsrdistralph}
\end{figure}

\clearpage
\begin{figure}
\includegraphics[scale=0.5]{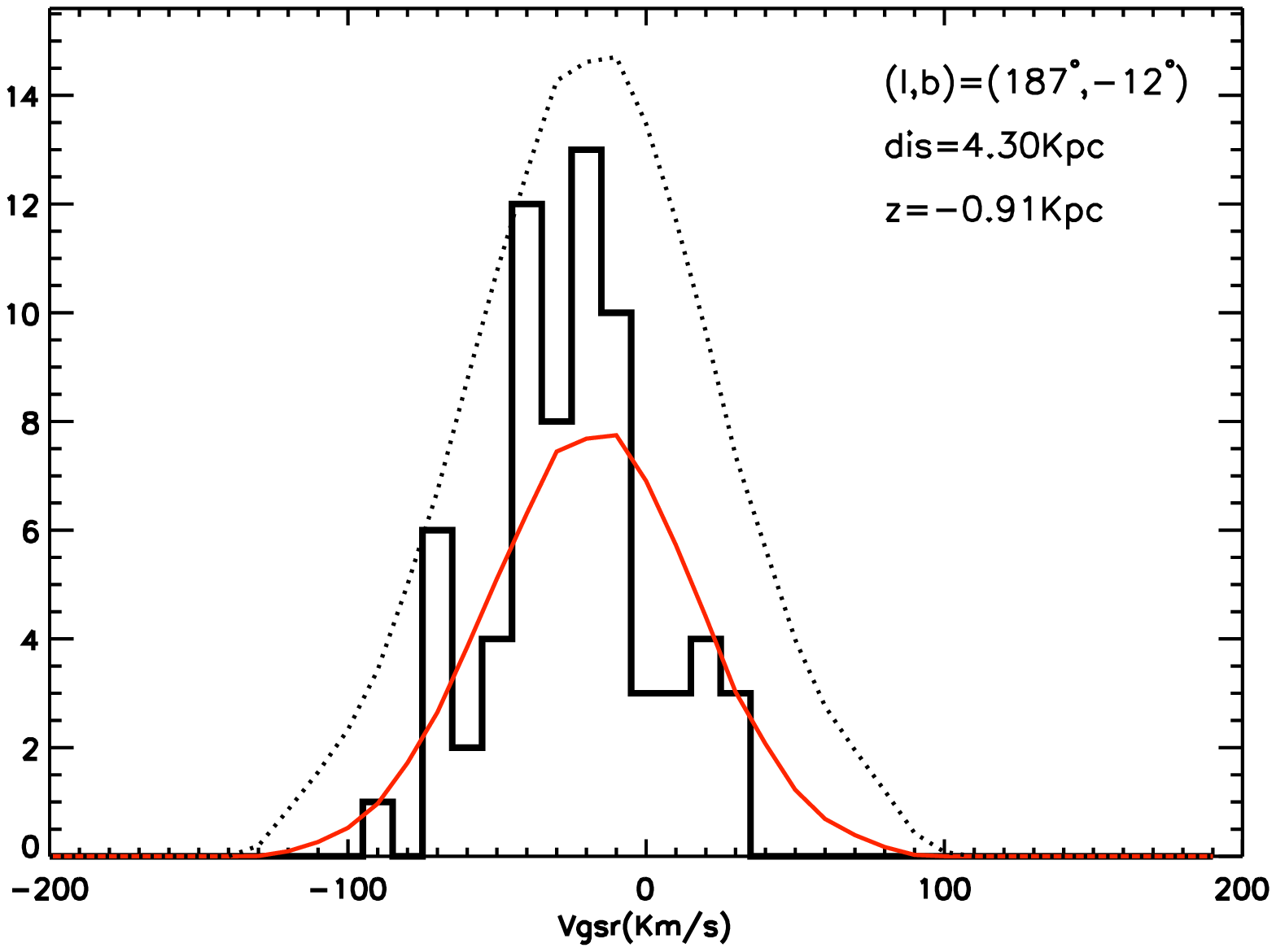}
\includegraphics[scale=0.5]{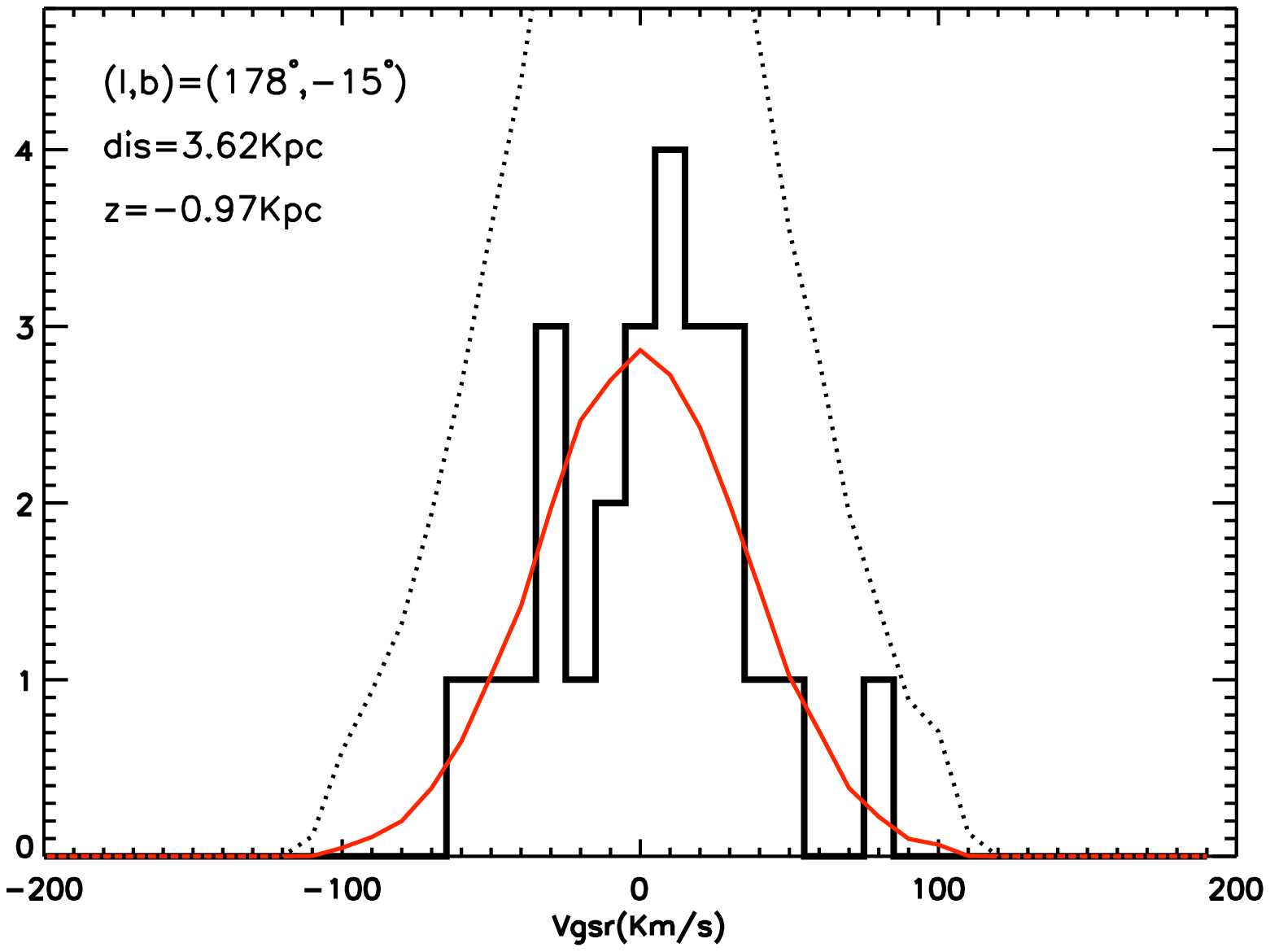}
\includegraphics[scale=0.5]{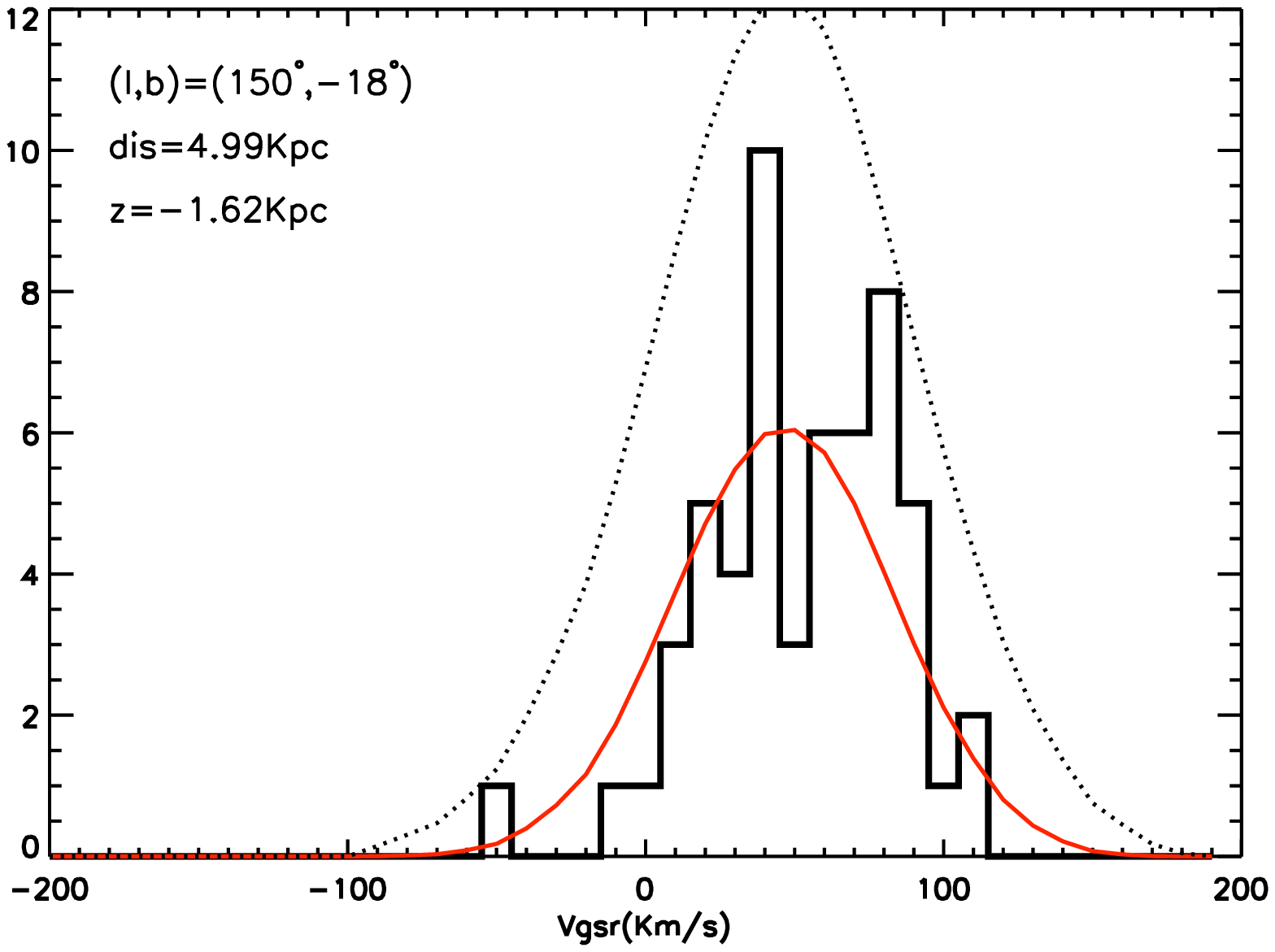}
\caption[vgsrdistralphsouth] {
\footnotesize
$V_{\rm gsr}$ distribution of stars with metallicities of disk stars in the south middle structure, fit  with the result of a dynamics equation from Sch{\"o}nrich \& Binney (2012), which includes the asymmetric drift.  There are only three fields in the south that are close enough to the plane that we can use the parameters in the published paper to generate a model to compare with the data.  Attempts to extrapolate the parameters to larger distances from the Galactic plane were not successful in fitting the observed velocity distributions.
}\label{vgsrdistralphsouth}
\end{figure}

\begin{figure}
\includegraphics[scale=1.0]{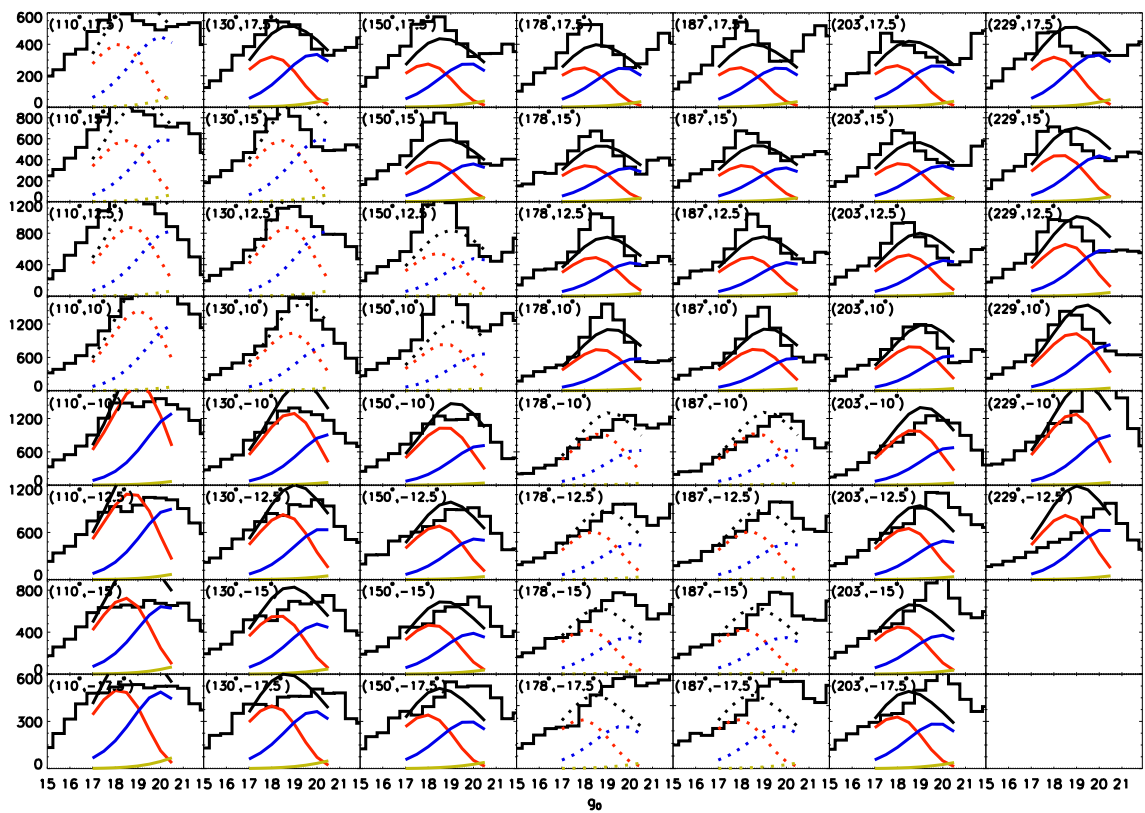}
\caption[modelcountsnw] {
\footnotesize
Apparent magnitude distrubution of early K-type stars compared to a standard Galactic model. The black histograms show the number of K stars as a function of apparent magnitude in each plate.  The red curve is theoretical thin disk star counts; the blue curve is theoretical thick disk star counts; the yellow curve is theoretical halo star counts; and the black curve is the sum of the three theoretical star counts.  Panels where these curves are shown as dashed lines were not used in finding the best fit disk parameters, since these regions are of relatively high extinction.  Note that this simple model is a very poor fit to the data.
}\label{modelcountsnw}
\end{figure}

\begin{figure}
\includegraphics[scale=1.0]{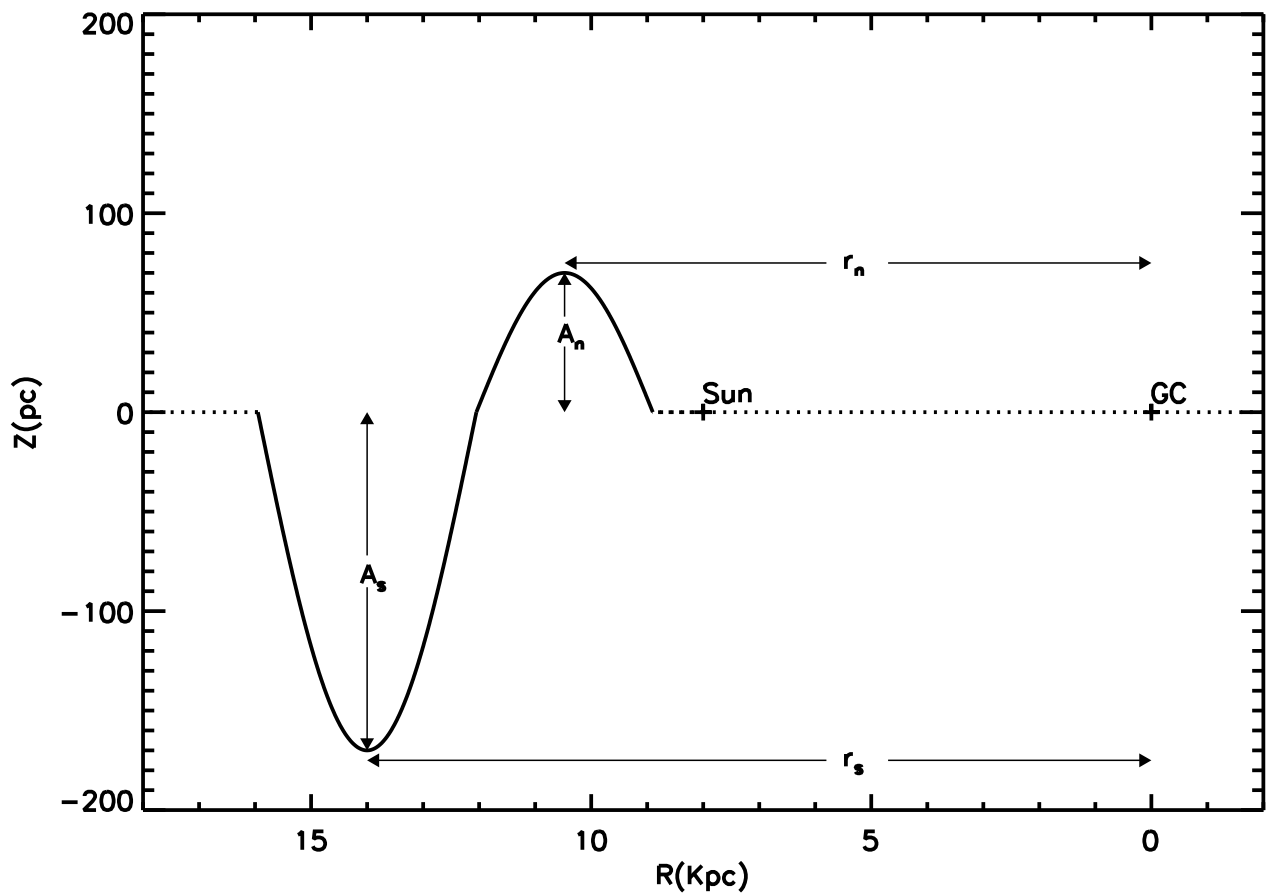}
\caption[toymodel] {
\footnotesize
Schematic of the ``oscillations" model which is used to fit star counts of K stars in Figure 19.  The disk interior to 8.9 kpc is unperturbed.  Between 8.9 kpc and 12.1 kpc from the Galactic center, the midplane disk is perturbed up in a sinusoidal pattern, to a maximum of 70 pc above its nominal position.  Between 12.1 kpc from the Galactic center and 16 kpc from the Galactic center, the midplane of the disk is perturbed down in a sinusoidal pattern, to a maximum of 170 pc below the plane.  Although the figure gives the impression that the Sun is located in the Galactic plane, the model included the location of the Sun 27 pc above the plane.  The data that is being fit is primarily in the range $8.9<d<16$ kpc from the Sun, in the perturbed section of the disk model.
}\label{toymodel}
\end{figure}

\begin{figure}
\includegraphics[scale=1.0]{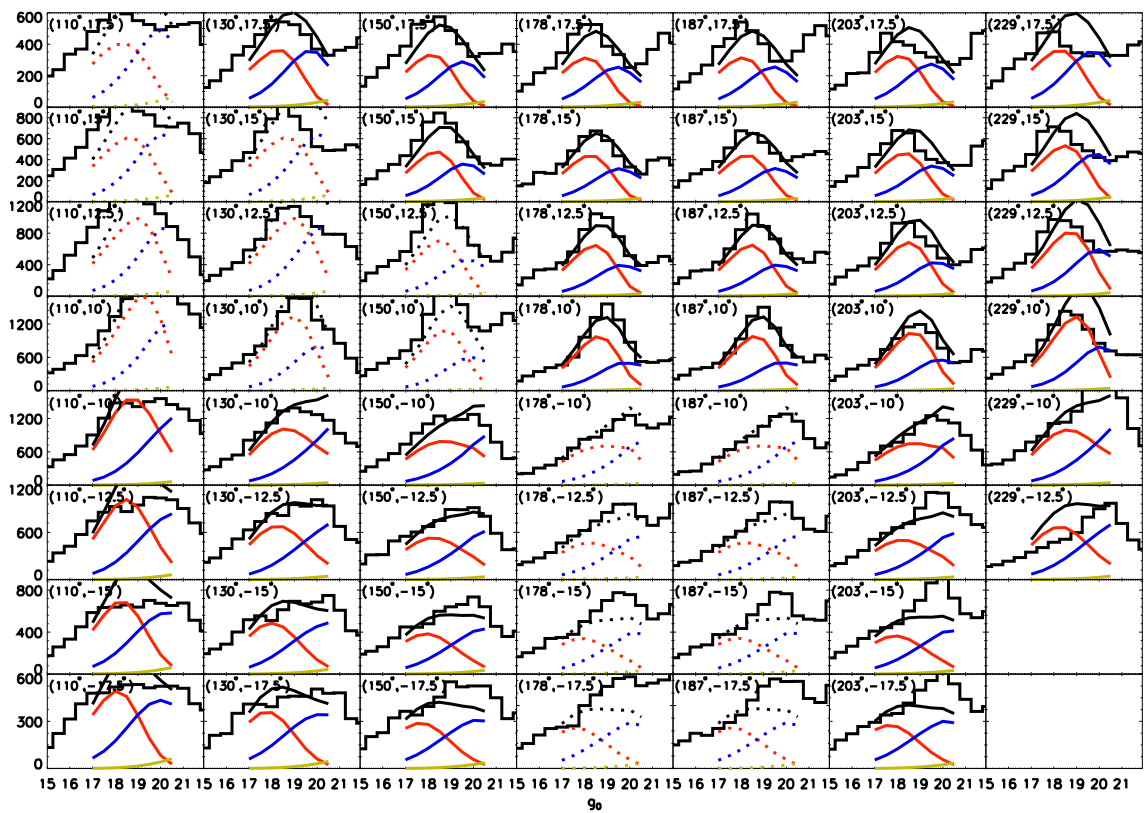}
\caption[modelcounts] {
\footnotesize
Star counts of early K-type stars compared to a perturbed Galactic model with the disk midplane oscillating up and down as shown in Figure 18.  The colored curves have same meaning as those of Figure 17.  Note that the stars near the anticenter in the north Galactic cap are now fit very well.  The histograms of star counts are very different for points that are at symmetric positions around the Galactic anticenter (with the same Galactic latitude).  This tells us why were are unable to obtain better fits; our model was manifestly axially symmetric.
}\label{modelcounts}
\end{figure}

\end{document}